\documentclass[prmaterials,psfig,pstricks,twocolumn,preprintnumbers,amsmath,amssymb,superscriptaddress]{revtex4}
\usepackage{graphicx}          %% Include figure files
\usepackage{dcolumn}           %% Align table columns on decimal point
\usepackage{bm}                %% bold math

\usepackage{array,booktabs,lmodern}
\usepackage[scaled=1.00]{helvet}
\usepackage[final]{listings,microtype}
\usepackage[T1]{fontenc}
\usepackage[version=3]{mhchem} % Formula subscripts using \ce{}
\usepackage{siunitx}
\usepackage[english]{babel}
\usepackage{float}
\usepackage{natbib}
\usepackage{setspace}
\usepackage{xkeyval}
\usepackage{natmove}
\usepackage{titlesec}

\usepackage{xfrac}
\usepackage{relsize}
\usepackage{amsmath}

\usepackage{braket}
\usepackage{amssymb}
\usepackage{hyperref}
\hypersetup{
    colorlinks=true,
    urlcolor= blue,
    citecolor=blue,
    linkcolor= blue,
    bookmarksopen=false,
    }

\begin{document}

\preprint{Page}

\title{Outstanding thermoelectric performance predicted for out-of-plane p-doped GeSe} 

\author{Anderson S. Chaves}
%\affiliation{John A. Paulson School of Engineering and Applied Sciences, Harvard University, Cambridge, Massachusetts, 02138, USA}
\affiliation{Gleb Wataghin Institute of Physics, University of Campinas, PO Box 13083-859, Campinas, SP, Brazil}
\affiliation{John A. Paulson School of Engineering and Applied Sciences, Harvard University, Cambridge, Massachusetts, 02138, USA}
\author{Daniel T. Larson}
\affiliation{Department of Physics, Harvard University, Cambridge, Massachusetts, 02138, USA}
\author{Efthimios Kaxiras}
\affiliation{Department of Physics, Harvard University, Cambridge, Massachusetts, 02138, USA}
\affiliation{John A. Paulson School of Engineering and Applied Sciences, Harvard University, Cambridge, Massachusetts, 02138, USA}
\author{Alex Antonelli}
\affiliation{Gleb Wataghin Institute of Physics and Center for Computing in 
Engineering \& Sciences, University of Campinas, PO Box 13083-859, Campinas, SP, Brazil}

\date{\today}

\begin{abstract}

%\keywords{Thermoelectrics, GeSe, first-principles calculations}

The record-breaking thermoelectric performance of tin selenide (SnSe) 
has motivated the investigation of analogue compounds with the same structure.  
A promising candidate that emerged recently is germanium selenide (GeSe). 
Here, using extensive first-principles calculations of the hole-phonon and 
hole-impurity scattering, we investigate the thermoelectric 
transport properties of the orthorhombic phase of p-doped GeSe. 
We predict outstanding thermoelectric performance 
for GeSe over a broad range of temperatures 
due to its  
high Seebeck coefficients, extremely low Lorenz numbers,
ultralow total thermal conductivity,
and relatively large band gap.
In particular, the out-of-plane direction in GeSe presents equivalent or even higher performance
than SnSe for temperatures above 500~K.
By extending the analysis to 900~K, we obtained an ultrahigh value for the thermoelectric figure of merit ($zT =$ 3.2) 
at the optimal hole density of 4$\times 10^{19}$~cm$^{-3}$.
Our work provides strong motivation for continued experimental work 
focusing on improving the GeSe doping efficiency 
in order to achieve this optimal hole density.

\end{abstract}

%%%%%%%%%%%%%%%%%%%%%%%%%%%%%%%%%%%%%%%%%%%%%%%%%%%%%%%%%%%%%%%%%%%%%
%% Start the main part of the manuscript here.
%%%%%%%%%%%%%%%%%%%%%%%%%%%%%%%%%%%%%%%%%%%%%%%%%%%%%%%%%%%%%%%%%%%%%

\maketitle

\section{Introduction}

High-efficiency thermoelectric (TE) materials 
have been systematically and comprehensively 
investigated during the past several decades, 
mainly due to their capability of functioning as 
all-solid-state modules for distributed spot-size refrigeration\cite{mao2020thermoelectric,mao2019high} or  
electric power generation from waste heat.\cite{disalvo1999thermoelectric,he2017advances} 
The key quantity to evaluate the efficiency of TE energy conversion 
is the dimensionless figure of merit, $zT = \sigma S^2 T/\kappa_{tot}$,
where $\sigma$, $S$, $T$ stand for the electrical conductivity, the Seebeck coefficient, and 
the absolute temperature and $\kappa_\mathrm{tot} = \kappa_\mathrm{latt} + \kappa_\mathrm{h}$
is the total thermal conductivity, comprised of lattice and
electronic carrier contributions, respectively.
Numerous TE materials have been discovered, and a few of them even have $zT$ values between 2 and 3\cite{lv2013optimized,venkatasubramanian2001thin,zhang2016pressure,zhong2014high,famili2018toward,xue2016laptsb,ohta2007giant,fu2016enhanced,zhao2014ultralow,zhao2015ultrahigh,chang20183d}. 
However, even those 
high-$zT$ materials do not have sufficient efficiency to be 
largely employed by industry\cite{disalvo1999thermoelectric,vining2009inconvenient,he2017advances}.
Indeed, it has been argued that materials with $zT > 3$ 
would represent a highly attractive prospect for applications, allowing TE refrigerators to compete with traditional 
compressor-based refrigerators\cite{mao2020thermoelectric,sales1996filled}. 
Such ultrahigh $zT$ values have not been measured in bulk materials until very recently.\cite{zhou2021polycrystalline}  
Thus, TE materials have so far only found niche applications where 
reliability is of higher priority than efficiency.

%% There are two ways to enhance $zT$, namely to minimize $\kappa_\mathrm{tot}$, 
%% or to maximize the power factor ($PF = \sigma S^2$). 
The figure of merit, $zT$, can be enhanced either by increasing the power factor ($PF=\sigma S^2$) or reducing the thermal conductivity, $\kappa_\mathrm{tot}$.
Ultimately, the main goal is to find TE materials that 
satisfy both of these conditions simultaneously, which is a challenge 
since the properties involved are interdependent.  
The maximization of $PF$ relies on band-structure
engineering\cite{pei2011convergence,pei2012band,dehkordi2015thermoelectric}
such as increasing band degeneracy through convergence of 
bands~\cite{zhao2013high,liu2012convergence} or taking advantage
of band structure anisotropy\cite{parker2015benefits} and non-parabolicity~\cite{chen2013importance}.
On the other hand, 
the main strategies to minimize $\kappa_\mathrm{tot}$ include  
identifying materials with intrinsically low $\kappa_\mathrm{latt}$\cite{gonzalez2018ultralow},
minimizing the electronic carrier contribution, $\kappa_\mathrm{h}$, through
the minimization of the Lorenz function\cite{mckinney2017search,chaves2021microscopic},
or by alloying or nanostructuring procedures\cite{hochbaum2008enhanced,boukai2008silicon,kanatzidis2009nanostructured,zhao2013high,vineis2010nanostructured}.
Despite the challenges, impressive achievements have been obtained 
on the basis of such strategies
\cite{he2017advances,biswas2012high,liu2012copper,fu2016enhanced,olvera2017partial,cheng2017new,ma2020alpha,roychowdhury2021enhanced}.
 
\begin{figure*}
        \centering
        \includegraphics[width=0.60\textwidth]{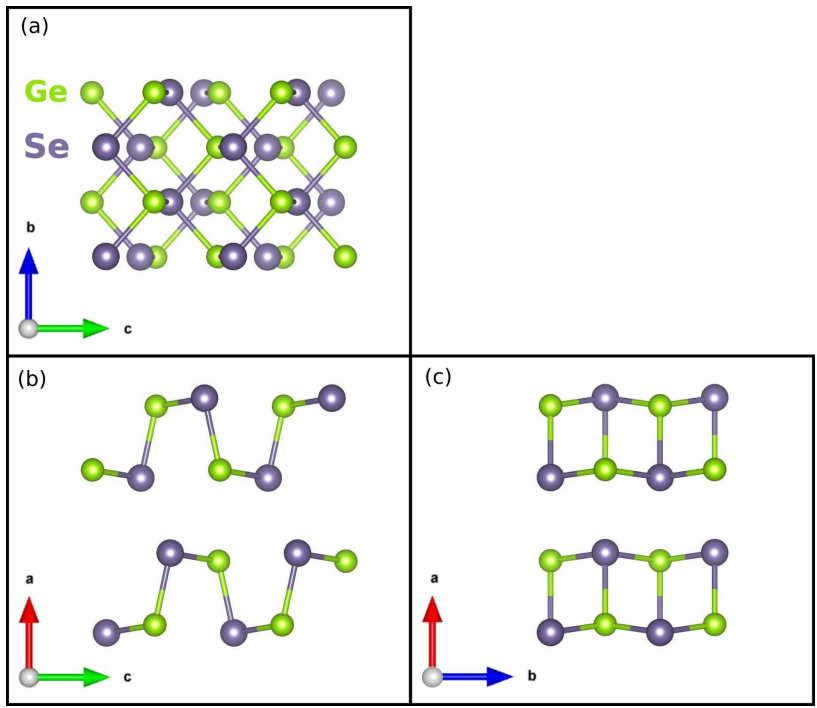}
        \caption{Crystal structure of \textit{Pnma} GeSe with views along each crystallographic direction: (a) perpendicular to 
                 the layer plane, (b,c) along the layer planes.  
                 Green and gray spheres represent Ge and Se, respectively.}
        \label{fig:structure}
\end{figure*}

The record-breaking TE performance of SnSe\cite{zhao2014ultralow,zhao2015ultrahigh,chang20183d}
has motivated the investigation of analogue
IV-VI compounds with the same puckered layer structure, in order to ascertain whether
such systems also possess inherently low $\kappa_\mathrm{latt}$ and high $zT$.
A promising candidate is germanium selenide (GeSe), which, like SnSe, crystalizes in the orthorhombic GeS-type structure shown in Fig.~\ref{fig:structure}, with a space group of $D_{2h}^{16}$ (\textit{Pnma})\cite{taniguchi1990core,okazaki1958crystal}.

Additional advantages that make GeSe very attractive for large applications
in thermoelectrics and photovoltaics include chemical stability, earth-abundance,
environmental compatibility and low toxicity (no lead).\cite{antunez2011tin,liu2017investigation,huang2017high,sarkar2020ferroelectric,sarkar2021metavalent}
Recently, theoretical work by Ding {\it{et al.}}\cite{ding2015high} put forward the possibility of achieving
large $S$ and $PF$ values by proper p- or n-type doping of GeSe. 
Due to low thermal conductivity
and multiband effects, Hao {\it{et al.}}\cite{hao2016computational}
predicted an ultrahigh peak $zT$ value of 2.5 along the in-plane (b-axis) direction of the orthorhombic phase of GeSe with a hole density of 
6.5$\times10^{19}$~cm$^{-3}$ at 800~K.
If this could be realized, it would outperform p-doped SnSe. 
%% This theoretical result was based on significant 
%% approximations that limit its predictive capability. In particular, 
%% were assumed 
Though highly suggestive, that prediction used
the same values for relaxation times and carrier densities as those reported for p-doped SnSe.
Despite the great potential of the orthorhombic phase of GeSe for
TE applications, there are still relatively few published experimental results\cite{zhang2016thermoelectric}.

In the present work we examine the thermoelectric performance of p-doped GeSe and SnSe
within the Boltzmann Transport equation (BTE) formalism,  
by explicitly calculating relaxation times due to hole-phonon (h-p) and hole-impurity couplings  
using a comprehensive first-principles approach. 
In particular, the hole-phonon coupling was calculated by using the dual interpolation scheme\cite{chaves2020boosting}
of the density functional theory (DFT) band structure\cite{hohenberg1964inhomogeneous,kohn1965self}.
The phonon dispersion and h-p matrix elements were determined by density functional perturbation theory (DFPT)\cite{baroni2001phonons}. 
The carrier density for the different axes was derived from the 
record-breaking transport data measured in p-doped SnSe\cite{zhao2015ultrahigh}. 
The calculated temperature and energy 
dependent relaxation times allow for a deeper understanding of the microscopic processes underlying 
the temperature-dependent transport phenomena in p-doped GeSe and SnSe.

Our results predict a very high figure of merit for both 
out-of-plane (a-axis) and in-plane (c-axis) GeSe in a broad range of 
temperatures. This can be attributed to several factors that synergistically influence the performance:
high Seebeck coefficients, extremely low Lorenz numbers, 
low hole thermal conductivities, very low lattice thermal conductivity,
%% intrinsic anharmonicity\cite{hao2016computational}, and 
and a relatively large band gap. 
In fact, for temperatures above 500~K, out-of-plane GeSe is predicted to potentially have a higher $zT$ 
than the record-breaking SnSe. 
By extending the analysis to 900~K, we obtain
the outstanding $zT$ values of 3.2 and 2.8 for the out-of-plane and in-plane directions with 
optimal carrier densities of 4$\times 10^{19}$~cm$^{-3}$ and 5$\times 10^{19}$~cm$^{-3}$, respectively. 
We find that the total relaxation time for the out-of-plane direction in GeSe
is much higher than the corresponding relaxation time in SnSe, demonstrating the importance of directly calculating the relaxation times for GeSe.
%% instead of approximating them with SnSe relaxation times\cite{hao2016computational}.
%crystallographic axis, our calculated TE properties for the out-of-plane direction in GeSe
%could not be foreseen if instead
%the relaxation times for SnSe were used in the calculations, as it has been previously done

\section{Theoretical Approach}

The hole-phonon (h-p) coupling and the scattering of holes
by ionized impurities are the microscopic processes that determine
the temperature-dependent p-type transport
phenomena in TE materials, such as GeSe and SnSe.
We calculate these TE transport properties from first-principles using the many-body perturbation
theory of the h-p interaction following the Fan-Migdal approach
and the Boltzmann transport formalism. %% beyond the constant relaxation time (RT).
The comprehensive theoretical framework for the calculation of the band ($n$) and
momentum ($\bf{k}$) resolved relaxation time (RT), $\tau_{n,\bf{k}}$, is described in
detail in our previous work~\cite{chaves2021microscopic} 
and summarized in the Supplemental Material (SM)\cite{SM} for easy reference. 
In brief, we calculate three contributions to the total relaxation
time. The nonpolar RT ($\tau_\mathrm{npol}$) comes from the short-range portion
of the hole coupling to acoustic and optical phonons, which can be calculated using dual interpolation.
The long-range portion of the hole coupling with optical phonons gives rise to the polar RT ($\tau_\mathrm{pol}$), which we
determine using the analytic Vogl formula~\cite{pellegrini2016physics,bostedt2016linac,vogl1976microscopic}
with the addition of Ehrenreich screening~\cite{ehrenreich1959screening}.
Finally, extrinsic scattering by ionized impurities ($\tau_\mathrm{imp}$) 
is calculated using the theory developed by Brooks and Herring (B-H)~\cite{brooks1955theory,chattopadhyay1981electron}, 
which has been extended to go beyond the parabolic band approximation.\cite{chaves2021investigating}

Assuming these scattering mechanisms can be treated independently,
the total RT is determined by Mathiessen's rule, where the dependence of scattering times on independent variables, including temperature ($T$) and chemical potential ($\mu$), is shown explicitly:
\begin{flalign}
\label{Mathiessen}
\frac{1}{\tau_\mathrm{tot}(n,{\bf{k}},\mu,T)} &= \frac{1}{\tau_\mathrm{npol}(n,{\bf{k}},T)} + \frac{1}{\tau_\mathrm{pol}(n,{\bf{k}},\mu,T)} \\
&+ \frac{1}{\tau_\mathrm{imp}(n,{\bf{k}},\mu,T)}~.&&\nonumber
\end{flalign}

From $\tau_\mathrm{tot}$ we calculate the TE transport coefficients using the
semiclassical BTE with the (non-constant) relaxation time approximation (RTA).\cite{madsen2006boltztrap,chaves2021investigating}
The key quantity is the momentum- and band-resolved transport distribution kernel, 
\begin{equation}
\label{sigmax}
\Sigma_{\alpha,\beta} (n,{\bf{k}},\mu,T) = e^2\tau_{n,{\bf{k}}}(\mu,T) v_{\alpha}(n,{\bf{k}}) v_{\beta} (n,{\bf{k}})~,
\end{equation}
where $\tau_{n,{\bf{k}}}(\mu,T) \equiv \tau_\mathrm{tot}$ is the total relaxation time and $v(n,{\bf{k}})$ is the
average group velocity. 
The energy projected transport function can then be defined as 
\begin{equation}
\label{kernel}
\Sigma_{\alpha,\beta} (\epsilon,\mu,T) = \frac{1}{N_{{\bf{k}}}} \sum_{\bf{k}}{\Sigma_{\alpha,\beta} (n,{\bf{k}},\mu,T)}\frac{\delta(\epsilon - \epsilon_{n,{\bf{k}}})}{d\epsilon}~,
\end{equation}
and is used to calculated the transport tensors
in terms of the different energy moments 
\begin{equation}
\label{tensor}
I^{(n)}_{\alpha,\beta}(T,\mu)= \frac{1}{\Omega}\int\Sigma_{\alpha,\beta}(\epsilon,\mu,T)(\epsilon - \mu)^{n}\left(-\frac{\partial f_{\mu}(\epsilon,T)}{\partial \epsilon}\right)d\epsilon~. 
\end{equation}
With the experimental conditions of zero temperature gradient ($\nabla T = 0$)
and zero electric current, the transport tensors
yield the electrical conductivity, 
\begin{equation}
\label{cond}
\sigma \equiv \sigma_{\alpha,\beta}(T,\mu) = I^{(0)}_{\alpha,\beta}(T,\mu)~,
\end{equation}
the Seebeck coefficient, 
\begin{equation}
\label{seebeck}
S \equiv S_{i,j}(T,\mu) = (eT)^{-1}I^{(1)}_{\alpha,i}(T,\mu)/I^{(0)}_{\alpha,j}(T,\mu)~,
\end{equation}
and the charge carrier contribution to the thermal conductivity,
%\begin{multline}
%\label{Mathiessen}
% \begin{split}
%& F_\mathrm{pol}(n,{\bf{k}}) = \left[1 -\frac{1}{2(r_{\infty}(n,{\bf{k}})\cdot{\bf{k}})^2}\right.
%\\
%& \left.\times \ln[1+4(r_{\infty}(n,{\bf{k}})\cdot{\bf{k}})^2] + \frac{1}{1+4(r_{\infty}(n,{\bf{k}})\cdot{\bf{k}})^2}\right]^{-1}~.
%\end{split} 

%\end{multline}
%\begin{multline}
%\label{kappa}
%\begin{split}
%& \kappa_\mathrm{h} \equiv \kappa^{h}_{i,j}(T,\mu) = (e^{2}T)^{-1} \left(I^{(2)}_{i,j}(T,\mu)\right.
%\\
%& \left. -~I^{(1)}_{i,\alpha}(T,\mu)\cdot{I^{(0)}_{\beta,\alpha}(T,\mu)^{-1}}\cdot{I^{(1)}_{\beta,j}(T,\mu)}\right)~.
%\end{split}
%\end{multline}

\begin{flalign}
\label{kappa}
\kappa_\mathrm{h} \equiv \kappa^{h}_{i,j}(T,\mu) &= (e^{2}T)^{-1} \left(I^{(2)}_{i,j}(T,\mu)\right. \\
 &\left. -~I^{(1)}_{i,\alpha}(T,\mu)\cdot{I^{(0)}_{\beta,\alpha}(T,\mu)^{-1}}\cdot{I^{(1)}_{\beta,j}(T,\mu)}\right)~.&&\nonumber
\end{flalign}

\section{Computational details}

%% Here we briefly describe the computational details of the calculations of the TE properties of GeSe, 
Below are the details for calculations involving GeSe;
the details for SnSe can be found in Ref.~\citenum{chaves2021microscopic}.
%, where the lattice parameters 
%are: $a = 11.79\:$\AA, $b = 4.22$ \AA{} and $c = 4.52$ \AA{}.   
The relaxed geometry and electronic structure of GeSe was calculated using DFT,
while the phonon dispersions and h-p matrix elements were calculated using DFPT, both implemented in the Quantum Espresso package\cite{giannozzi2009quantum}.
We employed fully-relativistic optimized norm-conserving Vanderbilt pseudopotentials\cite{hamann2013optimized,van2018pseudodojo}
within the generalized gradient approximation (GGA)
for the exchange-correlation functional according to the formulation of Perdew-Burke-Ernzerhof (PBE)\cite{perdew1996generalized}.
Monkhorst-Pack grids of 6$\times$18$\times$14 for ${\bf{k}}$-point sampling and
a kinetic energy cutoff of 80 Ry were employed to ensure the
convergence of the total energy in DFT calculations. 
The energy convergence threshold for the total energy difference  
between two successive self-consistency steps was 10$^{-11}$~Ry 
under the Davidson-type diagonalization method.
Because DFT-GGA calculations underestimate the GeSe band gap,
a scissor operator was used to rigidly shift the conduction bands upwards
in order to attain the experimental band gap of 1.1~eV~\cite{elkorashy1989photoconductivity,vaughn2010single}.
%% which is underestimated by DFT-GGA calculations.

At room temperature both GeSe and SnSe crystallize in a layered orthorhombic structure
with the \textit{Pnma} space group and 8 atoms in the unit cell, shown in Fig.~\ref{fig:structure}.
The melting point of GeSe occurs at 948$\pm$2~K,\cite{ross1969germanium,ipser1982germanium,bletskan2005phase}
but according to Wiedemeier {\it{et al.}}\cite{wiedemeier1975thermal}, at 924K a structural transition
takes place from the orthorhombic phase to the ideal structure of NaCl type. 
This is controversial since according to Sist {\it{et al.}}\cite{sist2017high} 
this structural phase transition occurs at the
lower temperature of 907~K.
For SnSe, a second-order phase transition to the higher symmetry \textit{Cmcm}
phase occurs at T$\sim$810~K\cite{li2015orbitally}.
In the present work we consider only \textit{Pnma} orthorhombic structures for
both materials and thus report their transport properties for temperatures
up to 807 and 900~K for SnSe and GeSe, respectively.

Both materials form covalently bonded layers with zig-zag chains along the b-axis
and significant corrugation along the c-axis. Those layers are held together by 
much weaker van der Waals interactions along the out-of-plane a-axis. In order to capture
such weak bonds between layers, we employed van der Waals corrections to DFT
according to the D3 approach of Grimme {\it{et al.}}~\cite{grimme2010consistent}.
For GeSe, we started from the \textit{Pnma} orthorhombic configuration
from the Materials Project\cite{jain2013commentary} (mp-700) and relaxed the lattice parameters and atomic positions until
all atomic force components were smaller in magnitude than 1~meV/\AA.
The relaxed lattice constants are $a = 11.02\:$\AA, $b = 3.58\:$\AA~and $c = 4.79\:$\AA,
which are in reasonable agreement with the corresponding experimental values\cite{wiedemeier1975thermal,wiedemeier1978refinement}.
Importantly, our DFT-D3 calculations accurately reproduce
the out-of-plane lattice constant (a-axis), differing from the experimental result
at 919~K by only $\sim$0.1\%.\cite{wiedemeier1975thermal}
\begin{figure*}[ht]
        \centering
        \includegraphics[width=0.40\textwidth]{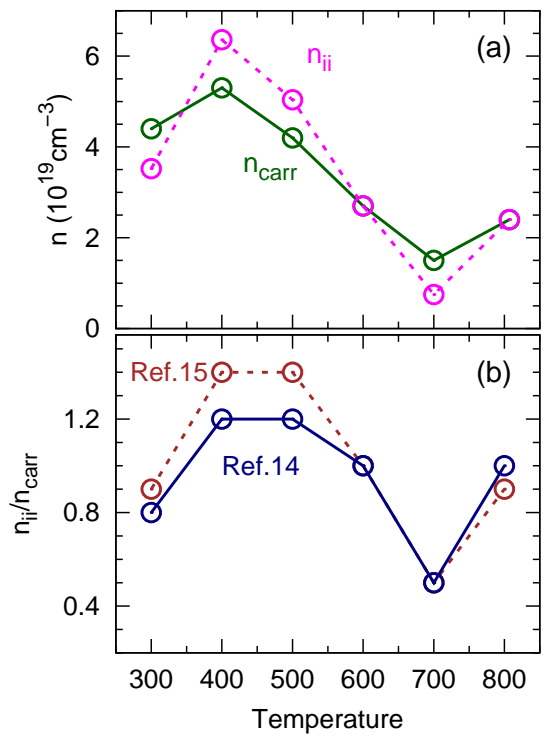}
        \caption{$(a)$ Carrier density, $n_\mathrm{carr}$ (green solid line), and ionized impurity concentration, $n_\mathrm{ii}$
          (dashed magenta line), determined by
          self-consistently matching calculations to the experimental results for a-axis p-doped SnSe reported in
          Zhao {\it{et al.}}\cite{zhao2015ultrahigh}. $(b)$ The ratio of values in (a), $n_\mathrm{ii}/n_\mathrm{carr}$
          (blue solid line), along with the same ratio as calculated in our previous work
          (brown dashed line)~\cite{chaves2021microscopic}, derived from experimental results reported in
          Chang {\it{et al.}}\cite{chang20183d} for the same axis.}
        \label{n_carr}
\end{figure*}

We used our Turbo-EPW implementation\cite{chaves2020boosting} 
to calculate the RTs limited by h-p coupling, including both contributions of
nonpolar and screened polar scatterings.
Turbo-EPW takes advantage of the dual interpolation technique
based on a first Wannier-Fourier interpolation\cite{giustino2007electron,ponce2016epw}, 
followed by a second interpolation using symmetry-adapted 
star functions, which allows for efficient interpolation of h-p
scattering matrix elements onto dense meshes of 
electron (${\bf{k}}$) and phonon (${\bf{q}}$) wave vectors.
In the present case, the first interpolation, using maximally localized Wannier functions determined
by Wannier90\cite{mostofi2008wannier90}, leads to a phonon grid of 10$\times$40$\times$20 ${\bf{q}}$ points. 
The calculated electron self energy only changes by $\sim2\%$ for a ninefold
denser ${\bf{q}}$ grid, 
indicating that our momentum sampling gives a well converged balance between accuracy and computational cost.   
Subsequently, $M=10$ star functions per ${\bf{k}}$ point were used
for the second interpolation, 
resulting in a denser grid of 27$\times$81$\times$64 ${\bf{k}}$ points. In total this results in over 1 billion ${\bf{k/q}}$ pairs.

The calculation of $\tau_\mathrm{imp}$ requires the static and high-frequency 
dielectric constants, $\zeta_0$ and $\zeta_\infty$. 
We have used the experimental values\cite{madelung2004semiconductors,chandrasekhar1977infrared} $\zeta_0 =$ 21.9, 30.4, 25.8 (45, 62, 42) 
and $\zeta_{\infty} =$ 18.7, 21.9, 14.4 (13, 17, 16) 
for the a-, b- and c-axis of GeSe (SnSe), respectively. 
We used the same value of $M=10$ star functions in the calculation of $\tau_\mathrm{imp}$ 
in order to obtain the same mesh for integration as in h-p calculations. 
Mathiessen's rule, Eq.~(\ref{Mathiessen}), yields the temperature and energy dependent $\tau_\mathrm{tot}$ that is used in our modified BoltzTraP code\cite{madsen2006boltztrap,chaves2021investigating} 
to carry out transport calculations and determine all the TE properties. 

\section{Results and discussion}

\subsection{Carrier density and ionized impurities concentration}

The thermoelectric (TE) transport properties of GeSe depend on the carrier density, $n_\mathrm{carr}$, and the
concentration of ionized impurities, $n_\mathrm{ii}$,
%% . Since these are material properties that
both of which depend on the
nonequilibrium growth process and can vary between samples.
In order to make meaningful predictions we need to
determine experimentally relevant values for $n_\mathrm{carr}$ and $n_\mathrm{ii}$, including
realistic temperature evolution. Due to the structural and chemical similarities between GeSe and SnSe,
and the dearth of experimental data on GeSe, we use the carrier and impurity
concentrations derived from SnSe experiments as a reasonable estimate for the values in GeSe samples.
This has the added benefit of allowing direct comparison of TE properties of two materials that differ only in chemical composition. 

Following the same procedure used in 
Ref.~\citenum{chaves2021investigating}, we determine $n_\mathrm{carr}$ and $n_\mathrm{ii}$ by self-consistently adjusting their values in order to reproduce,
 within our computational framework,
the experimentally measured values of $S$ and $\sigma$ in p-doped SnSe
reported by Zhao {\it{et al.}}\cite{zhao2015ultrahigh}.
Even though the carrier density of SnSe presents only weak anisotropy, 
as inferred by Hall measurements on SnSe\cite{zhao2014ultralow}, 
we considered different $n_\mathrm{carr}$ for the in-plane and the out-of-plane axes. 
We used the same temperature dependent $n_\mathrm{carr}$ and
$n_\mathrm{ii}$ derived for p-doped SnSe to calculate TE transport
properties of p-doped GeSe. 
For reasons of clarity, 
in the main text we report the results for a- and b-axis p-doped GeSe 
(referred to in the following as a-GeSe and b-GeSe, respectively) using 
$n_\mathrm{carr}$ and $n_\mathrm{ii}$ derived from a-axis p-doped SnSe. 
Results for c-axis GeSe (c-GeSe), as well as the properties calculated with 
$n_\mathrm{carr}$ and $n_\mathrm{ii}$ derived from b- or c-axis SnSe, are shown in Figures~S1-S3. 
%% For comparison, we also report 
%% the results for b-axis p-doped SnSe, which presents extraordinarily high TE performance. 
For reasons of expedience, we approximated $n_\mathrm{carr}$ and $n_\mathrm{ii}$ 
of GeSe at 800~K by the values obtained at 807~K for SnSe. 

Figure~\ref{n_carr}(a) shows the carrier and impurity concentrations derived for a-axis p-doped SnSe as a function
of $T$. The results derived for b- and c-axis p-doped SnSe are shown in the SM\cite{SM} (Figure~S4). 
At 300~K $n_\mathrm{carr}$ is approximately 4.4$\times 10^{19}$~cm$^{-3}$,
increases to 5.3$\times 10^{19}$~cm$^{-3}$ at 400~K,
and then decreases almost linearly down to 1.5$\times 10^{19}$~cm$^{-3}$ 
at 700~K, all consistent with Hall measurements~\cite{zhao2015ultrahigh}.
Above 700~K our results indicate that $n_\mathrm{carr}$ increases due to vacancy formation~\cite{chaves2021microscopic}
reaching 2.4$\times 10^{19}$~cm$^{-3}$ at 807~K.
Figure~\ref{n_carr}(b) shows the temperature dependence of the ratio $n_\mathrm{ii}/n_\mathrm{carr}$ compared to a
previous calculation of the same quantity\cite{chaves2021microscopic} that was based on the experimental data reported by Chang
{\it{et al.}}\cite{chang20183d} for a different a-axis SnSe sample with the same dopant.
The similarity between the ratios determined in this work and those reported in our previous work
demonstrates that these values of carrier and impurity concentrations are experimentally relevant and
approximately sample independent.

\subsection{Thermoelectric transport properties}

\begin{figure*}
        \centering
        \includegraphics[width=1\textwidth]{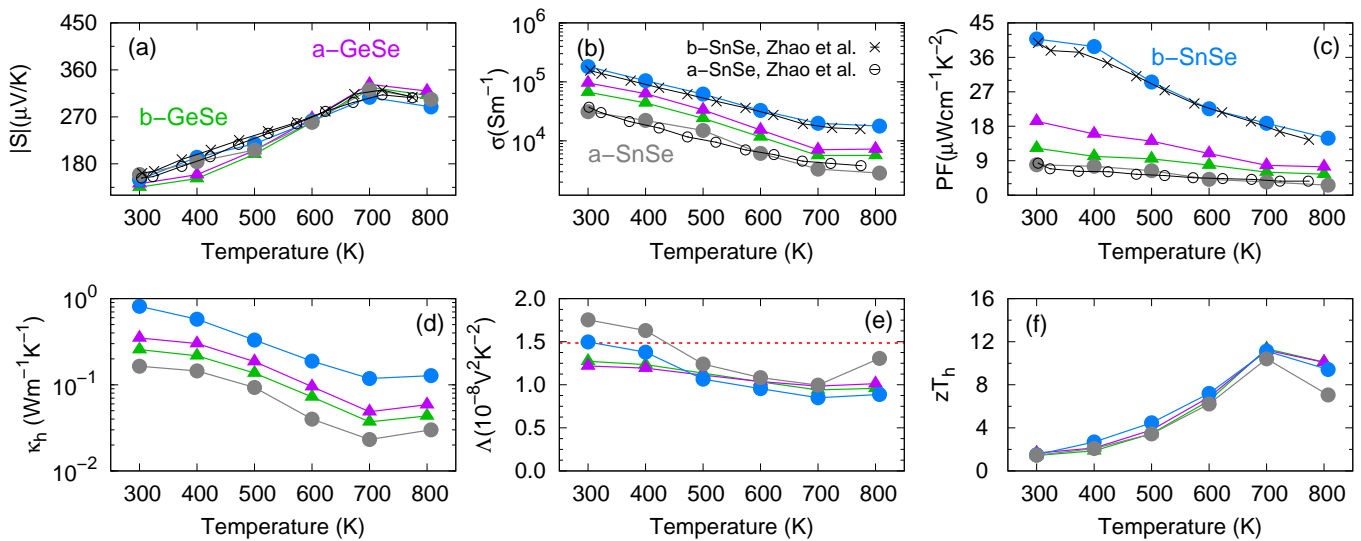}
        \caption{Calculated TE transport properties of p-doped GeSe and SnSe as functions of temperature, 
                 along with available experimental data on p-doped SnSe reported 
                 by Zhao \textit{et al.}\cite{zhao2015ultrahigh}.
                 $(a)$ Seebeck coefficient $S$, $(b)$ electrical conductivity, $\sigma$, $(c)$ power factor, PF, 
                 $(d)$ thermal conductivity due to the hole transport, ${\kappa}_\mathrm{h}$, $(e)$ Lorenz function, $\Lambda$, 
                 with a dashed red line at 
                 $\Lambda_\mathrm{nd}$ (see text),
                 %= 1.485\times 10^{-8}$V$^2$K$^{-2}$, 
                 and $(f)$ the hole figure of merit, $zT_\mathrm{h}$.}
        \label{therm_prop}
\end{figure*}

The calculated TE properties for p-doped out-of-plane (a-axis) and in-plane (b-axis) GeSe and SnSe  
are shown in Fig.~\ref{therm_prop}, along with available experimental data\cite{zhao2015ultrahigh}.  
All four systems show similar behavior of their Seebeck coefficients as a function of temperature,  
with $S$ increasing with $T$ up to 700~K, reaching 
332~$\mu V/K$ (326~$\mu V/K$) for a-GeSe (b-GeSe). 
Above that temperature the increase in $n_\mathrm{carr}$ causes $S$ to
decrease to 319$\mu V/K$ (311$\mu V/K$) at 800~K.
For temperatures above 600~K the calculated GeSe 
Seebeck coefficients are slightly higher 
than those of SnSe, in close agreement with previous theoretical findings.~\cite{ding2015high,hao2016computational}
As pointed out by Hao {\it{et al.}}\cite{hao2016computational}, p-doping in both materials induces a multiband effect that 
leads to an enhancement of $S$. 

The electrical conductivity, $\sigma$, of all four systems shows the expected exponential decrease
with temperature to 700~K, remaining nearly constant up to $\sim$800~K.
Both axes of GeSe present $\sigma$ values that are intermediate to those of
a- and b-SnSe. As has been found previously, the in-plane electrical conductivity of GeSe is much lower
than that of SnSe.\cite{hao2016computational} 
However, the out-of-plane $\sigma$ of GeSe is greater than its in-plane $\sigma$ and much higher than
the out-of-plane conductivity in SnSe.
This is a direct consequence of the low scattering rate
by ionized impurities for holes close to the valence band maximum (VBM) of a-GeSe, as will be discussed
further below. 

Similarly, the thermal conductivity due to hole transport, $\kappa_\mathrm{h}$,
also decreases with temperature up to 700~K 
and increases again for both materials between 700 and 800~K. 
The Lorenz function, defined as $\Lambda = \kappa_\mathrm{h}/(\sigma T)$, is shown in Fig.~\ref{therm_prop}(e), along with a red dashed line at the non-degenerate limit 
for semiconductors, $\Lambda_\mathrm{nd} = 2(k_B/e)^2 = 1.485\times 10^{-8}$V$^2$K$^{-2}$.
It has been shown previously that $\Lambda$ 
can be much smaller than $\Lambda_\mathrm{nd}$ when a rigorous first-principles approach is used instead of 
simplified band structures and scattering processes~\cite{wang2018calculation,putatunda2019lorenz}.   
Both axes of GeSe have very small values of $\Lambda$ that remain below $\Lambda_\mathrm{nd}$ throughout the temperature range.
The Lorenz function for in-plane SnSe shows higher (lower) values than both axes of GeSe 
for temperatures below (above) 500~K.
Even though a-SnSe also presents 
very low values of $\Lambda$, it is the highest of the four systems studied. Furthermore, 
it shows an abrupt enhancement above 700~K that is caused by a 
slight decrease in $\sigma$ accompanied by a considerable increase in $\kappa_\mathrm{h}$. 
Finally, the hole thermoelectric figure of merit, $zT_\mathrm{h} = S^2/\Lambda$, 
is shown in Fig.~\ref{therm_prop}(f).
$zT_\mathrm{h}$ values are quite similar for both materials 
throughout the entire temperature range, except for a-SnSe
that exhibits a sharp decrease at $800$~K mirroring the increase in $\Lambda$.

\subsection{Dominant Scattering mechanisms and Relaxation times}
\label{sec:RT}

\begin{figure*}[ht]
\centering
\includegraphics[width=0.7\textwidth]{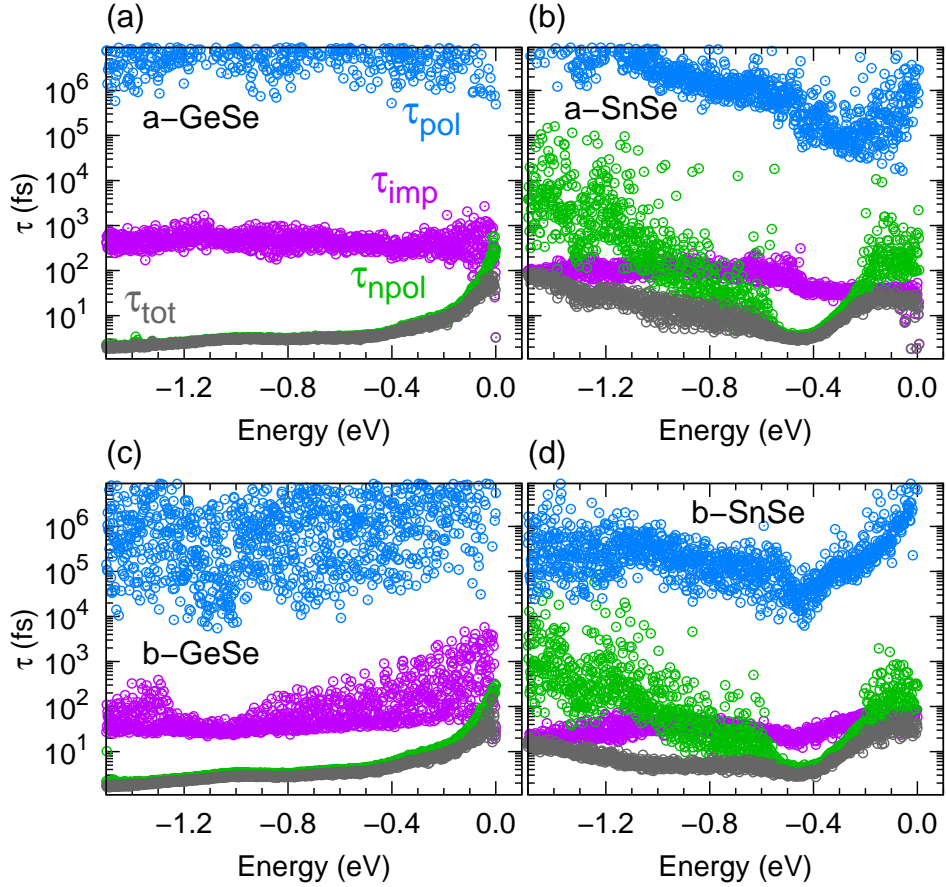}
\caption{Relaxation times (RTs) at 300~K as a function of hole energy for
  %for each scattering process in the
  $(a)$ p-doped GeSe (a-axis), $(b)$ p-doped SnSe (a-axis), $(c)$ p-doped GeSe (b-axis)
         and $(d)$ p-doped SnSe (b-axis). Each panel includes
         the screened Fr{\"o}hlich polar scattering of optical phonons ($\tau_\mathrm{pol}$, blue),
         nonpolar scattering of acoustic and optical phonons ($\tau_\mathrm{npol}$, green), scattering by ionized
         impurities ($\tau_\mathrm{imp}$, purple), and the total RT calculated with Mathiessen's rule ($\tau_\mathrm{tot}$, grey).
         The zero of the energy scale corresponds to the VBM.
}
\label{Tau}
\end{figure*}

In order to understand the temperature-dependent
transport phenomena in GeSe and SnSe, we extensively analyzed their carrier scattering mechanisms.
Figure~\ref{Tau} shows the relaxation times (RT) at 300~K due to nonpolar ($\tau_\mathrm{npol}$) and
screened Fr{\"o}hlich polar ($\tau_\mathrm{pol}$) scattering arising from the hole-phonon coupling, as well as scattering
by ionized impurities ($\tau_\mathrm{imp}$) and the total RT ($\tau_\mathrm{tot}$) based on Mathiessen's rule, Eq.~\ref{Mathiessen}.
These RTs are calculated as a function of the hole band and momentum, but plotted as a function of the hole energy
using the following conversion:
\begin{equation}
\label{Tau_eq}
\tau(\epsilon) = \frac{\sum_{n,{\bf{k}}} \tau_{n,{\bf{k}}} v_{n,{\bf{k}}} v_{n,{\bf{k}}} \delta (\epsilon-\epsilon_{n,{\bf{k}}})}{\sum_{n,{\bf{k}}} v_{n,{\bf{k}}} v_{n,{\bf{k}}} \delta (\epsilon-\epsilon_{n,{\bf{k}}})}~.
\end{equation}
In the SM\cite{SM} we provide additional details about
the temperature dependence of the RTs as well as comparisons between the RTs for different
systems and axes.

Due to the effectiveness of screening in these doped systems, $\tau_\mathrm{pol}$ is by far the largest RT,
demonstrating that the
Fr{\"o}hlich coupling does not contribute significantly to
the transport properties along either axis of p-doped GeSe. 
For both axes of GeSe we observe that 
$\tau_\mathrm{imp}$ is competitive with $\tau_\mathrm{npol}$ near the VBM at $E=0$. 
For energies well below the VBM $\tau_\mathrm{imp}$ is quickly overtaken by $\tau_\mathrm{npol}$. 
Comparing the results for the two axes in GeSe, we observe 
the total RTs are quite similar in magnitude and present similar energetic behavior,
%except for very few outliers.
as can also be clearly seen in Figure~S5.  

For SnSe the scenario is more complex because $\tau_\mathrm{npol}$
exhibits non-monotonic dependence on the hole energy with a minimum around $E=-0.5$ eV.
In this case, $\tau_\mathrm{imp}$ dominates
the carrier scattering near the VBM for a-SnSe, while for b-SnSe $\tau_\mathrm{npol}$ and $\tau_\mathrm{imp}$ 
compete with each other. $\tau_\mathrm{npol}$ becomes dominant in the range of $E=-0.8$ to $-0.2$ eV,
and the two mechanisms are comparable for lower energies.
This complicated energy dependence of the RTs strongly affects the TE transport properties.
In particular, it is responsible for the increase of $\kappa_\mathrm{h}$ at 807~K for a-SnSe, 
since the enhancement of $\tau_\mathrm{npol}$ and greater importance of scattering at higher energies increases the integral that appears in the calculation of $\kappa_\mathrm{h}$\cite{chaves2021microscopic}. 
A detailed comparison between the RTs for both axes in p-doped SnSe is presented 
in the SM\cite{SM} (Figs.~S6,~S7,~S8 and~S9), demonstrating that near the VBM $\tau_\mathrm{tot}$ 
is largely determined by $\tau_\mathrm{imp}$. Thus, it is the higher $\tau_\mathrm{imp}$ that causes b-SnSe to have a larger $\tau_\mathrm{tot}$ than a-SnSe throughout the full range of temperatures studied. 

By carefully comparing the total RTs of both materials, 
we observe that GeSe presents higher RT close to the VBM, which can be attributed mainly to 
the weaker scattering of 
holes by ionized impurities (see Figs.~S6 and~S10). 
Since all other contributions to $\tau_\mathrm{imp}$ are comparable in size, 
it must be the screening function $F_\mathrm{imp}$ that 
appears in the denominator of Eq.~(S8) which leads to the larger RT.
Hence, it is the greater effectiveness of the screening that raises the RT for GeSe. 
On the basis of Figures~S10-S13 the use of SnSe RTs to estimate the 
thermoelectric figure of merit for GeSe  
cannot be justified, since we see that
the SnSe RTs are generally smaller near the VBM.
Fig.~S11 clearly shows that such an approximation would significantly underestimate the RT for a-GeSe.
In addition, the clear variations in RT with hole energy is a strong argument against the use of 
the constant relaxation time approximation.

\subsection{Average hole group velocities and Transport distribution function}

\begin{figure*}
        \centering
        \includegraphics[width=0.7\textwidth]{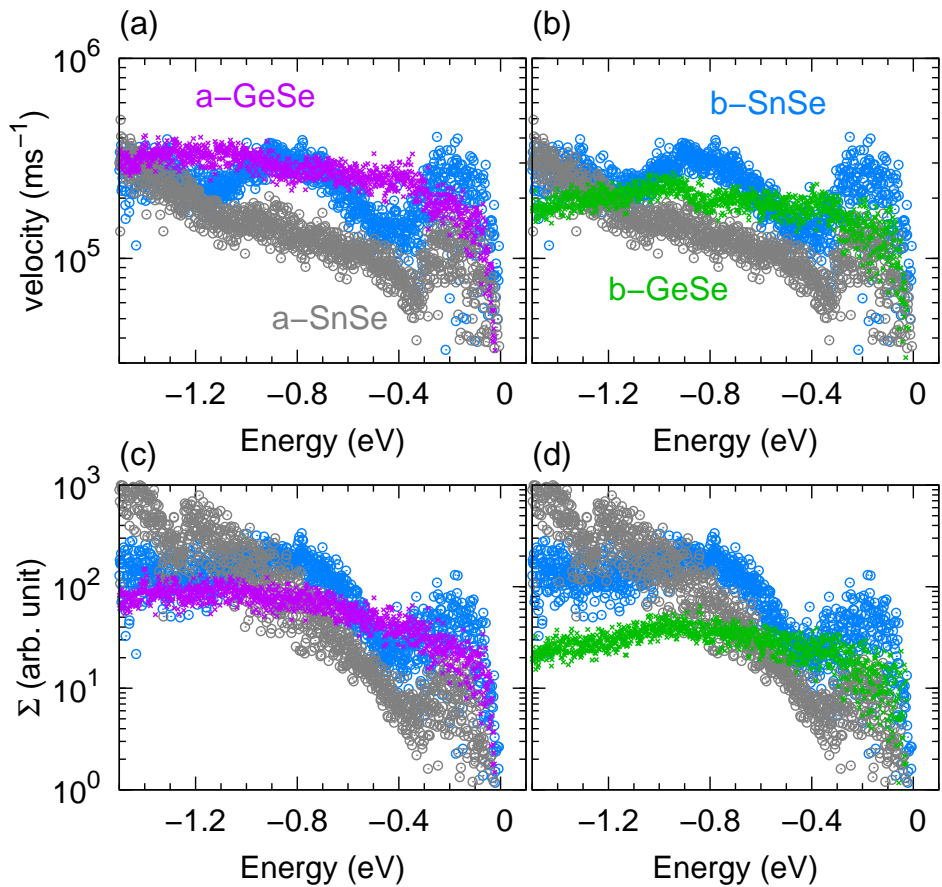}
        \caption{Average hole group velocity as a function of hole energy for 
                $(a)$ GeSe (a-axis, purple) and $(b)$ GeSe (b-axis, green). 
                Transport distribution function $\Sigma$ at 300~K as a function of hole energy for
         $(c)$ GeSe (a-axis, purple) and $(d)$ GeSe (b-axis, green). For comparison, each panel shows the values for SnSe, a-axis (grey) and b-axis (blue).}
        \label{veloc}
\end{figure*}

The average hole group velocities as a function of energy can be derived from the calculated band structure:
\begin{equation}
\label{vel}
v(\epsilon) = \sqrt{\left. \sum_{n,{\bf{k}}} \left|\frac{\partial \epsilon_{n,{\bf{k}}}}{\partial\bf{k}} \right|^2 \delta (\epsilon-\epsilon_{n,{\bf{k}}}) \right/ \sum_{n,{\bf{k}}} \delta (\epsilon-\epsilon_{n,{\bf{k}}})}~. 
\end{equation}
The group velocities, together with the energy dependent transport distribution function, $\Sigma(\epsilon)$, 
are shown in Fig.~\ref{veloc} 
for in-plane and out-of-plane GeSe and SnSe. Clearly, b-SnSe
has the highest velocities around the VBM
resulting in high electrical conductivity.
Along the same line, the lowest velocities in a-SnSe
are responsible for its inferior overall TE properties.
Close to the VBM, GeSe presents intermediate values for $v(\epsilon)$ and $\Sigma(\epsilon)$,
higher than a-SnSe and lower than b-SnSe.
Since the Seebeck coefficients of a-GeSe and b-GeSe are similar, it is the
higher hole velocity, which leads to higher electrical conductivity, that
%% has higher hole velocities than b-GeSe, 
%% resulting in higher $\sigma$ values. Since their $S$ values are quite similar, it is the hole velocity that
results in the larger $PF$ for a-GeSe.
For GeSe, the velocities smoothly increase as hole energy increases away from the VBM, 
except for highly energetic holes in b-GeSe.
On the other hand, for b-SnSe the velocities are high at the VBM,
but they decreases with hole energy, becoming smaller than
those of GeSe for holes between $-0.3$~eV to $-0.6$~eV. 
For GeSe, higher velocities at high hole energies
contribute to the increase of $\kappa_\mathrm{h}$ between 700 and 800~K. 
In general, the behavior of $\Sigma(\epsilon)$ follows that of the velocities. 
%However, it is
%important to note the steeper growth of $\Sigma$ for SnSe below $-0.4$~eV, which is explained by
%the increase in RT for those energies.

\subsection{Outstanding thermoelectric performance of GeSe}

In order to calculate the TE figure of merit, $zT$, we need an estimate of the total thermal conductivity, $\kappa_\mathrm{tot}$.
For SnSe we use the experimental values measured by Zhao {\it{et al.}}\cite{zhao2015ultrahigh}.
Since the necessary measurements have not yet been made for GeSe, we 
rely on theoretical results based on the
Debye-Callaway theory for lattice thermal conductivity~\cite{hao2016computational}, to which we add our 
calculated hole thermal conductivities. 
The resulting total thermal conductivity for both SnSe and GeSe is plotted in Fig.~S14. 
The thermal conductivity is nearly the same for a-axis GeSe and SnSe, though at the highest
temperatures it is slightly lower for a-GeSe. 
In-plane b-GeSe exhibits higher total thermal conductivity than out-of-plane GeSe throughout the temperature range, but   
it is comparatively lower than b-SnSe. 

Both contributions to the total thermal conductivity are extremely low in GeSe.
First, due to the relatively low electrical conductivity in GeSe and extremely low Lorenz numbers, carrier thermal conductivity is also
very small. Second, GeSe displays strong anharmonicity as quantified by its large 
Gr\"{u}neisen parameters that are comparable to or even 
larger than the Gr\"{u}neisen parameters for SnSe~\cite{hao2016computational}.
Such anomalously high Gr\"{u}neisen parameters of GeSe are a consequence of its hinge-like structure, 
distorted GeSe polyhedral, and van der Waals gaps in the out-of-plane direction 
that efficiently scatter phonons.
Since the measured $\kappa_\mathrm{tot}$ for SnSe is extremely low, %0.64 W/m/K, in the out-of-plane at 300~K).
we expect the lattice contribution, $\kappa_\mathrm{latt}$, to be extremely low for GeSe also. 

\begin{figure*}[ht]
\centering
\includegraphics[width=0.56\textwidth]{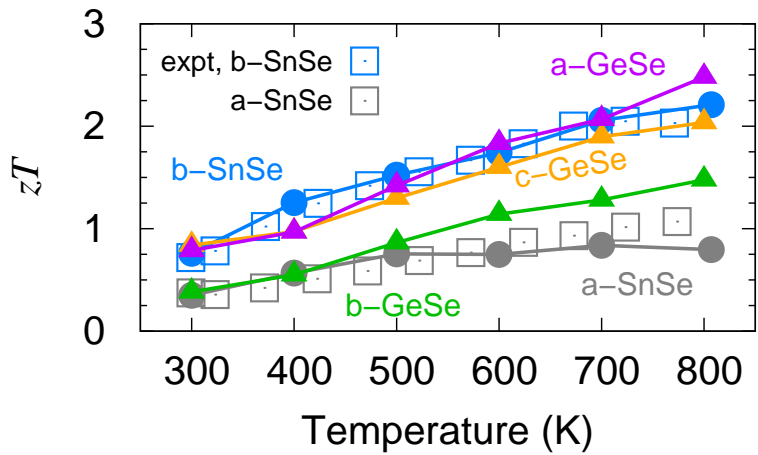}
%\vspace*{7mm}
\caption{Thermoelectric figure of merit ($zT$) as a function of temperature
        for p-doped a-GeSe (magenta), b-GeSe (green), c-GeSe (orange), a-SnSe (gray) and b-SnSe (blue).
        The experimental data reported by Zhao {\it{et al.}}\cite{zhao2015ultrahigh} for both axes of p-doped
        SnSe are shown for comparison.}
\label{zT_comp}
\end{figure*}

In view of its relatively high Seebeck coefficients 
and extremely low Lorenz numbers, coupled with an ultralow $\kappa_\mathrm{tot}$, we predict outstanding 
TE performance for out-of-plane GeSe, as can be seen in Fig.~\ref{zT_comp}. 
In particular, a-GeSe has $zT$ values that equal or
even exceed the record-breaking performance
of b-SnSe at temperatures above 500~K. This result would be missed if we assumed that the RTs for GeSe were the same as SnSe
along the corresponding crystallographic axis.  
Along with a-GeSe, high TE performance has also been obtained for c-GeSe 
throughout the whole temperature range, while 
b-GeSe has comparatively lower TE performance. 
It is important to emphasize that GeSe, just like SnSe, presents high $zT$ 
over a wide temperature range, which is a consequence 
of its relatively large band gap.\cite{xiao2020seeking}
Furthermore, we continue to find a high figure of merit for out-of-plane GeSe when using $n_\mathrm{carr}$ and $n_\mathrm{ii}$ derived from b- and c-SnSe data (see Figures~S15 and~S16). 

The figure of merit is increased for high power factors and low thermal conductivities. It is instructive to compare these
contributions to $zT$ for each of the systems studied here. For SnSe, the out-of-plane direction has a low power factor and
a low thermal conductivity, and the former dominates, leading to a (relatively) low $zT$ value. In contrast, the in-plane
direction has a high power factor and a high thermal conductivity, and once again it is the power factor that dominates, this time
yielding a high $zT$ value. Compared to its SnSe counterpart, b-GeSe has much lower power factor as well as a somewhat 
lower thermal conductivity,
producing a comparably low $zT$ value. 
The surprise, at least with reference to SnSe, is that a-GeSe maintains a very low $\kappa_\mathrm{latt}$
without the dramatic decrease in $PF$ shown by a-SnSe.

It is important to note that the calculation of $\kappa_\mathrm{latt}$ is very challenging.
For SnSe there is a long debate about $\kappa_\mathrm{latt}$ in the experimental 
literature\cite{wei2016intrinsic,wu2017direct,li2017ideal,chen2016understanding2} as well as questions regarding the comparison  
with first-principles calculations\cite{wu2017direct}. Similarly, the intrinsic thermal 
conductivity of GeSe is likely to be the subject of intense debate. Therefore, we re-calculated the value of $zT$ for GeSe using
values for $\kappa_\mathrm{latt}$ determined by Yuan et al.\cite{yuan2019tailoring} based on third-order force constants. Their values of $\kappa_\mathrm{latt}$ are noticeably higher, leading to a lower prediction for $zT$ as shown in the SM\cite{SM} (Figure~S17). 
Even with the larger $\kappa_\mathrm{latt}$, a-GeSe presents reasonable TE performance, reaching $zT=1.56$ at $800$~K for $n_\mathrm{carr}$ and $n_\mathrm{ii}$ 
derived from a-SnSe. 
Though the approach of Yuan et al.\cite{yuan2019tailoring} is more realistic than Debye-Callaway theory, 
the example of chalcogenides has shown that including additional factors such 
as thermal expansion, anharmonic phonon renormalization, four-phonon scattering, 
and impurity scatterings all generally serve to reduce  
the calculated values of $\kappa_\mathrm{latt}$\cite{wu2017direct,xia2018revisiting}, bringing 
them closer to the simpler Debye-Callaway approximation. Since 
doped GeSe is a strongly anharmonic material, it is not unreasonable that the Debye-Callaway method 
might yield reasonable results, as it has for other low-conductivity thermoelectric compounds.\cite{zhang2012first}

Since GeSe should maintain its $Pnma$ structure at 
higher temperatures than SnSe, we extend our analysis to 900~K.
At that high temperature we estimate $\kappa_\mathrm{latt}$ using a $1/T$
extrapolation~\cite{grimvall1999thermophysical} of the theoretical calculations done by Hao {\it{et al.}}~\cite{hao2016computational}
To that result we add $\kappa_h$ calculated within our current framework. 
However, due to its phase transition, there is no SnSe transport data at 900~K that we can use to determine $n_\mathrm{carr}$ 
and $n_\mathrm{ii}$. 
Instead, we scan over a range of $n_\mathrm{carr}$ and the ratio $n_\mathrm{ii}/n_\mathrm{carr}$ and carry out transport
calculations by solving the BTE for each pair of values. 
The density $n_\mathrm{carr}$ is varied between 1$\times 10^{19}$~cm$^{-3}$ and 10$\times 10^{19}$~cm$^{-3}$ in 10 equally spaced steps,
while for each value of $n_\mathrm{carr}$, $n_\mathrm{ii}/n_\mathrm{carr}$ was varied between 0.8 to $2.0$ in steps of 0.2.

\begin{figure*}[ht]
        \centering
        \includegraphics[width=0.66\textwidth]{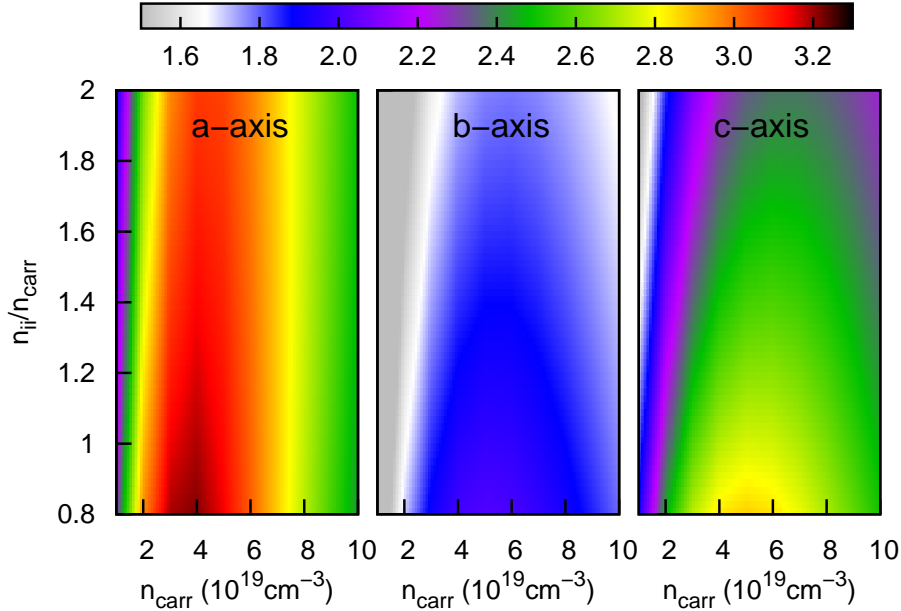}
        \caption{Colormap of the thermoelectric performance figure of merit $zT$ as a function of
                 the variation in $n_\mathrm{carr}$ and in the ratio $n_\mathrm{ii}/n_\mathrm{carr}$ for out-of-plane (a-axis)
                 and in-plane (b- and c-axis) GeSe at 900K.}
        \label{zT_map}
\end{figure*}

The calculated $zT$ values along all three crystallographic axes of GeSe at 900~K are shown
as a function of $n_\mathrm{carr}$ and $n_\mathrm{ii}/n_\mathrm{carr}$ in Fig.~\ref{zT_map}. 
The out-of-plane direction presents 
the highest performance in comparison to the other axes, reaching the ultrahigh value of $zT = 3.2$ at 
optimal $n_\mathrm{carr} = n_\mathrm{ii}$ = 4$\times 10^{19}$~cm$^{-3}$. It is important to note that 
$zT$ remains very high even if the ratio $n_\mathrm{ii}/n_\mathrm{carr}$ increases. For example, 
$zT = 3.06$ when $n_\mathrm{ii}/n_\mathrm{carr} = 2$, 
indicating a high performance with $zT$ larger than 3 even if $n_\mathrm{ii}$ 
is doubled. Horizontal line cuts at fixed ratios are shown in Fig.~S18, clearly demonstrating that a tenfold increase in the ratio  
$n_\mathrm{ii}/n_\mathrm{carr}$ can still lead to great performance provided $n_\mathrm{carr}$ is correspondingly increased.
For example, with $n_\mathrm{ii}/n_\mathrm{carr}=10$, an optimal value of $n_\mathrm{carr} =$ 6$\times 10^{19}$~cm$^{-3}$ yields $zT =2.7$. 
Fig.~S18 also shows the robustness of the high $zT$ values as $n_\mathrm{carr}$ is varied. 
For instance, $zT \geq$ 3 for $n_\mathrm{carr}$ 
between 3$\times 10^{19}$~cm$^{-3}$ and 6$\times 10^{19}$~cm$^{-3}$ when $n_\mathrm{ii}/n_\mathrm{carr} = 1$. 
Though not as impressive as the out-of-plane direction, the two in-plane directions still exhibit relatively high $zT$ values at 900~K, 
namely, $zT =$ 2.0 (2.8) for the optimal ratio  
$n_\mathrm{carr} = n_\mathrm{ii} =$ 5$\times 10^{19}$~cm$^{-3}$ for the b-axis (c-axis). 

\section{Conclusions}

In summary, we applied extensive first-principles calculations 
within the BTE framework to throughly investigate the temperature dependence of the 
TE transport properties of the orthorhombic \textit{Pnma} phase of p-doped GeSe and SnSe. 
These calculations were done for values of the carrier density that yielded the record-breaking 
TE performance of p-doped SnSe.\cite{zhao2015ultrahigh} 
We explicitly calculated the RTs due to nonpolar and screened Fr{\"o}hlich polar h-p scattering, 
as well as the RT associated with the scattering by ionized impurities. 
The obtained temperature and hole-energy dependent RTs provide insight into the microscopic origin of the
transport properties in p-doped GeSe and SnSe. 

Our results indicate that the calculated GeSe
Seebeck coefficients, $S$, are slightly higher
than those of SnSe at temperatures above 600~K, while 
both axes of GeSe present electrical conductivity values that are intermediate between
those of a- and b-SnSe. Importantly, both axes of GeSe exhibit Lorenz numbers below the non-degenerate limit
of semiconductors. In-plane SnSe also has very low values for the Lorenz function $\Lambda$, and at temperatures above 500~K they 
are the same or even smaller than those of in-plane GeSe. 
On the other hand, a-SnSe possesses the highest $\Lambda$ among all studied systems.  
Those results for $\Lambda$ are directly correlated with TE performance,
in which a low $\Lambda$ plays a beneficial role in obtaining a high $zT_\mathrm{h}$.

All axes of GeSe have quite low thermal conductivity $\kappa_\mathrm{tot}$, 
which is a consequence of relatively low 
$\sigma$ that yields low hole thermal conductivity, and high anharmonicity~\cite{hao2016computational} that leads to predictions of low $\kappa_\mathrm{latt}$. 
Combined with the high Seebeck coefficients, 
extremely low Lorenz numbers, and a relatively large band gap, 
our calculations predict an outstanding
TE performance for both out-of-plane (a-axis) and in-plane (b- and c-axis) GeSe throughout a wide range of 
temperatures. Above 500~K the out-of-plane direction has $zT$ values 
equal to or greater than the record-breaking performance
of b-SnSe. 
By extending the analysis of the TE performance of GeSe
to 900~K, we find that the out-of-plane direction still presents
the highest performance compared to other axes, reaching an ultrahigh $zT = 3.2$ at
the optimal carrier density of 4$\times 10^{19}$~cm$^{-3}$. In addition, 
the two in-plane axes also have impressive figures of merit, with
$zT =$ 2.0 (2.8) for the b-axis (c-axis) with optimal carrier density of 5$\times 10^{19}$~cm$^{-3}$. 
It is important to point out that the total RTs of out-of-plane GeSe 
are much higher than those of a-SnSe. Thus our results for the out-of-plane direction could not be anticipated by calculations using
the same RTs as those for SnSe.~\cite{hao2016computational}

Finally, it is also important to point out that intrinsic GeSe possesses low carrier density.
To date Ge substitution by Ag is the most effective method of doping,
enabling a hole density up to $\sim 10^{18}$~cm$^{-3}$
and $zT \approx 0.2$ at 700~K for polycrystalline GeSe.~\cite{zhang2016thermoelectric} This experimental result is far below 
our highest theoretically predicted value of $zT = 3.2$ at 900~K, which
can be attributed to the low carrier density %% limited by chemical doping
that is far below our predicted optimal
carrier density of $\sim10^{19}$~cm$^{-3}$.
Our results indicate that there is enormous room for further improvement in the TE 
performance of GeSe by increasing the doping to optimal carrier density.

\setcounter{secnumdepth}{0}

\vspace*{7 mm}

\section*{Supplemental Material}
Extra data and analyses are provided in the Supplemental Material.

\section*{Conflicts of interest}
There are no conflicts to declare.

\section*{Acknowledgements}

ASC and AA gratefully acknowledge support from the Brazilian agencies CNPq and FAPESP under
Grants \#2010/16970-0, \#2013/08293-7, \#2015/26434-2, \#2016/23891-6, \#2017/26105-4, and \#2019/26088-8. DTL acknowledges funding from the STC Center for Integrated Quantum Materials, NSF Grant No. DMR-1231319; NSF DMREF Award No. 1922172; and the Army Research Office under Cooperative Agreement Number W911NF-21-2-0147.
The calculations were performed at CCJDR-IFGW-UNICAMP in Brazil.

\vspace*{7 mm}
\section*{Computer Code Availability}
All computer implementations of the methodology developed in this project were written
in Fortran 90 and are available upon request.

\nocite{giustino2017electron}
\nocite{ponce2018towards}
\nocite{ponce2020first}
\nocite{frohlich1937h}
\nocite{verdi2015frohlich}
\nocite{marzari1997maximally}
\nocite{askerov2009thermodynamics}
\nocite{shankland1971interpolation}
\nocite{koelling1986interpolation}

\bibliographystyle{apsrev4-1}
{\footnotesize
\bibliography{Ref2.bib}}

%merlin.mbs apsrev4-1.bst 2010-07-25 4.21a (PWD, AO, DPC) hacked
%Control: key (0)
%Control: author (72) initials jnrlst
%Control: editor formatted (1) identically to author
%Control: production of article title (-1) disabled
%Control: page (0) single
%Control: year (1) truncated
%Control: production of eprint (0) enabled
\begin{thebibliography}{101}%
\makeatletter
\providecommand \@ifxundefined [1]{%
 \@ifx{#1\undefined}
}%
\providecommand \@ifnum [1]{%
 \ifnum #1\expandafter \@firstoftwo
 \else \expandafter \@secondoftwo
 \fi
}%
\providecommand \@ifx [1]{%
 \ifx #1\expandafter \@firstoftwo
 \else \expandafter \@secondoftwo
 \fi
}%
\providecommand \natexlab [1]{#1}%
\providecommand \enquote  [1]{``#1''}%
\providecommand \bibnamefont  [1]{#1}%
\providecommand \bibfnamefont [1]{#1}%
\providecommand \citenamefont [1]{#1}%
\providecommand \href@noop [0]{\@secondoftwo}%
\providecommand \href [0]{\begingroup \@sanitize@url \@href}%
\providecommand \@href[1]{\@@startlink{#1}\@@href}%
\providecommand \@@href[1]{\endgroup#1\@@endlink}%
\providecommand \@sanitize@url [0]{\catcode `\\12\catcode `\$12\catcode
  `\&12\catcode `\#12\catcode `\^12\catcode `\_12\catcode `\%12\relax}%
\providecommand \@@startlink[1]{}%
\providecommand \@@endlink[0]{}%
\providecommand \url  [0]{\begingroup\@sanitize@url \@url }%
\providecommand \@url [1]{\endgroup\@href {#1}{\urlprefix }}%
\providecommand \urlprefix  [0]{URL }%
\providecommand \Eprint [0]{\href }%
\providecommand \doibase [0]{http://dx.doi.org/}%
\providecommand \selectlanguage [0]{\@gobble}%
\providecommand \bibinfo  [0]{\@secondoftwo}%
\providecommand \bibfield  [0]{\@secondoftwo}%
\providecommand \translation [1]{[#1]}%
\providecommand \BibitemOpen [0]{}%
\providecommand \bibitemStop [0]{}%
\providecommand \bibitemNoStop [0]{.\EOS\space}%
\providecommand \EOS [0]{\spacefactor3000\relax}%
\providecommand \BibitemShut  [1]{\csname bibitem#1\endcsname}%
\let\auto@bib@innerbib\@empty
%</preamble>
\bibitem [{\citenamefont {Mao}\ \emph {et~al.}(2020)\citenamefont {Mao},
  \citenamefont {Chen},\ and\ \citenamefont {Ren}}]{mao2020thermoelectric}%
  \BibitemOpen
  \bibfield  {author} {\bibinfo {author} {\bibfnamefont {J.}~\bibnamefont
  {Mao}}, \bibinfo {author} {\bibfnamefont {G.}~\bibnamefont {Chen}}, \ and\
  \bibinfo {author} {\bibfnamefont {Z.}~\bibnamefont {Ren}},\ }\href@noop {}
  {\bibfield  {journal} {\bibinfo  {journal} {Nature Materials}\ ,\ \bibinfo
  {pages} {1}} (\bibinfo {year} {2020})}\BibitemShut {NoStop}%
\bibitem [{\citenamefont {Mao}\ \emph {et~al.}(2019)\citenamefont {Mao},
  \citenamefont {Zhu}, \citenamefont {Ding}, \citenamefont {Liu}, \citenamefont
  {Gamage}, \citenamefont {Chen},\ and\ \citenamefont {Ren}}]{mao2019high}%
  \BibitemOpen
  \bibfield  {author} {\bibinfo {author} {\bibfnamefont {J.}~\bibnamefont
  {Mao}}, \bibinfo {author} {\bibfnamefont {H.}~\bibnamefont {Zhu}}, \bibinfo
  {author} {\bibfnamefont {Z.}~\bibnamefont {Ding}}, \bibinfo {author}
  {\bibfnamefont {Z.}~\bibnamefont {Liu}}, \bibinfo {author} {\bibfnamefont
  {G.~A.}\ \bibnamefont {Gamage}}, \bibinfo {author} {\bibfnamefont
  {G.}~\bibnamefont {Chen}}, \ and\ \bibinfo {author} {\bibfnamefont
  {Z.}~\bibnamefont {Ren}},\ }\href@noop {} {\bibfield  {journal} {\bibinfo
  {journal} {Science}\ }\textbf {\bibinfo {volume} {365}},\ \bibinfo {pages}
  {495} (\bibinfo {year} {2019})}\BibitemShut {NoStop}%
\bibitem [{\citenamefont {DiSalvo}(1999)}]{disalvo1999thermoelectric}%
  \BibitemOpen
  \bibfield  {author} {\bibinfo {author} {\bibfnamefont {F.~J.}\ \bibnamefont
  {DiSalvo}},\ }\href@noop {} {\bibfield  {journal} {\bibinfo  {journal}
  {Science}\ }\textbf {\bibinfo {volume} {285}},\ \bibinfo {pages} {703}
  (\bibinfo {year} {1999})}\BibitemShut {NoStop}%
\bibitem [{\citenamefont {He}\ and\ \citenamefont
  {Tritt}(2017)}]{he2017advances}%
  \BibitemOpen
  \bibfield  {author} {\bibinfo {author} {\bibfnamefont {J.}~\bibnamefont
  {He}}\ and\ \bibinfo {author} {\bibfnamefont {T.~M.}\ \bibnamefont {Tritt}},\
  }\href@noop {} {\bibfield  {journal} {\bibinfo  {journal} {Science}\ }\textbf
  {\bibinfo {volume} {357}},\ \bibinfo {pages} {eaak9997} (\bibinfo {year}
  {2017})}\BibitemShut {NoStop}%
\bibitem [{\citenamefont {Lv}\ \emph {et~al.}(2013)\citenamefont {Lv},
  \citenamefont {Liu}, \citenamefont {Shi}, \citenamefont {Tang},\ and\
  \citenamefont {Uher}}]{lv2013optimized}%
  \BibitemOpen
  \bibfield  {author} {\bibinfo {author} {\bibfnamefont {H.}~\bibnamefont
  {Lv}}, \bibinfo {author} {\bibfnamefont {H.}~\bibnamefont {Liu}}, \bibinfo
  {author} {\bibfnamefont {J.}~\bibnamefont {Shi}}, \bibinfo {author}
  {\bibfnamefont {X.}~\bibnamefont {Tang}}, \ and\ \bibinfo {author}
  {\bibfnamefont {C.}~\bibnamefont {Uher}},\ }\href@noop {} {\bibfield
  {journal} {\bibinfo  {journal} {Journal of Materials Chemistry A}\ }\textbf
  {\bibinfo {volume} {1}},\ \bibinfo {pages} {6831} (\bibinfo {year}
  {2013})}\BibitemShut {NoStop}%
\bibitem [{\citenamefont {Venkatasubramanian}\ \emph
  {et~al.}(2001)\citenamefont {Venkatasubramanian}, \citenamefont {Siivola},
  \citenamefont {Colpitts},\ and\ \citenamefont
  {O'quinn}}]{venkatasubramanian2001thin}%
  \BibitemOpen
  \bibfield  {author} {\bibinfo {author} {\bibfnamefont {R.}~\bibnamefont
  {Venkatasubramanian}}, \bibinfo {author} {\bibfnamefont {E.}~\bibnamefont
  {Siivola}}, \bibinfo {author} {\bibfnamefont {T.}~\bibnamefont {Colpitts}}, \
  and\ \bibinfo {author} {\bibfnamefont {B.}~\bibnamefont {O'quinn}},\
  }\href@noop {} {\bibfield  {journal} {\bibinfo  {journal} {Nature}\ }\textbf
  {\bibinfo {volume} {413}},\ \bibinfo {pages} {597} (\bibinfo {year}
  {2001})}\BibitemShut {NoStop}%
\bibitem [{\citenamefont {Zhang}\ \emph
  {et~al.}(2016{\natexlab{a}})\citenamefont {Zhang}, \citenamefont {Hao},
  \citenamefont {Zhao}, \citenamefont {Wolverton},\ and\ \citenamefont
  {Zeng}}]{zhang2016pressure}%
  \BibitemOpen
  \bibfield  {author} {\bibinfo {author} {\bibfnamefont {Y.}~\bibnamefont
  {Zhang}}, \bibinfo {author} {\bibfnamefont {S.}~\bibnamefont {Hao}}, \bibinfo
  {author} {\bibfnamefont {L.-D.}\ \bibnamefont {Zhao}}, \bibinfo {author}
  {\bibfnamefont {C.}~\bibnamefont {Wolverton}}, \ and\ \bibinfo {author}
  {\bibfnamefont {Z.}~\bibnamefont {Zeng}},\ }\href@noop {} {\bibfield
  {journal} {\bibinfo  {journal} {Journal of Materials Chemistry A}\ }\textbf
  {\bibinfo {volume} {4}},\ \bibinfo {pages} {12073} (\bibinfo {year}
  {2016}{\natexlab{a}})}\BibitemShut {NoStop}%
\bibitem [{\citenamefont {Zhong}\ \emph {et~al.}(2014)\citenamefont {Zhong},
  \citenamefont {Zhang}, \citenamefont {Li}, \citenamefont {Chen},
  \citenamefont {Cui}, \citenamefont {Li}, \citenamefont {Xie}, \citenamefont
  {Hao},\ and\ \citenamefont {He}}]{zhong2014high}%
  \BibitemOpen
  \bibfield  {author} {\bibinfo {author} {\bibfnamefont {B.}~\bibnamefont
  {Zhong}}, \bibinfo {author} {\bibfnamefont {Y.}~\bibnamefont {Zhang}},
  \bibinfo {author} {\bibfnamefont {W.}~\bibnamefont {Li}}, \bibinfo {author}
  {\bibfnamefont {Z.}~\bibnamefont {Chen}}, \bibinfo {author} {\bibfnamefont
  {J.}~\bibnamefont {Cui}}, \bibinfo {author} {\bibfnamefont {W.}~\bibnamefont
  {Li}}, \bibinfo {author} {\bibfnamefont {Y.}~\bibnamefont {Xie}}, \bibinfo
  {author} {\bibfnamefont {Q.}~\bibnamefont {Hao}}, \ and\ \bibinfo {author}
  {\bibfnamefont {Q.}~\bibnamefont {He}},\ }\href@noop {} {\bibfield  {journal}
  {\bibinfo  {journal} {Applied Physics Letters}\ }\textbf {\bibinfo {volume}
  {105}},\ \bibinfo {pages} {123902} (\bibinfo {year} {2014})}\BibitemShut
  {NoStop}%
\bibitem [{\citenamefont {Famili}\ \emph {et~al.}(2018)\citenamefont {Famili},
  \citenamefont {Grace}, \citenamefont {Al-Galiby}, \citenamefont {Sadeghi},\
  and\ \citenamefont {Lambert}}]{famili2018toward}%
  \BibitemOpen
  \bibfield  {author} {\bibinfo {author} {\bibfnamefont {M.}~\bibnamefont
  {Famili}}, \bibinfo {author} {\bibfnamefont {I.~M.}\ \bibnamefont {Grace}},
  \bibinfo {author} {\bibfnamefont {Q.}~\bibnamefont {Al-Galiby}}, \bibinfo
  {author} {\bibfnamefont {H.}~\bibnamefont {Sadeghi}}, \ and\ \bibinfo
  {author} {\bibfnamefont {C.~J.}\ \bibnamefont {Lambert}},\ }\href@noop {}
  {\bibfield  {journal} {\bibinfo  {journal} {Advanced Functional Materials}\
  }\textbf {\bibinfo {volume} {28}},\ \bibinfo {pages} {1703135} (\bibinfo
  {year} {2018})}\BibitemShut {NoStop}%
\bibitem [{\citenamefont {Xue}\ \emph {et~al.}(2016)\citenamefont {Xue},
  \citenamefont {Liu}, \citenamefont {Fan}, \citenamefont {Cheng},
  \citenamefont {Zhao},\ and\ \citenamefont {Shi}}]{xue2016laptsb}%
  \BibitemOpen
  \bibfield  {author} {\bibinfo {author} {\bibfnamefont {Q.}~\bibnamefont
  {Xue}}, \bibinfo {author} {\bibfnamefont {H.}~\bibnamefont {Liu}}, \bibinfo
  {author} {\bibfnamefont {D.}~\bibnamefont {Fan}}, \bibinfo {author}
  {\bibfnamefont {L.}~\bibnamefont {Cheng}}, \bibinfo {author} {\bibfnamefont
  {B.}~\bibnamefont {Zhao}}, \ and\ \bibinfo {author} {\bibfnamefont
  {J.}~\bibnamefont {Shi}},\ }\href@noop {} {\bibfield  {journal} {\bibinfo
  {journal} {Physical Chemistry Chemical Physics}\ }\textbf {\bibinfo {volume}
  {18}},\ \bibinfo {pages} {17912} (\bibinfo {year} {2016})}\BibitemShut
  {NoStop}%
\bibitem [{\citenamefont {Ohta}\ \emph {et~al.}(2007)\citenamefont {Ohta},
  \citenamefont {Kim}, \citenamefont {Mune}, \citenamefont {Mizoguchi},
  \citenamefont {Nomura}, \citenamefont {Ohta}, \citenamefont {Nomura},
  \citenamefont {Nakanishi}, \citenamefont {Ikuhara}, \citenamefont {Hirano},
  \citenamefont {Hosono},\ and\ \citenamefont {Koumoto}}]{ohta2007giant}%
  \BibitemOpen
  \bibfield  {author} {\bibinfo {author} {\bibfnamefont {H.}~\bibnamefont
  {Ohta}}, \bibinfo {author} {\bibfnamefont {S.}~\bibnamefont {Kim}}, \bibinfo
  {author} {\bibfnamefont {Y.}~\bibnamefont {Mune}}, \bibinfo {author}
  {\bibfnamefont {T.}~\bibnamefont {Mizoguchi}}, \bibinfo {author}
  {\bibfnamefont {K.}~\bibnamefont {Nomura}}, \bibinfo {author} {\bibfnamefont
  {S.}~\bibnamefont {Ohta}}, \bibinfo {author} {\bibfnamefont {T.}~\bibnamefont
  {Nomura}}, \bibinfo {author} {\bibfnamefont {Y.}~\bibnamefont {Nakanishi}},
  \bibinfo {author} {\bibfnamefont {Y.}~\bibnamefont {Ikuhara}}, \bibinfo
  {author} {\bibfnamefont {M.}~\bibnamefont {Hirano}}, \bibinfo {author}
  {\bibfnamefont {H.}~\bibnamefont {Hosono}}, \ and\ \bibinfo {author}
  {\bibfnamefont {K.}~\bibnamefont {Koumoto}},\ }\href@noop {} {\bibfield
  {journal} {\bibinfo  {journal} {Nature Materials}\ }\textbf {\bibinfo
  {volume} {6}},\ \bibinfo {pages} {129} (\bibinfo {year} {2007})}\BibitemShut
  {NoStop}%
\bibitem [{\citenamefont {Fu}\ \emph {et~al.}(2016)\citenamefont {Fu},
  \citenamefont {Yue}, \citenamefont {Wu}, \citenamefont {Fu}, \citenamefont
  {Zhu}, \citenamefont {Liu}, \citenamefont {Hu}, \citenamefont {Ying},
  \citenamefont {He},\ and\ \citenamefont {Zhao}}]{fu2016enhanced}%
  \BibitemOpen
  \bibfield  {author} {\bibinfo {author} {\bibfnamefont {T.}~\bibnamefont
  {Fu}}, \bibinfo {author} {\bibfnamefont {X.}~\bibnamefont {Yue}}, \bibinfo
  {author} {\bibfnamefont {H.}~\bibnamefont {Wu}}, \bibinfo {author}
  {\bibfnamefont {C.}~\bibnamefont {Fu}}, \bibinfo {author} {\bibfnamefont
  {T.}~\bibnamefont {Zhu}}, \bibinfo {author} {\bibfnamefont {X.}~\bibnamefont
  {Liu}}, \bibinfo {author} {\bibfnamefont {L.}~\bibnamefont {Hu}}, \bibinfo
  {author} {\bibfnamefont {P.}~\bibnamefont {Ying}}, \bibinfo {author}
  {\bibfnamefont {J.}~\bibnamefont {He}}, \ and\ \bibinfo {author}
  {\bibfnamefont {X.}~\bibnamefont {Zhao}},\ }\href@noop {} {\bibfield
  {journal} {\bibinfo  {journal} {Journal of Materiomics}\ }\textbf {\bibinfo
  {volume} {2}},\ \bibinfo {pages} {141} (\bibinfo {year} {2016})}\BibitemShut
  {NoStop}%
\bibitem [{\citenamefont {Zhao}\ \emph {et~al.}(2014)\citenamefont {Zhao},
  \citenamefont {Lo}, \citenamefont {Zhang}, \citenamefont {Sun}, \citenamefont
  {Tan}, \citenamefont {Uher}, \citenamefont {Wolverton}, \citenamefont
  {Dravid},\ and\ \citenamefont {Kanatzidis}}]{zhao2014ultralow}%
  \BibitemOpen
  \bibfield  {author} {\bibinfo {author} {\bibfnamefont {L.-D.}\ \bibnamefont
  {Zhao}}, \bibinfo {author} {\bibfnamefont {S.-H.}\ \bibnamefont {Lo}},
  \bibinfo {author} {\bibfnamefont {Y.}~\bibnamefont {Zhang}}, \bibinfo
  {author} {\bibfnamefont {H.}~\bibnamefont {Sun}}, \bibinfo {author}
  {\bibfnamefont {G.}~\bibnamefont {Tan}}, \bibinfo {author} {\bibfnamefont
  {C.}~\bibnamefont {Uher}}, \bibinfo {author} {\bibfnamefont {C.}~\bibnamefont
  {Wolverton}}, \bibinfo {author} {\bibfnamefont {V.~P.}\ \bibnamefont
  {Dravid}}, \ and\ \bibinfo {author} {\bibfnamefont {M.~G.}\ \bibnamefont
  {Kanatzidis}},\ }\href@noop {} {\bibfield  {journal} {\bibinfo  {journal}
  {Nature}\ }\textbf {\bibinfo {volume} {508}},\ \bibinfo {pages} {373}
  (\bibinfo {year} {2014})}\BibitemShut {NoStop}%
\bibitem [{\citenamefont {Zhao}\ \emph {et~al.}(2016)\citenamefont {Zhao},
  \citenamefont {Tan}, \citenamefont {Hao}, \citenamefont {He}, \citenamefont
  {Pei}, \citenamefont {Chi}, \citenamefont {Wang}, \citenamefont {Gong},
  \citenamefont {Xu}, \citenamefont {Dravid}, \citenamefont {Uher},
  \citenamefont {Snyder}, \citenamefont {Wolverton},\ and\ \citenamefont
  {Kanatzidis}}]{zhao2015ultrahigh}%
  \BibitemOpen
  \bibfield  {author} {\bibinfo {author} {\bibfnamefont {L.-D.}\ \bibnamefont
  {Zhao}}, \bibinfo {author} {\bibfnamefont {G.}~\bibnamefont {Tan}}, \bibinfo
  {author} {\bibfnamefont {S.}~\bibnamefont {Hao}}, \bibinfo {author}
  {\bibfnamefont {J.}~\bibnamefont {He}}, \bibinfo {author} {\bibfnamefont
  {Y.}~\bibnamefont {Pei}}, \bibinfo {author} {\bibfnamefont {H.}~\bibnamefont
  {Chi}}, \bibinfo {author} {\bibfnamefont {H.}~\bibnamefont {Wang}}, \bibinfo
  {author} {\bibfnamefont {S.}~\bibnamefont {Gong}}, \bibinfo {author}
  {\bibfnamefont {H.}~\bibnamefont {Xu}}, \bibinfo {author} {\bibfnamefont
  {V.~P.}\ \bibnamefont {Dravid}}, \bibinfo {author} {\bibfnamefont
  {C.}~\bibnamefont {Uher}}, \bibinfo {author} {\bibfnamefont {G.~J.}\
  \bibnamefont {Snyder}}, \bibinfo {author} {\bibfnamefont {C.}~\bibnamefont
  {Wolverton}}, \ and\ \bibinfo {author} {\bibfnamefont {M.~G.}\ \bibnamefont
  {Kanatzidis}},\ }\href@noop {} {\bibfield  {journal} {\bibinfo  {journal}
  {Science}\ }\textbf {\bibinfo {volume} {351}},\ \bibinfo {pages} {141}
  (\bibinfo {year} {2016})}\BibitemShut {NoStop}%
\bibitem [{\citenamefont {Chang}\ \emph {et~al.}(2018)\citenamefont {Chang},
  \citenamefont {Wu}, \citenamefont {He}, \citenamefont {Pei}, \citenamefont
  {Wu}, \citenamefont {Wu}, \citenamefont {Yu}, \citenamefont {Zhu},
  \citenamefont {Wang}, \citenamefont {Chen}, \citenamefont {Huang},
  \citenamefont {Li}, \citenamefont {He},\ and\ \citenamefont
  {Zhao}}]{chang20183d}%
  \BibitemOpen
  \bibfield  {author} {\bibinfo {author} {\bibfnamefont {C.}~\bibnamefont
  {Chang}}, \bibinfo {author} {\bibfnamefont {M.}~\bibnamefont {Wu}}, \bibinfo
  {author} {\bibfnamefont {D.}~\bibnamefont {He}}, \bibinfo {author}
  {\bibfnamefont {Y.}~\bibnamefont {Pei}}, \bibinfo {author} {\bibfnamefont
  {C.-F.}\ \bibnamefont {Wu}}, \bibinfo {author} {\bibfnamefont
  {X.}~\bibnamefont {Wu}}, \bibinfo {author} {\bibfnamefont {H.}~\bibnamefont
  {Yu}}, \bibinfo {author} {\bibfnamefont {F.}~\bibnamefont {Zhu}}, \bibinfo
  {author} {\bibfnamefont {K.}~\bibnamefont {Wang}}, \bibinfo {author}
  {\bibfnamefont {Y.}~\bibnamefont {Chen}}, \bibinfo {author} {\bibfnamefont
  {L.}~\bibnamefont {Huang}}, \bibinfo {author} {\bibfnamefont {J.-F.}\
  \bibnamefont {Li}}, \bibinfo {author} {\bibfnamefont {J.}~\bibnamefont {He}},
  \ and\ \bibinfo {author} {\bibfnamefont {L.-D.}\ \bibnamefont {Zhao}},\
  }\href@noop {} {\bibfield  {journal} {\bibinfo  {journal} {Science}\ }\textbf
  {\bibinfo {volume} {360}},\ \bibinfo {pages} {778} (\bibinfo {year}
  {2018})}\BibitemShut {NoStop}%
\bibitem [{\citenamefont {Vining}(2009)}]{vining2009inconvenient}%
  \BibitemOpen
  \bibfield  {author} {\bibinfo {author} {\bibfnamefont {C.~B.}\ \bibnamefont
  {Vining}},\ }\href@noop {} {\bibfield  {journal} {\bibinfo  {journal} {Nature
  Materials}\ }\textbf {\bibinfo {volume} {8}},\ \bibinfo {pages} {83}
  (\bibinfo {year} {2009})}\BibitemShut {NoStop}%
\bibitem [{\citenamefont {Sales}\ \emph {et~al.}(1996)\citenamefont {Sales},
  \citenamefont {Mandrus},\ and\ \citenamefont {Williams}}]{sales1996filled}%
  \BibitemOpen
  \bibfield  {author} {\bibinfo {author} {\bibfnamefont {B.}~\bibnamefont
  {Sales}}, \bibinfo {author} {\bibfnamefont {D.}~\bibnamefont {Mandrus}}, \
  and\ \bibinfo {author} {\bibfnamefont {R.~K.}\ \bibnamefont {Williams}},\
  }\href@noop {} {\bibfield  {journal} {\bibinfo  {journal} {Science}\ }\textbf
  {\bibinfo {volume} {272}},\ \bibinfo {pages} {1325} (\bibinfo {year}
  {1996})}\BibitemShut {NoStop}%
\bibitem [{\citenamefont {Zhou}\ \emph {et~al.}(2021)\citenamefont {Zhou},
  \citenamefont {Lee}, \citenamefont {Yu}, \citenamefont {Byun}, \citenamefont
  {Luo}, \citenamefont {Lee}, \citenamefont {Ge}, \citenamefont {Lee},
  \citenamefont {Chen}, \citenamefont {Lee}, \citenamefont {Oana},
  \citenamefont {Chang}, \citenamefont {Im}, \citenamefont {Cho}, \citenamefont
  {Wuttig}, \citenamefont {Dravid}, \citenamefont {Kanatzidis},\ and\
  \citenamefont {Chung}}]{zhou2021polycrystalline}%
  \BibitemOpen
  \bibfield  {author} {\bibinfo {author} {\bibfnamefont {C.}~\bibnamefont
  {Zhou}}, \bibinfo {author} {\bibfnamefont {Y.~K.}\ \bibnamefont {Lee}},
  \bibinfo {author} {\bibfnamefont {Y.}~\bibnamefont {Yu}}, \bibinfo {author}
  {\bibfnamefont {S.}~\bibnamefont {Byun}}, \bibinfo {author} {\bibfnamefont
  {Z.-Z.}\ \bibnamefont {Luo}}, \bibinfo {author} {\bibfnamefont
  {H.}~\bibnamefont {Lee}}, \bibinfo {author} {\bibfnamefont {B.}~\bibnamefont
  {Ge}}, \bibinfo {author} {\bibfnamefont {Y.-L.}\ \bibnamefont {Lee}},
  \bibinfo {author} {\bibfnamefont {X.}~\bibnamefont {Chen}}, \bibinfo {author}
  {\bibfnamefont {J.~Y.}\ \bibnamefont {Lee}}, \bibinfo {author} {\bibfnamefont
  {C.-M.}\ \bibnamefont {Oana}}, \bibinfo {author} {\bibfnamefont
  {H.}~\bibnamefont {Chang}}, \bibinfo {author} {\bibfnamefont
  {J.}~\bibnamefont {Im}}, \bibinfo {author} {\bibfnamefont {S.-P.}\
  \bibnamefont {Cho}}, \bibinfo {author} {\bibfnamefont {M.}~\bibnamefont
  {Wuttig}}, \bibinfo {author} {\bibfnamefont {V.~P.}\ \bibnamefont {Dravid}},
  \bibinfo {author} {\bibfnamefont {M.~G.}\ \bibnamefont {Kanatzidis}}, \ and\
  \bibinfo {author} {\bibfnamefont {I.}~\bibnamefont {Chung}},\ }\href@noop {}
  {\bibfield  {journal} {\bibinfo  {journal} {Nature materials}\ ,\ \bibinfo
  {pages} {1}} (\bibinfo {year} {2021})}\BibitemShut {NoStop}%
\bibitem [{\citenamefont {Pei}\ \emph {et~al.}(2011)\citenamefont {Pei},
  \citenamefont {Shi}, \citenamefont {LaLonde}, \citenamefont {Wang},
  \citenamefont {Chen},\ and\ \citenamefont {Snyder}}]{pei2011convergence}%
  \BibitemOpen
  \bibfield  {author} {\bibinfo {author} {\bibfnamefont {Y.}~\bibnamefont
  {Pei}}, \bibinfo {author} {\bibfnamefont {X.}~\bibnamefont {Shi}}, \bibinfo
  {author} {\bibfnamefont {A.}~\bibnamefont {LaLonde}}, \bibinfo {author}
  {\bibfnamefont {H.}~\bibnamefont {Wang}}, \bibinfo {author} {\bibfnamefont
  {L.}~\bibnamefont {Chen}}, \ and\ \bibinfo {author} {\bibfnamefont {G.~J.}\
  \bibnamefont {Snyder}},\ }\href@noop {} {\bibfield  {journal} {\bibinfo
  {journal} {Nature}\ }\textbf {\bibinfo {volume} {473}},\ \bibinfo {pages}
  {66} (\bibinfo {year} {2011})}\BibitemShut {NoStop}%
\bibitem [{\citenamefont {Pei}\ \emph {et~al.}(2012)\citenamefont {Pei},
  \citenamefont {Wang},\ and\ \citenamefont {Snyder}}]{pei2012band}%
  \BibitemOpen
  \bibfield  {author} {\bibinfo {author} {\bibfnamefont {Y.}~\bibnamefont
  {Pei}}, \bibinfo {author} {\bibfnamefont {H.}~\bibnamefont {Wang}}, \ and\
  \bibinfo {author} {\bibfnamefont {G.~J.}\ \bibnamefont {Snyder}},\
  }\href@noop {} {\bibfield  {journal} {\bibinfo  {journal} {Advanced
  Materials}\ }\textbf {\bibinfo {volume} {24}},\ \bibinfo {pages} {6125}
  (\bibinfo {year} {2012})}\BibitemShut {NoStop}%
\bibitem [{\citenamefont {Dehkordi}\ \emph {et~al.}(2015)\citenamefont
  {Dehkordi}, \citenamefont {Zebarjadi}, \citenamefont {He},\ and\
  \citenamefont {Tritt}}]{dehkordi2015thermoelectric}%
  \BibitemOpen
  \bibfield  {author} {\bibinfo {author} {\bibfnamefont {A.~M.}\ \bibnamefont
  {Dehkordi}}, \bibinfo {author} {\bibfnamefont {M.}~\bibnamefont {Zebarjadi}},
  \bibinfo {author} {\bibfnamefont {J.}~\bibnamefont {He}}, \ and\ \bibinfo
  {author} {\bibfnamefont {T.~M.}\ \bibnamefont {Tritt}},\ }\href@noop {}
  {\bibfield  {journal} {\bibinfo  {journal} {Materials Science and
  Engineering: R: Reports}\ }\textbf {\bibinfo {volume} {97}},\ \bibinfo
  {pages} {1} (\bibinfo {year} {2015})}\BibitemShut {NoStop}%
\bibitem [{\citenamefont {Zhao}\ \emph {et~al.}(2013)\citenamefont {Zhao},
  \citenamefont {Hao}, \citenamefont {Lo}, \citenamefont {Wu}, \citenamefont
  {Zhou}, \citenamefont {Lee}, \citenamefont {Li}, \citenamefont {Biswas},
  \citenamefont {Hogan}, \citenamefont {Uher}, \citenamefont {Wolverton},
  \citenamefont {Dravid},\ and\ \citenamefont {G}}]{zhao2013high}%
  \BibitemOpen
  \bibfield  {author} {\bibinfo {author} {\bibfnamefont {L.-D.}\ \bibnamefont
  {Zhao}}, \bibinfo {author} {\bibfnamefont {S.}~\bibnamefont {Hao}}, \bibinfo
  {author} {\bibfnamefont {S.-H.}\ \bibnamefont {Lo}}, \bibinfo {author}
  {\bibfnamefont {C.-I.}\ \bibnamefont {Wu}}, \bibinfo {author} {\bibfnamefont
  {X.}~\bibnamefont {Zhou}}, \bibinfo {author} {\bibfnamefont {Y.}~\bibnamefont
  {Lee}}, \bibinfo {author} {\bibfnamefont {H.}~\bibnamefont {Li}}, \bibinfo
  {author} {\bibfnamefont {K.}~\bibnamefont {Biswas}}, \bibinfo {author}
  {\bibfnamefont {T.~P.}\ \bibnamefont {Hogan}}, \bibinfo {author}
  {\bibfnamefont {C.}~\bibnamefont {Uher}}, \bibinfo {author} {\bibfnamefont
  {C.}~\bibnamefont {Wolverton}}, \bibinfo {author} {\bibfnamefont {V.~P.}\
  \bibnamefont {Dravid}}, \ and\ \bibinfo {author} {\bibfnamefont {K.~M.}\
  \bibnamefont {G}},\ }\href@noop {} {\bibfield  {journal} {\bibinfo  {journal}
  {Journal of the American Chemical Society}\ }\textbf {\bibinfo {volume}
  {135}},\ \bibinfo {pages} {7364} (\bibinfo {year} {2013})}\BibitemShut
  {NoStop}%
\bibitem [{\citenamefont {Liu}\ \emph {et~al.}(2012{\natexlab{a}})\citenamefont
  {Liu}, \citenamefont {Tan}, \citenamefont {Yin}, \citenamefont {Liu},
  \citenamefont {Tang}, \citenamefont {Shi}, \citenamefont {Zhang},\ and\
  \citenamefont {Uher}}]{liu2012convergence}%
  \BibitemOpen
  \bibfield  {author} {\bibinfo {author} {\bibfnamefont {W.}~\bibnamefont
  {Liu}}, \bibinfo {author} {\bibfnamefont {X.}~\bibnamefont {Tan}}, \bibinfo
  {author} {\bibfnamefont {K.}~\bibnamefont {Yin}}, \bibinfo {author}
  {\bibfnamefont {H.}~\bibnamefont {Liu}}, \bibinfo {author} {\bibfnamefont
  {X.}~\bibnamefont {Tang}}, \bibinfo {author} {\bibfnamefont {J.}~\bibnamefont
  {Shi}}, \bibinfo {author} {\bibfnamefont {Q.}~\bibnamefont {Zhang}}, \ and\
  \bibinfo {author} {\bibfnamefont {C.}~\bibnamefont {Uher}},\ }\href@noop {}
  {\bibfield  {journal} {\bibinfo  {journal} {Physical Review Letters}\
  }\textbf {\bibinfo {volume} {108}},\ \bibinfo {pages} {166601} (\bibinfo
  {year} {2012}{\natexlab{a}})}\BibitemShut {NoStop}%
\bibitem [{\citenamefont {Parker}\ \emph {et~al.}(2015)\citenamefont {Parker},
  \citenamefont {May},\ and\ \citenamefont {Singh}}]{parker2015benefits}%
  \BibitemOpen
  \bibfield  {author} {\bibinfo {author} {\bibfnamefont {D.~S.}\ \bibnamefont
  {Parker}}, \bibinfo {author} {\bibfnamefont {A.~F.}\ \bibnamefont {May}}, \
  and\ \bibinfo {author} {\bibfnamefont {D.~J.}\ \bibnamefont {Singh}},\
  }\href@noop {} {\bibfield  {journal} {\bibinfo  {journal} {Physical Review
  Applied}\ }\textbf {\bibinfo {volume} {3}},\ \bibinfo {pages} {064003}
  (\bibinfo {year} {2015})}\BibitemShut {NoStop}%
\bibitem [{\citenamefont {Chen}\ \emph {et~al.}(2013)\citenamefont {Chen},
  \citenamefont {Parker},\ and\ \citenamefont {Singh}}]{chen2013importance}%
  \BibitemOpen
  \bibfield  {author} {\bibinfo {author} {\bibfnamefont {X.}~\bibnamefont
  {Chen}}, \bibinfo {author} {\bibfnamefont {D.}~\bibnamefont {Parker}}, \ and\
  \bibinfo {author} {\bibfnamefont {D.~J.}\ \bibnamefont {Singh}},\ }\href@noop
  {} {\bibfield  {journal} {\bibinfo  {journal} {Scientific Reports}\ }\textbf
  {\bibinfo {volume} {3}},\ \bibinfo {pages} {3168} (\bibinfo {year}
  {2013})}\BibitemShut {NoStop}%
\bibitem [{\citenamefont {Gonz{\'a}lez-Romero}\ \emph
  {et~al.}(2018)\citenamefont {Gonz{\'a}lez-Romero}, \citenamefont {Antonelli},
  \citenamefont {Chaves},\ and\ \citenamefont
  {Mel{\'e}ndez}}]{gonzalez2018ultralow}%
  \BibitemOpen
  \bibfield  {author} {\bibinfo {author} {\bibfnamefont {R.~L.}\ \bibnamefont
  {Gonz{\'a}lez-Romero}}, \bibinfo {author} {\bibfnamefont {A.}~\bibnamefont
  {Antonelli}}, \bibinfo {author} {\bibfnamefont {A.~S.}\ \bibnamefont
  {Chaves}}, \ and\ \bibinfo {author} {\bibfnamefont {J.~J.}\ \bibnamefont
  {Mel{\'e}ndez}},\ }\href@noop {} {\bibfield  {journal} {\bibinfo  {journal}
  {Physical Chemistry Chemical Physics}\ }\textbf {\bibinfo {volume} {20}},\
  \bibinfo {pages} {1809} (\bibinfo {year} {2018})}\BibitemShut {NoStop}%
\bibitem [{\citenamefont {McKinney}\ \emph {et~al.}(2017)\citenamefont
  {McKinney}, \citenamefont {Gorai}, \citenamefont {Stevanovi{\'c}},\ and\
  \citenamefont {Toberer}}]{mckinney2017search}%
  \BibitemOpen
  \bibfield  {author} {\bibinfo {author} {\bibfnamefont {R.~W.}\ \bibnamefont
  {McKinney}}, \bibinfo {author} {\bibfnamefont {P.}~\bibnamefont {Gorai}},
  \bibinfo {author} {\bibfnamefont {V.}~\bibnamefont {Stevanovi{\'c}}}, \ and\
  \bibinfo {author} {\bibfnamefont {E.~S.}\ \bibnamefont {Toberer}},\
  }\href@noop {} {\bibfield  {journal} {\bibinfo  {journal} {Journal of
  Materials Chemistry A}\ }\textbf {\bibinfo {volume} {5}},\ \bibinfo {pages}
  {17302} (\bibinfo {year} {2017})}\BibitemShut {NoStop}%
\bibitem [{\citenamefont {Chaves}\ \emph
  {et~al.}(2021{\natexlab{a}})\citenamefont {Chaves}, \citenamefont {Larson},
  \citenamefont {Kaxiras},\ and\ \citenamefont
  {Antonelli}}]{chaves2021microscopic}%
  \BibitemOpen
  \bibfield  {author} {\bibinfo {author} {\bibfnamefont {A.~S.}\ \bibnamefont
  {Chaves}}, \bibinfo {author} {\bibfnamefont {D.~T.}\ \bibnamefont {Larson}},
  \bibinfo {author} {\bibfnamefont {E.}~\bibnamefont {Kaxiras}}, \ and\
  \bibinfo {author} {\bibfnamefont {A.}~\bibnamefont {Antonelli}},\ }\href@noop
  {} {\bibfield  {journal} {\bibinfo  {journal} {Physical Review B}\ }\textbf
  {\bibinfo {volume} {104}},\ \bibinfo {pages} {115204} (\bibinfo {year}
  {2021}{\natexlab{a}})}\BibitemShut {NoStop}%
\bibitem [{\citenamefont {Hochbaum}\ \emph {et~al.}(2008)\citenamefont
  {Hochbaum}, \citenamefont {Chen}, \citenamefont {Delgado}, \citenamefont
  {Liang}, \citenamefont {Garnett}, \citenamefont {Najarian}, \citenamefont
  {Majumdar},\ and\ \citenamefont {Yang}}]{hochbaum2008enhanced}%
  \BibitemOpen
  \bibfield  {author} {\bibinfo {author} {\bibfnamefont {A.~I.}\ \bibnamefont
  {Hochbaum}}, \bibinfo {author} {\bibfnamefont {R.}~\bibnamefont {Chen}},
  \bibinfo {author} {\bibfnamefont {R.~D.}\ \bibnamefont {Delgado}}, \bibinfo
  {author} {\bibfnamefont {W.}~\bibnamefont {Liang}}, \bibinfo {author}
  {\bibfnamefont {E.~C.}\ \bibnamefont {Garnett}}, \bibinfo {author}
  {\bibfnamefont {M.}~\bibnamefont {Najarian}}, \bibinfo {author}
  {\bibfnamefont {A.}~\bibnamefont {Majumdar}}, \ and\ \bibinfo {author}
  {\bibfnamefont {P.}~\bibnamefont {Yang}},\ }\href@noop {} {\bibfield
  {journal} {\bibinfo  {journal} {Nature}\ }\textbf {\bibinfo {volume} {451}},\
  \bibinfo {pages} {163} (\bibinfo {year} {2008})}\BibitemShut {NoStop}%
\bibitem [{\citenamefont {Boukai}\ \emph {et~al.}(2008)\citenamefont {Boukai},
  \citenamefont {Bunimovich}, \citenamefont {Tahir-Kheli}, \citenamefont {Yu},
  \citenamefont {Goddard~III},\ and\ \citenamefont
  {Heath}}]{boukai2008silicon}%
  \BibitemOpen
  \bibfield  {author} {\bibinfo {author} {\bibfnamefont {A.~I.}\ \bibnamefont
  {Boukai}}, \bibinfo {author} {\bibfnamefont {Y.}~\bibnamefont {Bunimovich}},
  \bibinfo {author} {\bibfnamefont {J.}~\bibnamefont {Tahir-Kheli}}, \bibinfo
  {author} {\bibfnamefont {J.-K.}\ \bibnamefont {Yu}}, \bibinfo {author}
  {\bibfnamefont {W.~A.}\ \bibnamefont {Goddard~III}}, \ and\ \bibinfo {author}
  {\bibfnamefont {J.~R.}\ \bibnamefont {Heath}},\ }\href@noop {} {\bibfield
  {journal} {\bibinfo  {journal} {Nature}\ }\textbf {\bibinfo {volume} {451}},\
  \bibinfo {pages} {168} (\bibinfo {year} {2008})}\BibitemShut {NoStop}%
\bibitem [{\citenamefont {Kanatzidis}(2009)}]{kanatzidis2009nanostructured}%
  \BibitemOpen
  \bibfield  {author} {\bibinfo {author} {\bibfnamefont {M.~G.}\ \bibnamefont
  {Kanatzidis}},\ }\href@noop {} {\bibfield  {journal} {\bibinfo  {journal}
  {Chemistry of Materials}\ }\textbf {\bibinfo {volume} {22}},\ \bibinfo
  {pages} {648} (\bibinfo {year} {2009})}\BibitemShut {NoStop}%
\bibitem [{\citenamefont {Vineis}\ \emph {et~al.}(2010)\citenamefont {Vineis},
  \citenamefont {Shakouri}, \citenamefont {Majumdar},\ and\ \citenamefont
  {Kanatzidis}}]{vineis2010nanostructured}%
  \BibitemOpen
  \bibfield  {author} {\bibinfo {author} {\bibfnamefont {C.~J.}\ \bibnamefont
  {Vineis}}, \bibinfo {author} {\bibfnamefont {A.}~\bibnamefont {Shakouri}},
  \bibinfo {author} {\bibfnamefont {A.}~\bibnamefont {Majumdar}}, \ and\
  \bibinfo {author} {\bibfnamefont {M.~G.}\ \bibnamefont {Kanatzidis}},\
  }\href@noop {} {\bibfield  {journal} {\bibinfo  {journal} {Advanced
  Materials}\ }\textbf {\bibinfo {volume} {22}},\ \bibinfo {pages} {3970}
  (\bibinfo {year} {2010})}\BibitemShut {NoStop}%
\bibitem [{\citenamefont {Biswas}\ \emph {et~al.}(2012)\citenamefont {Biswas},
  \citenamefont {He}, \citenamefont {Blum}, \citenamefont {Wu}, \citenamefont
  {Hogan}, \citenamefont {Seidman}, \citenamefont {Dravid},\ and\ \citenamefont
  {Kanatzidis}}]{biswas2012high}%
  \BibitemOpen
  \bibfield  {author} {\bibinfo {author} {\bibfnamefont {K.}~\bibnamefont
  {Biswas}}, \bibinfo {author} {\bibfnamefont {J.}~\bibnamefont {He}}, \bibinfo
  {author} {\bibfnamefont {I.~D.}\ \bibnamefont {Blum}}, \bibinfo {author}
  {\bibfnamefont {C.-I.}\ \bibnamefont {Wu}}, \bibinfo {author} {\bibfnamefont
  {T.~P.}\ \bibnamefont {Hogan}}, \bibinfo {author} {\bibfnamefont {D.~N.}\
  \bibnamefont {Seidman}}, \bibinfo {author} {\bibfnamefont {V.~P.}\
  \bibnamefont {Dravid}}, \ and\ \bibinfo {author} {\bibfnamefont {M.~G.}\
  \bibnamefont {Kanatzidis}},\ }\href@noop {} {\bibfield  {journal} {\bibinfo
  {journal} {Nature}\ }\textbf {\bibinfo {volume} {489}},\ \bibinfo {pages}
  {414} (\bibinfo {year} {2012})}\BibitemShut {NoStop}%
\bibitem [{\citenamefont {Liu}\ \emph {et~al.}(2012{\natexlab{b}})\citenamefont
  {Liu}, \citenamefont {Shi}, \citenamefont {Xu}, \citenamefont {Zhang},
  \citenamefont {Zhang}, \citenamefont {Chen}, \citenamefont {Li},
  \citenamefont {Uher}, \citenamefont {Day},\ and\ \citenamefont
  {Snyder}}]{liu2012copper}%
  \BibitemOpen
  \bibfield  {author} {\bibinfo {author} {\bibfnamefont {H.}~\bibnamefont
  {Liu}}, \bibinfo {author} {\bibfnamefont {X.}~\bibnamefont {Shi}}, \bibinfo
  {author} {\bibfnamefont {F.}~\bibnamefont {Xu}}, \bibinfo {author}
  {\bibfnamefont {L.}~\bibnamefont {Zhang}}, \bibinfo {author} {\bibfnamefont
  {W.}~\bibnamefont {Zhang}}, \bibinfo {author} {\bibfnamefont
  {L.}~\bibnamefont {Chen}}, \bibinfo {author} {\bibfnamefont {Q.}~\bibnamefont
  {Li}}, \bibinfo {author} {\bibfnamefont {C.}~\bibnamefont {Uher}}, \bibinfo
  {author} {\bibfnamefont {T.}~\bibnamefont {Day}}, \ and\ \bibinfo {author}
  {\bibfnamefont {G.~J.}\ \bibnamefont {Snyder}},\ }\href@noop {} {\bibfield
  {journal} {\bibinfo  {journal} {Nature Materials}\ }\textbf {\bibinfo
  {volume} {11}},\ \bibinfo {pages} {422} (\bibinfo {year}
  {2012}{\natexlab{b}})}\BibitemShut {NoStop}%
\bibitem [{\citenamefont {Olvera}\ \emph {et~al.}(2017)\citenamefont {Olvera},
  \citenamefont {Moroz}, \citenamefont {Sahoo}, \citenamefont {Ren},
  \citenamefont {Bailey}, \citenamefont {Page}, \citenamefont {Uher},\ and\
  \citenamefont {Poudeu}}]{olvera2017partial}%
  \BibitemOpen
  \bibfield  {author} {\bibinfo {author} {\bibfnamefont {A.}~\bibnamefont
  {Olvera}}, \bibinfo {author} {\bibfnamefont {N.}~\bibnamefont {Moroz}},
  \bibinfo {author} {\bibfnamefont {P.}~\bibnamefont {Sahoo}}, \bibinfo
  {author} {\bibfnamefont {P.}~\bibnamefont {Ren}}, \bibinfo {author}
  {\bibfnamefont {T.}~\bibnamefont {Bailey}}, \bibinfo {author} {\bibfnamefont
  {A.}~\bibnamefont {Page}}, \bibinfo {author} {\bibfnamefont {C.}~\bibnamefont
  {Uher}}, \ and\ \bibinfo {author} {\bibfnamefont {P.}~\bibnamefont
  {Poudeu}},\ }\href@noop {} {\bibfield  {journal} {\bibinfo  {journal} {Energy
  \& Environmental Science}\ }\textbf {\bibinfo {volume} {10}},\ \bibinfo
  {pages} {1668} (\bibinfo {year} {2017})}\BibitemShut {NoStop}%
\bibitem [{\citenamefont {Cheng}\ \emph {et~al.}(2017)\citenamefont {Cheng},
  \citenamefont {Yang}, \citenamefont {Jiang}, \citenamefont {He},
  \citenamefont {He}, \citenamefont {Luo}, \citenamefont {Zhang}, \citenamefont
  {Zhou}, \citenamefont {Ren},\ and\ \citenamefont {Xin}}]{cheng2017new}%
  \BibitemOpen
  \bibfield  {author} {\bibinfo {author} {\bibfnamefont {Y.}~\bibnamefont
  {Cheng}}, \bibinfo {author} {\bibfnamefont {J.}~\bibnamefont {Yang}},
  \bibinfo {author} {\bibfnamefont {Q.}~\bibnamefont {Jiang}}, \bibinfo
  {author} {\bibfnamefont {D.}~\bibnamefont {He}}, \bibinfo {author}
  {\bibfnamefont {J.}~\bibnamefont {He}}, \bibinfo {author} {\bibfnamefont
  {Y.}~\bibnamefont {Luo}}, \bibinfo {author} {\bibfnamefont {D.}~\bibnamefont
  {Zhang}}, \bibinfo {author} {\bibfnamefont {Z.}~\bibnamefont {Zhou}},
  \bibinfo {author} {\bibfnamefont {Y.}~\bibnamefont {Ren}}, \ and\ \bibinfo
  {author} {\bibfnamefont {J.}~\bibnamefont {Xin}},\ }\href@noop {} {\bibfield
  {journal} {\bibinfo  {journal} {Journal of Materials Chemistry A}\ }\textbf
  {\bibinfo {volume} {5}},\ \bibinfo {pages} {5163} (\bibinfo {year}
  {2017})}\BibitemShut {NoStop}%
\bibitem [{\citenamefont {Ma}\ \emph {et~al.}(2020)\citenamefont {Ma},
  \citenamefont {Li}, \citenamefont {Chen},\ and\ \citenamefont
  {Wu}}]{ma2020alpha}%
  \BibitemOpen
  \bibfield  {author} {\bibinfo {author} {\bibfnamefont {N.}~\bibnamefont
  {Ma}}, \bibinfo {author} {\bibfnamefont {Y.-Y.}\ \bibnamefont {Li}}, \bibinfo
  {author} {\bibfnamefont {L.}~\bibnamefont {Chen}}, \ and\ \bibinfo {author}
  {\bibfnamefont {L.-M.}\ \bibnamefont {Wu}},\ }\href@noop {} {\bibfield
  {journal} {\bibinfo  {journal} {Journal of the American Chemical Society}\
  }\textbf {\bibinfo {volume} {142}},\ \bibinfo {pages} {5293} (\bibinfo {year}
  {2020})}\BibitemShut {NoStop}%
\bibitem [{\citenamefont {Roychowdhury}\ \emph {et~al.}(2021)\citenamefont
  {Roychowdhury}, \citenamefont {Ghosh}, \citenamefont {Arora}, \citenamefont
  {Samanta}, \citenamefont {Xie}, \citenamefont {Singh}, \citenamefont {Soni},
  \citenamefont {He}, \citenamefont {Waghmare},\ and\ \citenamefont
  {Biswas}}]{roychowdhury2021enhanced}%
  \BibitemOpen
  \bibfield  {author} {\bibinfo {author} {\bibfnamefont {S.}~\bibnamefont
  {Roychowdhury}}, \bibinfo {author} {\bibfnamefont {T.}~\bibnamefont {Ghosh}},
  \bibinfo {author} {\bibfnamefont {R.}~\bibnamefont {Arora}}, \bibinfo
  {author} {\bibfnamefont {M.}~\bibnamefont {Samanta}}, \bibinfo {author}
  {\bibfnamefont {L.}~\bibnamefont {Xie}}, \bibinfo {author} {\bibfnamefont
  {N.~K.}\ \bibnamefont {Singh}}, \bibinfo {author} {\bibfnamefont
  {A.}~\bibnamefont {Soni}}, \bibinfo {author} {\bibfnamefont {J.}~\bibnamefont
  {He}}, \bibinfo {author} {\bibfnamefont {U.~V.}\ \bibnamefont {Waghmare}}, \
  and\ \bibinfo {author} {\bibfnamefont {K.}~\bibnamefont {Biswas}},\
  }\href@noop {} {\bibfield  {journal} {\bibinfo  {journal} {Science}\ }\textbf
  {\bibinfo {volume} {371}},\ \bibinfo {pages} {722} (\bibinfo {year}
  {2021})}\BibitemShut {NoStop}%
\bibitem [{\citenamefont {Taniguchi}\ \emph {et~al.}(1990)\citenamefont
  {Taniguchi}, \citenamefont {Johnson}, \citenamefont {Ghijsen},\ and\
  \citenamefont {Cardona}}]{taniguchi1990core}%
  \BibitemOpen
  \bibfield  {author} {\bibinfo {author} {\bibfnamefont {M.}~\bibnamefont
  {Taniguchi}}, \bibinfo {author} {\bibfnamefont {R.}~\bibnamefont {Johnson}},
  \bibinfo {author} {\bibfnamefont {J.}~\bibnamefont {Ghijsen}}, \ and\
  \bibinfo {author} {\bibfnamefont {M.}~\bibnamefont {Cardona}},\ }\href@noop
  {} {\bibfield  {journal} {\bibinfo  {journal} {Physical Review B}\ }\textbf
  {\bibinfo {volume} {42}},\ \bibinfo {pages} {3634} (\bibinfo {year}
  {1990})}\BibitemShut {NoStop}%
\bibitem [{\citenamefont {Okazaki}(1958)}]{okazaki1958crystal}%
  \BibitemOpen
  \bibfield  {author} {\bibinfo {author} {\bibfnamefont {A.}~\bibnamefont
  {Okazaki}},\ }\href@noop {} {\bibfield  {journal} {\bibinfo  {journal}
  {Journal of the Physical Society of Japan}\ }\textbf {\bibinfo {volume}
  {13}},\ \bibinfo {pages} {1151} (\bibinfo {year} {1958})}\BibitemShut
  {NoStop}%
\bibitem [{\citenamefont {Antunez}\ \emph {et~al.}(2011)\citenamefont
  {Antunez}, \citenamefont {Buckley},\ and\ \citenamefont
  {Brutchey}}]{antunez2011tin}%
  \BibitemOpen
  \bibfield  {author} {\bibinfo {author} {\bibfnamefont {P.~D.}\ \bibnamefont
  {Antunez}}, \bibinfo {author} {\bibfnamefont {J.~J.}\ \bibnamefont
  {Buckley}}, \ and\ \bibinfo {author} {\bibfnamefont {R.~L.}\ \bibnamefont
  {Brutchey}},\ }\href@noop {} {\bibfield  {journal} {\bibinfo  {journal}
  {Nanoscale}\ }\textbf {\bibinfo {volume} {3}},\ \bibinfo {pages} {2399}
  (\bibinfo {year} {2011})}\BibitemShut {NoStop}%
\bibitem [{\citenamefont {Liu}\ \emph {et~al.}(2017)\citenamefont {Liu},
  \citenamefont {Mi}, \citenamefont {Xue}, \citenamefont {Chen}, \citenamefont
  {He}, \citenamefont {Liu}, \citenamefont {Hu},\ and\ \citenamefont
  {Wan}}]{liu2017investigation}%
  \BibitemOpen
  \bibfield  {author} {\bibinfo {author} {\bibfnamefont {S.-C.}\ \bibnamefont
  {Liu}}, \bibinfo {author} {\bibfnamefont {Y.}~\bibnamefont {Mi}}, \bibinfo
  {author} {\bibfnamefont {D.-J.}\ \bibnamefont {Xue}}, \bibinfo {author}
  {\bibfnamefont {Y.-X.}\ \bibnamefont {Chen}}, \bibinfo {author}
  {\bibfnamefont {C.}~\bibnamefont {He}}, \bibinfo {author} {\bibfnamefont
  {X.}~\bibnamefont {Liu}}, \bibinfo {author} {\bibfnamefont {J.-S.}\
  \bibnamefont {Hu}}, \ and\ \bibinfo {author} {\bibfnamefont {L.-J.}\
  \bibnamefont {Wan}},\ }\href@noop {} {\bibfield  {journal} {\bibinfo
  {journal} {Advanced Electronic Materials}\ }\textbf {\bibinfo {volume} {3}},\
  \bibinfo {pages} {1700141} (\bibinfo {year} {2017})}\BibitemShut {NoStop}%
\bibitem [{\citenamefont {Huang}\ \emph {et~al.}(2017)\citenamefont {Huang},
  \citenamefont {Miller}, \citenamefont {Ge}, \citenamefont {Yan},
  \citenamefont {Anand}, \citenamefont {Wu}, \citenamefont {Nan}, \citenamefont
  {Zhu}, \citenamefont {Zhuang}, \citenamefont {Snyder}, \citenamefont
  {Jiang},\ and\ \citenamefont {Bao}}]{huang2017high}%
  \BibitemOpen
  \bibfield  {author} {\bibinfo {author} {\bibfnamefont {Z.}~\bibnamefont
  {Huang}}, \bibinfo {author} {\bibfnamefont {S.~A.}\ \bibnamefont {Miller}},
  \bibinfo {author} {\bibfnamefont {B.}~\bibnamefont {Ge}}, \bibinfo {author}
  {\bibfnamefont {M.}~\bibnamefont {Yan}}, \bibinfo {author} {\bibfnamefont
  {S.}~\bibnamefont {Anand}}, \bibinfo {author} {\bibfnamefont
  {T.}~\bibnamefont {Wu}}, \bibinfo {author} {\bibfnamefont {P.}~\bibnamefont
  {Nan}}, \bibinfo {author} {\bibfnamefont {Y.}~\bibnamefont {Zhu}}, \bibinfo
  {author} {\bibfnamefont {W.}~\bibnamefont {Zhuang}}, \bibinfo {author}
  {\bibfnamefont {G.~J.}\ \bibnamefont {Snyder}}, \bibinfo {author}
  {\bibfnamefont {P.}~\bibnamefont {Jiang}}, \ and\ \bibinfo {author}
  {\bibfnamefont {X.}~\bibnamefont {Bao}},\ }\href@noop {} {\bibfield
  {journal} {\bibinfo  {journal} {Angewandte Chemie International Edition}\
  }\textbf {\bibinfo {volume} {56}},\ \bibinfo {pages} {14113} (\bibinfo {year}
  {2017})}\BibitemShut {NoStop}%
\bibitem [{\citenamefont {Sarkar}\ \emph {et~al.}(2020)\citenamefont {Sarkar},
  \citenamefont {Ghosh}, \citenamefont {Roychowdhury}, \citenamefont {Arora},
  \citenamefont {Sajan}, \citenamefont {Sheet}, \citenamefont {Waghmare},\ and\
  \citenamefont {Biswas}}]{sarkar2020ferroelectric}%
  \BibitemOpen
  \bibfield  {author} {\bibinfo {author} {\bibfnamefont {D.}~\bibnamefont
  {Sarkar}}, \bibinfo {author} {\bibfnamefont {T.}~\bibnamefont {Ghosh}},
  \bibinfo {author} {\bibfnamefont {S.}~\bibnamefont {Roychowdhury}}, \bibinfo
  {author} {\bibfnamefont {R.}~\bibnamefont {Arora}}, \bibinfo {author}
  {\bibfnamefont {S.}~\bibnamefont {Sajan}}, \bibinfo {author} {\bibfnamefont
  {G.}~\bibnamefont {Sheet}}, \bibinfo {author} {\bibfnamefont {U.~V.}\
  \bibnamefont {Waghmare}}, \ and\ \bibinfo {author} {\bibfnamefont
  {K.}~\bibnamefont {Biswas}},\ }\href@noop {} {\bibfield  {journal} {\bibinfo
  {journal} {Journal of the American Chemical Society}\ }\textbf {\bibinfo
  {volume} {142}},\ \bibinfo {pages} {12237} (\bibinfo {year}
  {2020})}\BibitemShut {NoStop}%
\bibitem [{\citenamefont {Sarkar}\ \emph {et~al.}(2021)\citenamefont {Sarkar},
  \citenamefont {Roychowdhury}, \citenamefont {Arora}, \citenamefont {Ghosh},
  \citenamefont {Vasdev}, \citenamefont {Joseph}, \citenamefont {Sheet},
  \citenamefont {Waghmare},\ and\ \citenamefont
  {Biswas}}]{sarkar2021metavalent}%
  \BibitemOpen
  \bibfield  {author} {\bibinfo {author} {\bibfnamefont {D.}~\bibnamefont
  {Sarkar}}, \bibinfo {author} {\bibfnamefont {S.}~\bibnamefont
  {Roychowdhury}}, \bibinfo {author} {\bibfnamefont {R.}~\bibnamefont {Arora}},
  \bibinfo {author} {\bibfnamefont {T.}~\bibnamefont {Ghosh}}, \bibinfo
  {author} {\bibfnamefont {A.}~\bibnamefont {Vasdev}}, \bibinfo {author}
  {\bibfnamefont {B.}~\bibnamefont {Joseph}}, \bibinfo {author} {\bibfnamefont
  {G.}~\bibnamefont {Sheet}}, \bibinfo {author} {\bibfnamefont {U.~V.}\
  \bibnamefont {Waghmare}}, \ and\ \bibinfo {author} {\bibfnamefont
  {K.}~\bibnamefont {Biswas}},\ }\href@noop {} {\bibfield  {journal} {\bibinfo
  {journal} {Angewandte Chemie}\ } (\bibinfo {year} {2021})}\BibitemShut
  {NoStop}%
\bibitem [{\citenamefont {Ding}\ \emph {et~al.}(2015)\citenamefont {Ding},
  \citenamefont {Gao},\ and\ \citenamefont {Yao}}]{ding2015high}%
  \BibitemOpen
  \bibfield  {author} {\bibinfo {author} {\bibfnamefont {G.}~\bibnamefont
  {Ding}}, \bibinfo {author} {\bibfnamefont {G.}~\bibnamefont {Gao}}, \ and\
  \bibinfo {author} {\bibfnamefont {K.}~\bibnamefont {Yao}},\ }\href@noop {}
  {\bibfield  {journal} {\bibinfo  {journal} {Scientific reports}\ }\textbf
  {\bibinfo {volume} {5}},\ \bibinfo {pages} {1} (\bibinfo {year}
  {2015})}\BibitemShut {NoStop}%
\bibitem [{\citenamefont {Hao}\ \emph {et~al.}(2016)\citenamefont {Hao},
  \citenamefont {Shi}, \citenamefont {Dravid}, \citenamefont {Kanatzidis},\
  and\ \citenamefont {Wolverton}}]{hao2016computational}%
  \BibitemOpen
  \bibfield  {author} {\bibinfo {author} {\bibfnamefont {S.}~\bibnamefont
  {Hao}}, \bibinfo {author} {\bibfnamefont {F.}~\bibnamefont {Shi}}, \bibinfo
  {author} {\bibfnamefont {V.~P.}\ \bibnamefont {Dravid}}, \bibinfo {author}
  {\bibfnamefont {M.~G.}\ \bibnamefont {Kanatzidis}}, \ and\ \bibinfo {author}
  {\bibfnamefont {C.}~\bibnamefont {Wolverton}},\ }\href@noop {} {\bibfield
  {journal} {\bibinfo  {journal} {Chemistry of Materials}\ }\textbf {\bibinfo
  {volume} {28}},\ \bibinfo {pages} {3218} (\bibinfo {year}
  {2016})}\BibitemShut {NoStop}%
\bibitem [{\citenamefont {Zhang}\ \emph
  {et~al.}(2016{\natexlab{b}})\citenamefont {Zhang}, \citenamefont {Shen},
  \citenamefont {Lin}, \citenamefont {Li}, \citenamefont {Chen}, \citenamefont
  {Li},\ and\ \citenamefont {Pei}}]{zhang2016thermoelectric}%
  \BibitemOpen
  \bibfield  {author} {\bibinfo {author} {\bibfnamefont {X.}~\bibnamefont
  {Zhang}}, \bibinfo {author} {\bibfnamefont {J.}~\bibnamefont {Shen}},
  \bibinfo {author} {\bibfnamefont {S.}~\bibnamefont {Lin}}, \bibinfo {author}
  {\bibfnamefont {J.}~\bibnamefont {Li}}, \bibinfo {author} {\bibfnamefont
  {Z.}~\bibnamefont {Chen}}, \bibinfo {author} {\bibfnamefont {W.}~\bibnamefont
  {Li}}, \ and\ \bibinfo {author} {\bibfnamefont {Y.}~\bibnamefont {Pei}},\
  }\href@noop {} {\bibfield  {journal} {\bibinfo  {journal} {Journal of
  Materiomics}\ }\textbf {\bibinfo {volume} {2}},\ \bibinfo {pages} {331}
  (\bibinfo {year} {2016}{\natexlab{b}})}\BibitemShut {NoStop}%
\bibitem [{\citenamefont {Chaves}\ \emph {et~al.}(2020)\citenamefont {Chaves},
  \citenamefont {Antonelli}, \citenamefont {Larson},\ and\ \citenamefont
  {Kaxiras}}]{chaves2020boosting}%
  \BibitemOpen
  \bibfield  {author} {\bibinfo {author} {\bibfnamefont {A.~S.}\ \bibnamefont
  {Chaves}}, \bibinfo {author} {\bibfnamefont {A.}~\bibnamefont {Antonelli}},
  \bibinfo {author} {\bibfnamefont {D.~T.}\ \bibnamefont {Larson}}, \ and\
  \bibinfo {author} {\bibfnamefont {E.}~\bibnamefont {Kaxiras}},\ }\href@noop
  {} {\bibfield  {journal} {\bibinfo  {journal} {Physical Review B}\ }\textbf
  {\bibinfo {volume} {102}},\ \bibinfo {pages} {125116} (\bibinfo {year}
  {2020})}\BibitemShut {NoStop}%
\bibitem [{\citenamefont {Hohenberg}\ and\ \citenamefont
  {Kohn}(1964)}]{hohenberg1964inhomogeneous}%
  \BibitemOpen
  \bibfield  {author} {\bibinfo {author} {\bibfnamefont {P.}~\bibnamefont
  {Hohenberg}}\ and\ \bibinfo {author} {\bibfnamefont {W.}~\bibnamefont
  {Kohn}},\ }\href@noop {} {\bibfield  {journal} {\bibinfo  {journal} {Physical
  Review}\ }\textbf {\bibinfo {volume} {136}},\ \bibinfo {pages} {B864}
  (\bibinfo {year} {1964})}\BibitemShut {NoStop}%
\bibitem [{\citenamefont {Kohn}\ and\ \citenamefont
  {Sham}(1965)}]{kohn1965self}%
  \BibitemOpen
  \bibfield  {author} {\bibinfo {author} {\bibfnamefont {W.}~\bibnamefont
  {Kohn}}\ and\ \bibinfo {author} {\bibfnamefont {L.~J.}\ \bibnamefont
  {Sham}},\ }\href@noop {} {\bibfield  {journal} {\bibinfo  {journal} {Physical
  Review}\ }\textbf {\bibinfo {volume} {140}},\ \bibinfo {pages} {A1133}
  (\bibinfo {year} {1965})}\BibitemShut {NoStop}%
\bibitem [{\citenamefont {Baroni}\ \emph {et~al.}(2001)\citenamefont {Baroni},
  \citenamefont {De~Gironcoli}, \citenamefont {Dal~Corso},\ and\ \citenamefont
  {Giannozzi}}]{baroni2001phonons}%
  \BibitemOpen
  \bibfield  {author} {\bibinfo {author} {\bibfnamefont {S.}~\bibnamefont
  {Baroni}}, \bibinfo {author} {\bibfnamefont {S.}~\bibnamefont
  {De~Gironcoli}}, \bibinfo {author} {\bibfnamefont {A.}~\bibnamefont
  {Dal~Corso}}, \ and\ \bibinfo {author} {\bibfnamefont {P.}~\bibnamefont
  {Giannozzi}},\ }\href@noop {} {\bibfield  {journal} {\bibinfo  {journal}
  {Reviews of Modern Physics}\ }\textbf {\bibinfo {volume} {73}},\ \bibinfo
  {pages} {515} (\bibinfo {year} {2001})}\BibitemShut {NoStop}%
\bibitem [{SM()}]{SM}%
  \BibitemOpen
  \href@noop {} {}\bibinfo {note} {See Supplemental Material (SM) at for a more
  comprehensive theoretical framework, additional details about the temperature
  dependence of the relaxation times (RTs), comparisons between the RTs for
  different systems and axes, additional analysis of the carrier densities,
  concentration of ionized impurities and thermoelectric figure of merit for
  different axes and conditions. SM contains additional references
  [93-101]}\BibitemShut {NoStop}%
\bibitem [{\citenamefont {Pellegrini}\ \emph {et~al.}(2016)\citenamefont
  {Pellegrini}, \citenamefont {Marinelli},\ and\ \citenamefont
  {Reiche}}]{pellegrini2016physics}%
  \BibitemOpen
  \bibfield  {author} {\bibinfo {author} {\bibfnamefont {C.}~\bibnamefont
  {Pellegrini}}, \bibinfo {author} {\bibfnamefont {A.}~\bibnamefont
  {Marinelli}}, \ and\ \bibinfo {author} {\bibfnamefont {S.}~\bibnamefont
  {Reiche}},\ }\href@noop {} {\bibfield  {journal} {\bibinfo  {journal}
  {Reviews of Modern Physics}\ }\textbf {\bibinfo {volume} {88}},\ \bibinfo
  {pages} {015006} (\bibinfo {year} {2016})}\BibitemShut {NoStop}%
\bibitem [{\citenamefont {Bostedt}\ \emph {et~al.}(2016)\citenamefont
  {Bostedt}, \citenamefont {Boutet}, \citenamefont {Fritz}, \citenamefont
  {Huang}, \citenamefont {Lee}, \citenamefont {Lemke}, \citenamefont {Robert},
  \citenamefont {Schlotter}, \citenamefont {Turner},\ and\ \citenamefont
  {Williams}}]{bostedt2016linac}%
  \BibitemOpen
  \bibfield  {author} {\bibinfo {author} {\bibfnamefont {C.}~\bibnamefont
  {Bostedt}}, \bibinfo {author} {\bibfnamefont {S.}~\bibnamefont {Boutet}},
  \bibinfo {author} {\bibfnamefont {D.~M.}\ \bibnamefont {Fritz}}, \bibinfo
  {author} {\bibfnamefont {Z.}~\bibnamefont {Huang}}, \bibinfo {author}
  {\bibfnamefont {H.~J.}\ \bibnamefont {Lee}}, \bibinfo {author} {\bibfnamefont
  {H.~T.}\ \bibnamefont {Lemke}}, \bibinfo {author} {\bibfnamefont
  {A.}~\bibnamefont {Robert}}, \bibinfo {author} {\bibfnamefont {W.~F.}\
  \bibnamefont {Schlotter}}, \bibinfo {author} {\bibfnamefont {J.~J.}\
  \bibnamefont {Turner}}, \ and\ \bibinfo {author} {\bibfnamefont {G.~J.}\
  \bibnamefont {Williams}},\ }\href@noop {} {\bibfield  {journal} {\bibinfo
  {journal} {Reviews of Modern Physics}\ }\textbf {\bibinfo {volume} {88}},\
  \bibinfo {pages} {015007} (\bibinfo {year} {2016})}\BibitemShut {NoStop}%
\bibitem [{\citenamefont {Vogl}(1976)}]{vogl1976microscopic}%
  \BibitemOpen
  \bibfield  {author} {\bibinfo {author} {\bibfnamefont {P.}~\bibnamefont
  {Vogl}},\ }\href@noop {} {\bibfield  {journal} {\bibinfo  {journal} {Physical
  Review B}\ }\textbf {\bibinfo {volume} {13}},\ \bibinfo {pages} {694}
  (\bibinfo {year} {1976})}\BibitemShut {NoStop}%
\bibitem [{\citenamefont {Ehrenreich}(1959)}]{ehrenreich1959screening}%
  \BibitemOpen
  \bibfield  {author} {\bibinfo {author} {\bibfnamefont {H.}~\bibnamefont
  {Ehrenreich}},\ }\href@noop {} {\bibfield  {journal} {\bibinfo  {journal}
  {Journal of Physics and Chemistry of Solids}\ }\textbf {\bibinfo {volume}
  {8}},\ \bibinfo {pages} {130} (\bibinfo {year} {1959})}\BibitemShut {NoStop}%
\bibitem [{\citenamefont {Brooks}(1955)}]{brooks1955theory}%
  \BibitemOpen
  \bibfield  {author} {\bibinfo {author} {\bibfnamefont {H.}~\bibnamefont
  {Brooks}},\ }in\ \href@noop {} {\emph {\bibinfo {booktitle} {Advances in
  electronics and electron physics}}},\ Vol.~\bibinfo {volume} {7}\ (\bibinfo
  {publisher} {Elsevier},\ \bibinfo {year} {1955})\ pp.\ \bibinfo {pages}
  {85--182}\BibitemShut {NoStop}%
\bibitem [{\citenamefont {Chattopadhyay}\ and\ \citenamefont
  {Queisser}(1981)}]{chattopadhyay1981electron}%
  \BibitemOpen
  \bibfield  {author} {\bibinfo {author} {\bibfnamefont {D.}~\bibnamefont
  {Chattopadhyay}}\ and\ \bibinfo {author} {\bibfnamefont {H.}~\bibnamefont
  {Queisser}},\ }\href@noop {} {\bibfield  {journal} {\bibinfo  {journal}
  {Reviews of Modern Physics}\ }\textbf {\bibinfo {volume} {53}},\ \bibinfo
  {pages} {745} (\bibinfo {year} {1981})}\BibitemShut {NoStop}%
\bibitem [{\citenamefont {Chaves}\ \emph
  {et~al.}(2021{\natexlab{b}})\citenamefont {Chaves}, \citenamefont
  {Gonz{\'a}lez-Romero}, \citenamefont {Mel{\'e}ndez},\ and\ \citenamefont
  {Antonelli}}]{chaves2021investigating}%
  \BibitemOpen
  \bibfield  {author} {\bibinfo {author} {\bibfnamefont {A.~S.}\ \bibnamefont
  {Chaves}}, \bibinfo {author} {\bibfnamefont {R.~L.}\ \bibnamefont
  {Gonz{\'a}lez-Romero}}, \bibinfo {author} {\bibfnamefont {J.~J.}\
  \bibnamefont {Mel{\'e}ndez}}, \ and\ \bibinfo {author} {\bibfnamefont
  {A.}~\bibnamefont {Antonelli}},\ }\href@noop {} {\bibfield  {journal}
  {\bibinfo  {journal} {Physical Chemistry Chemical Physics}\ }\textbf
  {\bibinfo {volume} {23}},\ \bibinfo {pages} {900} (\bibinfo {year}
  {2021}{\natexlab{b}})}\BibitemShut {NoStop}%
\bibitem [{\citenamefont {Madsen}\ and\ \citenamefont
  {Singh}(2006)}]{madsen2006boltztrap}%
  \BibitemOpen
  \bibfield  {author} {\bibinfo {author} {\bibfnamefont {G.~K.}\ \bibnamefont
  {Madsen}}\ and\ \bibinfo {author} {\bibfnamefont {D.~J.}\ \bibnamefont
  {Singh}},\ }\href@noop {} {\bibfield  {journal} {\bibinfo  {journal}
  {Computer Physics Communications}\ }\textbf {\bibinfo {volume} {175}},\
  \bibinfo {pages} {67} (\bibinfo {year} {2006})}\BibitemShut {NoStop}%
\bibitem [{\citenamefont {Giannozzi}\ \emph {et~al.}(2009)\citenamefont
  {Giannozzi}, \citenamefont {Baroni}, \citenamefont {Bonini}, \citenamefont
  {Calandra}, \citenamefont {Car}, \citenamefont {Cavazzoni}, \citenamefont
  {Ceresoli}, \citenamefont {Chiarotti}, \citenamefont {Cococcioni},
  \citenamefont {Dabo}, \citenamefont {Dal~Corso}, \citenamefont
  {de~Gironcoli}, \citenamefont {Fabris}, \citenamefont {Fratesi},
  \citenamefont {Gebauer}, \citenamefont {Gerstmann}, \citenamefont
  {Gougoussis}, \citenamefont {Kokalj}, \citenamefont {Lazzeri}, \citenamefont
  {Martin-Samos}, \citenamefont {Marzari}, \citenamefont {Mauri}, \citenamefont
  {Mazzarello}, \citenamefont {Paolini}, \citenamefont {Pasquarello},
  \citenamefont {Paulatto}, \citenamefont {Sbraccia}, \citenamefont {Scandolo},
  \citenamefont {Sclauzero}, \citenamefont {Seitsonen}, \citenamefont
  {Smogunov}, \citenamefont {Umari},\ and\ \citenamefont
  {Wentzcovitch}}]{giannozzi2009quantum}%
  \BibitemOpen
  \bibfield  {author} {\bibinfo {author} {\bibfnamefont {P.}~\bibnamefont
  {Giannozzi}}, \bibinfo {author} {\bibfnamefont {S.}~\bibnamefont {Baroni}},
  \bibinfo {author} {\bibfnamefont {N.}~\bibnamefont {Bonini}}, \bibinfo
  {author} {\bibfnamefont {M.}~\bibnamefont {Calandra}}, \bibinfo {author}
  {\bibfnamefont {R.}~\bibnamefont {Car}}, \bibinfo {author} {\bibfnamefont
  {C.}~\bibnamefont {Cavazzoni}}, \bibinfo {author} {\bibfnamefont
  {D.}~\bibnamefont {Ceresoli}}, \bibinfo {author} {\bibfnamefont {G.~L.}\
  \bibnamefont {Chiarotti}}, \bibinfo {author} {\bibfnamefont {M.}~\bibnamefont
  {Cococcioni}}, \bibinfo {author} {\bibfnamefont {I.}~\bibnamefont {Dabo}},
  \bibinfo {author} {\bibfnamefont {A.}~\bibnamefont {Dal~Corso}}, \bibinfo
  {author} {\bibfnamefont {S.}~\bibnamefont {de~Gironcoli}}, \bibinfo {author}
  {\bibfnamefont {S.}~\bibnamefont {Fabris}}, \bibinfo {author} {\bibfnamefont
  {G.}~\bibnamefont {Fratesi}}, \bibinfo {author} {\bibfnamefont
  {R.}~\bibnamefont {Gebauer}}, \bibinfo {author} {\bibfnamefont
  {U.}~\bibnamefont {Gerstmann}}, \bibinfo {author} {\bibfnamefont
  {C.}~\bibnamefont {Gougoussis}}, \bibinfo {author} {\bibfnamefont
  {A.}~\bibnamefont {Kokalj}}, \bibinfo {author} {\bibfnamefont
  {M.}~\bibnamefont {Lazzeri}}, \bibinfo {author} {\bibfnamefont
  {L.}~\bibnamefont {Martin-Samos}}, \bibinfo {author} {\bibfnamefont
  {N.}~\bibnamefont {Marzari}}, \bibinfo {author} {\bibfnamefont
  {F.}~\bibnamefont {Mauri}}, \bibinfo {author} {\bibfnamefont
  {R.}~\bibnamefont {Mazzarello}}, \bibinfo {author} {\bibfnamefont
  {S.}~\bibnamefont {Paolini}}, \bibinfo {author} {\bibfnamefont
  {A.}~\bibnamefont {Pasquarello}}, \bibinfo {author} {\bibfnamefont
  {L.}~\bibnamefont {Paulatto}}, \bibinfo {author} {\bibfnamefont
  {C.}~\bibnamefont {Sbraccia}}, \bibinfo {author} {\bibfnamefont
  {S.}~\bibnamefont {Scandolo}}, \bibinfo {author} {\bibfnamefont
  {G.}~\bibnamefont {Sclauzero}}, \bibinfo {author} {\bibfnamefont {A.~P.}\
  \bibnamefont {Seitsonen}}, \bibinfo {author} {\bibfnamefont {A.}~\bibnamefont
  {Smogunov}}, \bibinfo {author} {\bibfnamefont {P.}~\bibnamefont {Umari}}, \
  and\ \bibinfo {author} {\bibfnamefont {R.~M.}\ \bibnamefont {Wentzcovitch}},\
  }\href@noop {} {\bibfield  {journal} {\bibinfo  {journal} {Journal of
  Physics: Condensed Matter}\ }\textbf {\bibinfo {volume} {21}},\ \bibinfo
  {pages} {395502} (\bibinfo {year} {2009})}\BibitemShut {NoStop}%
\bibitem [{\citenamefont {Hamann}(2013)}]{hamann2013optimized}%
  \BibitemOpen
  \bibfield  {author} {\bibinfo {author} {\bibfnamefont {D.}~\bibnamefont
  {Hamann}},\ }\href@noop {} {\bibfield  {journal} {\bibinfo  {journal}
  {Physical Review B}\ }\textbf {\bibinfo {volume} {88}},\ \bibinfo {pages}
  {085117} (\bibinfo {year} {2013})}\BibitemShut {NoStop}%
\bibitem [{\citenamefont {Van~Setten}\ \emph {et~al.}(2018)\citenamefont
  {Van~Setten}, \citenamefont {Giantomassi}, \citenamefont {Bousquet},
  \citenamefont {Verstraete}, \citenamefont {Hamann}, \citenamefont {Gonze},\
  and\ \citenamefont {Rignanese}}]{van2018pseudodojo}%
  \BibitemOpen
  \bibfield  {author} {\bibinfo {author} {\bibfnamefont {M.}~\bibnamefont
  {Van~Setten}}, \bibinfo {author} {\bibfnamefont {M.}~\bibnamefont
  {Giantomassi}}, \bibinfo {author} {\bibfnamefont {E.}~\bibnamefont
  {Bousquet}}, \bibinfo {author} {\bibfnamefont {M.~J.}\ \bibnamefont
  {Verstraete}}, \bibinfo {author} {\bibfnamefont {D.~R.}\ \bibnamefont
  {Hamann}}, \bibinfo {author} {\bibfnamefont {X.}~\bibnamefont {Gonze}}, \
  and\ \bibinfo {author} {\bibfnamefont {G.-M.}\ \bibnamefont {Rignanese}},\
  }\href@noop {} {\bibfield  {journal} {\bibinfo  {journal} {Computer Physics
  Communications}\ }\textbf {\bibinfo {volume} {226}},\ \bibinfo {pages} {39}
  (\bibinfo {year} {2018})}\BibitemShut {NoStop}%
\bibitem [{\citenamefont {Perdew}\ \emph {et~al.}(1996)\citenamefont {Perdew},
  \citenamefont {Burke},\ and\ \citenamefont
  {Ernzerhof}}]{perdew1996generalized}%
  \BibitemOpen
  \bibfield  {author} {\bibinfo {author} {\bibfnamefont {J.~P.}\ \bibnamefont
  {Perdew}}, \bibinfo {author} {\bibfnamefont {K.}~\bibnamefont {Burke}}, \
  and\ \bibinfo {author} {\bibfnamefont {M.}~\bibnamefont {Ernzerhof}},\
  }\href@noop {} {\bibfield  {journal} {\bibinfo  {journal} {Physical Review
  Letters}\ }\textbf {\bibinfo {volume} {77}},\ \bibinfo {pages} {3865}
  (\bibinfo {year} {1996})}\BibitemShut {NoStop}%
\bibitem [{\citenamefont {Elkorashy}(1989)}]{elkorashy1989photoconductivity}%
  \BibitemOpen
  \bibfield  {author} {\bibinfo {author} {\bibfnamefont {A.}~\bibnamefont
  {Elkorashy}},\ }\href@noop {} {\bibfield  {journal} {\bibinfo  {journal}
  {physica status solidi (b)}\ }\textbf {\bibinfo {volume} {152}},\ \bibinfo
  {pages} {249} (\bibinfo {year} {1989})}\BibitemShut {NoStop}%
\bibitem [{\citenamefont {Vaughn}\ \emph {et~al.}(2010)\citenamefont {Vaughn},
  \citenamefont {Patel}, \citenamefont {Hickner},\ and\ \citenamefont
  {Schaak}}]{vaughn2010single}%
  \BibitemOpen
  \bibfield  {author} {\bibinfo {author} {\bibfnamefont {D.~D.}\ \bibnamefont
  {Vaughn}}, \bibinfo {author} {\bibfnamefont {R.~J.}\ \bibnamefont {Patel}},
  \bibinfo {author} {\bibfnamefont {M.~A.}\ \bibnamefont {Hickner}}, \ and\
  \bibinfo {author} {\bibfnamefont {R.~E.}\ \bibnamefont {Schaak}},\
  }\href@noop {} {\bibfield  {journal} {\bibinfo  {journal} {Journal of the
  American Chemical Society}\ }\textbf {\bibinfo {volume} {132}},\ \bibinfo
  {pages} {15170} (\bibinfo {year} {2010})}\BibitemShut {NoStop}%
\bibitem [{\citenamefont {Ross}\ and\ \citenamefont
  {Bourgon}(1969)}]{ross1969germanium}%
  \BibitemOpen
  \bibfield  {author} {\bibinfo {author} {\bibfnamefont {L.}~\bibnamefont
  {Ross}}\ and\ \bibinfo {author} {\bibfnamefont {M.}~\bibnamefont {Bourgon}},\
  }\href@noop {} {\bibfield  {journal} {\bibinfo  {journal} {Canadian Journal
  of Chemistry}\ }\textbf {\bibinfo {volume} {47}},\ \bibinfo {pages} {2555}
  (\bibinfo {year} {1969})}\BibitemShut {NoStop}%
\bibitem [{\citenamefont {Ipser}\ \emph {et~al.}(1982)\citenamefont {Ipser},
  \citenamefont {Gambino},\ and\ \citenamefont
  {Schuster}}]{ipser1982germanium}%
  \BibitemOpen
  \bibfield  {author} {\bibinfo {author} {\bibfnamefont {H.}~\bibnamefont
  {Ipser}}, \bibinfo {author} {\bibfnamefont {M.}~\bibnamefont {Gambino}}, \
  and\ \bibinfo {author} {\bibfnamefont {W.}~\bibnamefont {Schuster}},\
  }\href@noop {} {\bibfield  {journal} {\bibinfo  {journal} {Monatshefte
  f{\"u}r Chemie/Chemical Monthly}\ }\textbf {\bibinfo {volume} {113}},\
  \bibinfo {pages} {389} (\bibinfo {year} {1982})}\BibitemShut {NoStop}%
\bibitem [{\citenamefont {Bletskan}(2005)}]{bletskan2005phase}%
  \BibitemOpen
  \bibfield  {author} {\bibinfo {author} {\bibfnamefont {D.}~\bibnamefont
  {Bletskan}},\ }\href@noop {} {\bibfield  {journal} {\bibinfo  {journal}
  {Journal of Ovonic Research}\ }\textbf {\bibinfo {volume} {1}},\ \bibinfo
  {pages} {53} (\bibinfo {year} {2005})}\BibitemShut {NoStop}%
\bibitem [{\citenamefont {Wiedemeier}\ and\ \citenamefont
  {Siemers}(1975)}]{wiedemeier1975thermal}%
  \BibitemOpen
  \bibfield  {author} {\bibinfo {author} {\bibfnamefont {H.}~\bibnamefont
  {Wiedemeier}}\ and\ \bibinfo {author} {\bibfnamefont {P.}~\bibnamefont
  {Siemers}},\ }\href@noop {} {\bibfield  {journal} {\bibinfo  {journal}
  {Zeitschrift f{\"u}r anorganische und allgemeine Chemie}\ }\textbf {\bibinfo
  {volume} {411}},\ \bibinfo {pages} {90} (\bibinfo {year} {1975})}\BibitemShut
  {NoStop}%
\bibitem [{\citenamefont {Sist}\ \emph {et~al.}(2017)\citenamefont {Sist},
  \citenamefont {Gatti}, \citenamefont {N{\o}rby}, \citenamefont {Cenedese},
  \citenamefont {Kasai}, \citenamefont {Kato},\ and\ \citenamefont
  {Iversen}}]{sist2017high}%
  \BibitemOpen
  \bibfield  {author} {\bibinfo {author} {\bibfnamefont {M.}~\bibnamefont
  {Sist}}, \bibinfo {author} {\bibfnamefont {C.}~\bibnamefont {Gatti}},
  \bibinfo {author} {\bibfnamefont {P.}~\bibnamefont {N{\o}rby}}, \bibinfo
  {author} {\bibfnamefont {S.}~\bibnamefont {Cenedese}}, \bibinfo {author}
  {\bibfnamefont {H.}~\bibnamefont {Kasai}}, \bibinfo {author} {\bibfnamefont
  {K.}~\bibnamefont {Kato}}, \ and\ \bibinfo {author} {\bibfnamefont {B.~B.}\
  \bibnamefont {Iversen}},\ }\href@noop {} {\bibfield  {journal} {\bibinfo
  {journal} {Chemistry--A European Journal}\ }\textbf {\bibinfo {volume}
  {23}},\ \bibinfo {pages} {6888} (\bibinfo {year} {2017})}\BibitemShut
  {NoStop}%
\bibitem [{\citenamefont {Li}\ \emph {et~al.}(2015)\citenamefont {Li},
  \citenamefont {Hong}, \citenamefont {May}, \citenamefont {Bansal},
  \citenamefont {Chi}, \citenamefont {Hong}, \citenamefont {Ehlers},\ and\
  \citenamefont {Delaire}}]{li2015orbitally}%
  \BibitemOpen
  \bibfield  {author} {\bibinfo {author} {\bibfnamefont {C.~W.}\ \bibnamefont
  {Li}}, \bibinfo {author} {\bibfnamefont {J.}~\bibnamefont {Hong}}, \bibinfo
  {author} {\bibfnamefont {A.~F.}\ \bibnamefont {May}}, \bibinfo {author}
  {\bibfnamefont {D.}~\bibnamefont {Bansal}}, \bibinfo {author} {\bibfnamefont
  {S.}~\bibnamefont {Chi}}, \bibinfo {author} {\bibfnamefont {T.}~\bibnamefont
  {Hong}}, \bibinfo {author} {\bibfnamefont {G.}~\bibnamefont {Ehlers}}, \ and\
  \bibinfo {author} {\bibfnamefont {O.}~\bibnamefont {Delaire}},\ }\href@noop
  {} {\bibfield  {journal} {\bibinfo  {journal} {Nature Physics}\ }\textbf
  {\bibinfo {volume} {11}},\ \bibinfo {pages} {1063} (\bibinfo {year}
  {2015})}\BibitemShut {NoStop}%
\bibitem [{\citenamefont {Grimme}\ \emph {et~al.}(2010)\citenamefont {Grimme},
  \citenamefont {Antony}, \citenamefont {Ehrlich},\ and\ \citenamefont
  {Krieg}}]{grimme2010consistent}%
  \BibitemOpen
  \bibfield  {author} {\bibinfo {author} {\bibfnamefont {S.}~\bibnamefont
  {Grimme}}, \bibinfo {author} {\bibfnamefont {J.}~\bibnamefont {Antony}},
  \bibinfo {author} {\bibfnamefont {S.}~\bibnamefont {Ehrlich}}, \ and\
  \bibinfo {author} {\bibfnamefont {H.}~\bibnamefont {Krieg}},\ }\href@noop {}
  {\bibfield  {journal} {\bibinfo  {journal} {The Journal of Chemical Physics}\
  }\textbf {\bibinfo {volume} {132}},\ \bibinfo {pages} {154104} (\bibinfo
  {year} {2010})}\BibitemShut {NoStop}%
\bibitem [{\citenamefont {Jain}\ \emph {et~al.}(2013)\citenamefont {Jain},
  \citenamefont {Ong}, \citenamefont {Hautier}, \citenamefont {Chen},
  \citenamefont {Richards}, \citenamefont {Dacek}, \citenamefont {Cholia},
  \citenamefont {Gunter}, \citenamefont {Skinner}, \citenamefont {Ceder},\ and\
  \citenamefont {Persson}}]{jain2013commentary}%
  \BibitemOpen
  \bibfield  {author} {\bibinfo {author} {\bibfnamefont {A.}~\bibnamefont
  {Jain}}, \bibinfo {author} {\bibfnamefont {S.~P.}\ \bibnamefont {Ong}},
  \bibinfo {author} {\bibfnamefont {G.}~\bibnamefont {Hautier}}, \bibinfo
  {author} {\bibfnamefont {W.}~\bibnamefont {Chen}}, \bibinfo {author}
  {\bibfnamefont {W.~D.}\ \bibnamefont {Richards}}, \bibinfo {author}
  {\bibfnamefont {S.}~\bibnamefont {Dacek}}, \bibinfo {author} {\bibfnamefont
  {S.}~\bibnamefont {Cholia}}, \bibinfo {author} {\bibfnamefont
  {D.}~\bibnamefont {Gunter}}, \bibinfo {author} {\bibfnamefont
  {D.}~\bibnamefont {Skinner}}, \bibinfo {author} {\bibfnamefont
  {G.}~\bibnamefont {Ceder}}, \ and\ \bibinfo {author} {\bibfnamefont {K.~A.}\
  \bibnamefont {Persson}},\ }\href@noop {} {\bibfield  {journal} {\bibinfo
  {journal} {APL materials}\ }\textbf {\bibinfo {volume} {1}},\ \bibinfo
  {pages} {011002} (\bibinfo {year} {2013})}\BibitemShut {NoStop}%
\bibitem [{\citenamefont {Wiedemeier}\ and\ \citenamefont {von
  Schnering}(1978)}]{wiedemeier1978refinement}%
  \BibitemOpen
  \bibfield  {author} {\bibinfo {author} {\bibfnamefont {H.}~\bibnamefont
  {Wiedemeier}}\ and\ \bibinfo {author} {\bibfnamefont {H.~G.}\ \bibnamefont
  {von Schnering}},\ }\href@noop {} {\bibfield  {journal} {\bibinfo  {journal}
  {Z. Kristallogr}\ }\textbf {\bibinfo {volume} {148}},\ \bibinfo {pages} {3}
  (\bibinfo {year} {1978})}\BibitemShut {NoStop}%
\bibitem [{\citenamefont {Giustino}\ \emph {et~al.}(2007)\citenamefont
  {Giustino}, \citenamefont {Cohen},\ and\ \citenamefont
  {Louie}}]{giustino2007electron}%
  \BibitemOpen
  \bibfield  {author} {\bibinfo {author} {\bibfnamefont {F.}~\bibnamefont
  {Giustino}}, \bibinfo {author} {\bibfnamefont {M.~L.}\ \bibnamefont {Cohen}},
  \ and\ \bibinfo {author} {\bibfnamefont {S.~G.}\ \bibnamefont {Louie}},\
  }\href@noop {} {\bibfield  {journal} {\bibinfo  {journal} {Physical Review
  B}\ }\textbf {\bibinfo {volume} {76}},\ \bibinfo {pages} {165108} (\bibinfo
  {year} {2007})}\BibitemShut {NoStop}%
\bibitem [{\citenamefont {Ponc{\'e}}\ \emph {et~al.}(2016)\citenamefont
  {Ponc{\'e}}, \citenamefont {Margine}, \citenamefont {Verdi},\ and\
  \citenamefont {Giustino}}]{ponce2016epw}%
  \BibitemOpen
  \bibfield  {author} {\bibinfo {author} {\bibfnamefont {S.}~\bibnamefont
  {Ponc{\'e}}}, \bibinfo {author} {\bibfnamefont {E.~R.}\ \bibnamefont
  {Margine}}, \bibinfo {author} {\bibfnamefont {C.}~\bibnamefont {Verdi}}, \
  and\ \bibinfo {author} {\bibfnamefont {F.}~\bibnamefont {Giustino}},\
  }\href@noop {} {\bibfield  {journal} {\bibinfo  {journal} {Computer Physics
  Communications}\ }\textbf {\bibinfo {volume} {209}},\ \bibinfo {pages} {116}
  (\bibinfo {year} {2016})}\BibitemShut {NoStop}%
\bibitem [{\citenamefont {Mostofi}\ \emph {et~al.}(2008)\citenamefont
  {Mostofi}, \citenamefont {Yates}, \citenamefont {Lee}, \citenamefont {Souza},
  \citenamefont {Vanderbilt},\ and\ \citenamefont
  {Marzari}}]{mostofi2008wannier90}%
  \BibitemOpen
  \bibfield  {author} {\bibinfo {author} {\bibfnamefont {A.~A.}\ \bibnamefont
  {Mostofi}}, \bibinfo {author} {\bibfnamefont {J.~R.}\ \bibnamefont {Yates}},
  \bibinfo {author} {\bibfnamefont {Y.-S.}\ \bibnamefont {Lee}}, \bibinfo
  {author} {\bibfnamefont {I.}~\bibnamefont {Souza}}, \bibinfo {author}
  {\bibfnamefont {D.}~\bibnamefont {Vanderbilt}}, \ and\ \bibinfo {author}
  {\bibfnamefont {N.}~\bibnamefont {Marzari}},\ }\href@noop {} {\bibfield
  {journal} {\bibinfo  {journal} {Computer Physics Communications}\ }\textbf
  {\bibinfo {volume} {178}},\ \bibinfo {pages} {685} (\bibinfo {year}
  {2008})}\BibitemShut {NoStop}%
\bibitem [{\citenamefont {Madelung}(2004)}]{madelung2004semiconductors}%
  \BibitemOpen
  \bibfield  {author} {\bibinfo {author} {\bibfnamefont {O.}~\bibnamefont
  {Madelung}},\ }\href@noop {} {\emph {\bibinfo {title} {Semiconductors: data
  handbook}}}\ (\bibinfo  {publisher} {Springer Science \& Business Media},\
  \bibinfo {year} {2004})\BibitemShut {NoStop}%
\bibitem [{\citenamefont {Chandrasekhar}\ \emph {et~al.}(1977)\citenamefont
  {Chandrasekhar}, \citenamefont {Humphreys}, \citenamefont {Zwick},\ and\
  \citenamefont {Cardona}}]{chandrasekhar1977infrared}%
  \BibitemOpen
  \bibfield  {author} {\bibinfo {author} {\bibfnamefont {H.}~\bibnamefont
  {Chandrasekhar}}, \bibinfo {author} {\bibfnamefont {R.}~\bibnamefont
  {Humphreys}}, \bibinfo {author} {\bibfnamefont {U.}~\bibnamefont {Zwick}}, \
  and\ \bibinfo {author} {\bibfnamefont {M.}~\bibnamefont {Cardona}},\
  }\href@noop {} {\bibfield  {journal} {\bibinfo  {journal} {Physical Review
  B}\ }\textbf {\bibinfo {volume} {15}},\ \bibinfo {pages} {2177} (\bibinfo
  {year} {1977})}\BibitemShut {NoStop}%
\bibitem [{\citenamefont {Wang}\ \emph {et~al.}(2018)\citenamefont {Wang},
  \citenamefont {Askarpour}, \citenamefont {Maassen},\ and\ \citenamefont
  {Lundstrom}}]{wang2018calculation}%
  \BibitemOpen
  \bibfield  {author} {\bibinfo {author} {\bibfnamefont {X.}~\bibnamefont
  {Wang}}, \bibinfo {author} {\bibfnamefont {V.}~\bibnamefont {Askarpour}},
  \bibinfo {author} {\bibfnamefont {J.}~\bibnamefont {Maassen}}, \ and\
  \bibinfo {author} {\bibfnamefont {M.}~\bibnamefont {Lundstrom}},\ }\href@noop
  {} {\bibfield  {journal} {\bibinfo  {journal} {Journal of Applied Physics}\
  }\textbf {\bibinfo {volume} {123}},\ \bibinfo {pages} {055104} (\bibinfo
  {year} {2018})}\BibitemShut {NoStop}%
\bibitem [{\citenamefont {Putatunda}\ and\ \citenamefont
  {Singh}(2019)}]{putatunda2019lorenz}%
  \BibitemOpen
  \bibfield  {author} {\bibinfo {author} {\bibfnamefont {A.}~\bibnamefont
  {Putatunda}}\ and\ \bibinfo {author} {\bibfnamefont {D.~J.}\ \bibnamefont
  {Singh}},\ }\href@noop {} {\bibfield  {journal} {\bibinfo  {journal}
  {Materials Today Physics}\ }\textbf {\bibinfo {volume} {8}},\ \bibinfo
  {pages} {49} (\bibinfo {year} {2019})}\BibitemShut {NoStop}%
\bibitem [{\citenamefont {Xiao}\ and\ \citenamefont
  {Zhao}(2020)}]{xiao2020seeking}%
  \BibitemOpen
  \bibfield  {author} {\bibinfo {author} {\bibfnamefont {Y.}~\bibnamefont
  {Xiao}}\ and\ \bibinfo {author} {\bibfnamefont {L.-D.}\ \bibnamefont
  {Zhao}},\ }\href@noop {} {\bibfield  {journal} {\bibinfo  {journal}
  {Science}\ }\textbf {\bibinfo {volume} {367}},\ \bibinfo {pages} {1196}
  (\bibinfo {year} {2020})}\BibitemShut {NoStop}%
\bibitem [{\citenamefont {Wei}\ \emph {et~al.}(2016)\citenamefont {Wei},
  \citenamefont {Bhattacharya}, \citenamefont {He}, \citenamefont {Neeleshwar},
  \citenamefont {Podila}, \citenamefont {Chen},\ and\ \citenamefont
  {Rao}}]{wei2016intrinsic}%
  \BibitemOpen
  \bibfield  {author} {\bibinfo {author} {\bibfnamefont {P.-C.}\ \bibnamefont
  {Wei}}, \bibinfo {author} {\bibfnamefont {S.}~\bibnamefont {Bhattacharya}},
  \bibinfo {author} {\bibfnamefont {J.}~\bibnamefont {He}}, \bibinfo {author}
  {\bibfnamefont {S.}~\bibnamefont {Neeleshwar}}, \bibinfo {author}
  {\bibfnamefont {R.}~\bibnamefont {Podila}}, \bibinfo {author} {\bibfnamefont
  {Y.}~\bibnamefont {Chen}}, \ and\ \bibinfo {author} {\bibfnamefont
  {A.}~\bibnamefont {Rao}},\ }\href@noop {} {\bibfield  {journal} {\bibinfo
  {journal} {Nature}\ }\textbf {\bibinfo {volume} {539}},\ \bibinfo {pages}
  {E1} (\bibinfo {year} {2016})}\BibitemShut {NoStop}%
\bibitem [{\citenamefont {Wu}\ \emph {et~al.}(2017)\citenamefont {Wu},
  \citenamefont {Wu}, \citenamefont {He}, \citenamefont {Zhao}, \citenamefont
  {Li}, \citenamefont {Wu}, \citenamefont {Jin}, \citenamefont {Xu},
  \citenamefont {Jiang}, \citenamefont {Huang}, \citenamefont {Zhu},
  \citenamefont {Kanatzidis},\ and\ \citenamefont {He}}]{wu2017direct}%
  \BibitemOpen
  \bibfield  {author} {\bibinfo {author} {\bibfnamefont {D.}~\bibnamefont
  {Wu}}, \bibinfo {author} {\bibfnamefont {L.}~\bibnamefont {Wu}}, \bibinfo
  {author} {\bibfnamefont {D.}~\bibnamefont {He}}, \bibinfo {author}
  {\bibfnamefont {L.-D.}\ \bibnamefont {Zhao}}, \bibinfo {author}
  {\bibfnamefont {W.}~\bibnamefont {Li}}, \bibinfo {author} {\bibfnamefont
  {M.}~\bibnamefont {Wu}}, \bibinfo {author} {\bibfnamefont {M.}~\bibnamefont
  {Jin}}, \bibinfo {author} {\bibfnamefont {J.}~\bibnamefont {Xu}}, \bibinfo
  {author} {\bibfnamefont {J.}~\bibnamefont {Jiang}}, \bibinfo {author}
  {\bibfnamefont {L.}~\bibnamefont {Huang}}, \bibinfo {author} {\bibfnamefont
  {Y.}~\bibnamefont {Zhu}}, \bibinfo {author} {\bibfnamefont {M.~G.}\
  \bibnamefont {Kanatzidis}}, \ and\ \bibinfo {author} {\bibfnamefont
  {J.}~\bibnamefont {He}},\ }\href@noop {} {\bibfield  {journal} {\bibinfo
  {journal} {Nano Energy}\ }\textbf {\bibinfo {volume} {35}},\ \bibinfo {pages}
  {321} (\bibinfo {year} {2017})}\BibitemShut {NoStop}%
\bibitem [{\citenamefont {Li}\ \emph {et~al.}(2017)\citenamefont {Li},
  \citenamefont {Aydemir}, \citenamefont {Wood}, \citenamefont {Goddard~III},
  \citenamefont {Zhai}, \citenamefont {Zhang},\ and\ \citenamefont
  {Snyder}}]{li2017ideal}%
  \BibitemOpen
  \bibfield  {author} {\bibinfo {author} {\bibfnamefont {G.}~\bibnamefont
  {Li}}, \bibinfo {author} {\bibfnamefont {U.}~\bibnamefont {Aydemir}},
  \bibinfo {author} {\bibfnamefont {M.}~\bibnamefont {Wood}}, \bibinfo {author}
  {\bibfnamefont {W.~A.}\ \bibnamefont {Goddard~III}}, \bibinfo {author}
  {\bibfnamefont {P.}~\bibnamefont {Zhai}}, \bibinfo {author} {\bibfnamefont
  {Q.}~\bibnamefont {Zhang}}, \ and\ \bibinfo {author} {\bibfnamefont {G.~J.}\
  \bibnamefont {Snyder}},\ }\href@noop {} {\bibfield  {journal} {\bibinfo
  {journal} {Chemistry of Materials}\ }\textbf {\bibinfo {volume} {29}},\
  \bibinfo {pages} {2382} (\bibinfo {year} {2017})}\BibitemShut {NoStop}%
\bibitem [{\citenamefont {Chen}\ \emph {et~al.}(2016)\citenamefont {Chen},
  \citenamefont {Ge}, \citenamefont {Yin}, \citenamefont {Feng}, \citenamefont
  {Huang}, \citenamefont {Zhao},\ and\ \citenamefont
  {He}}]{chen2016understanding2}%
  \BibitemOpen
  \bibfield  {author} {\bibinfo {author} {\bibfnamefont {Y.-X.}\ \bibnamefont
  {Chen}}, \bibinfo {author} {\bibfnamefont {Z.-H.}\ \bibnamefont {Ge}},
  \bibinfo {author} {\bibfnamefont {M.}~\bibnamefont {Yin}}, \bibinfo {author}
  {\bibfnamefont {D.}~\bibnamefont {Feng}}, \bibinfo {author} {\bibfnamefont
  {X.-Q.}\ \bibnamefont {Huang}}, \bibinfo {author} {\bibfnamefont
  {W.}~\bibnamefont {Zhao}}, \ and\ \bibinfo {author} {\bibfnamefont
  {J.}~\bibnamefont {He}},\ }\href@noop {} {\bibfield  {journal} {\bibinfo
  {journal} {Advanced Functional Materials}\ }\textbf {\bibinfo {volume}
  {26}},\ \bibinfo {pages} {6836} (\bibinfo {year} {2016})}\BibitemShut
  {NoStop}%
\bibitem [{\citenamefont {Yuan}\ \emph {et~al.}(2019)\citenamefont {Yuan},
  \citenamefont {Sun}, \citenamefont {Zhang},\ and\ \citenamefont
  {Tang}}]{yuan2019tailoring}%
  \BibitemOpen
  \bibfield  {author} {\bibinfo {author} {\bibfnamefont {K.}~\bibnamefont
  {Yuan}}, \bibinfo {author} {\bibfnamefont {Z.}~\bibnamefont {Sun}}, \bibinfo
  {author} {\bibfnamefont {X.}~\bibnamefont {Zhang}}, \ and\ \bibinfo {author}
  {\bibfnamefont {D.}~\bibnamefont {Tang}},\ }\href@noop {} {\bibfield
  {journal} {\bibinfo  {journal} {Scientific reports}\ }\textbf {\bibinfo
  {volume} {9}},\ \bibinfo {pages} {1} (\bibinfo {year} {2019})}\BibitemShut
  {NoStop}%
\bibitem [{\citenamefont {Xia}(2018)}]{xia2018revisiting}%
  \BibitemOpen
  \bibfield  {author} {\bibinfo {author} {\bibfnamefont {Y.}~\bibnamefont
  {Xia}},\ }\href@noop {} {\bibfield  {journal} {\bibinfo  {journal} {Applied
  Physics Letters}\ }\textbf {\bibinfo {volume} {113}},\ \bibinfo {pages}
  {073901} (\bibinfo {year} {2018})}\BibitemShut {NoStop}%
\bibitem [{\citenamefont {Zhang}\ \emph {et~al.}(2012)\citenamefont {Zhang},
  \citenamefont {Skoug}, \citenamefont {Cain}, \citenamefont
  {Ozoli{\c{n}}{\v{s}}}, \citenamefont {Morelli},\ and\ \citenamefont
  {Wolverton}}]{zhang2012first}%
  \BibitemOpen
  \bibfield  {author} {\bibinfo {author} {\bibfnamefont {Y.}~\bibnamefont
  {Zhang}}, \bibinfo {author} {\bibfnamefont {E.}~\bibnamefont {Skoug}},
  \bibinfo {author} {\bibfnamefont {J.}~\bibnamefont {Cain}}, \bibinfo {author}
  {\bibfnamefont {V.}~\bibnamefont {Ozoli{\c{n}}{\v{s}}}}, \bibinfo {author}
  {\bibfnamefont {D.}~\bibnamefont {Morelli}}, \ and\ \bibinfo {author}
  {\bibfnamefont {C.}~\bibnamefont {Wolverton}},\ }\href@noop {} {\bibfield
  {journal} {\bibinfo  {journal} {Physical Review B}\ }\textbf {\bibinfo
  {volume} {85}},\ \bibinfo {pages} {054306} (\bibinfo {year}
  {2012})}\BibitemShut {NoStop}%
\bibitem [{\citenamefont {Grimvall}(1999)}]{grimvall1999thermophysical}%
  \BibitemOpen
  \bibfield  {author} {\bibinfo {author} {\bibfnamefont {G.}~\bibnamefont
  {Grimvall}},\ }\href {https://books.google.com.br/books?id=TCWZlgbB3EEC}
  {\emph {\bibinfo {title} {Thermophysical Properties of Materials}}}\
  (\bibinfo  {publisher} {Elsevier},\ \bibinfo {year} {1999})\BibitemShut
  {NoStop}%
\bibitem [{\citenamefont {Giustino}(2017)}]{giustino2017electron}%
  \BibitemOpen
  \bibfield  {author} {\bibinfo {author} {\bibfnamefont {F.}~\bibnamefont
  {Giustino}},\ }\href@noop {} {\bibfield  {journal} {\bibinfo  {journal}
  {Reviews of Modern Physics}\ }\textbf {\bibinfo {volume} {89}},\ \bibinfo
  {pages} {015003} (\bibinfo {year} {2017})}\BibitemShut {NoStop}%
\bibitem [{\citenamefont {Ponc{\'e}}\ \emph {et~al.}(2018)\citenamefont
  {Ponc{\'e}}, \citenamefont {Margine},\ and\ \citenamefont
  {Giustino}}]{ponce2018towards}%
  \BibitemOpen
  \bibfield  {author} {\bibinfo {author} {\bibfnamefont {S.}~\bibnamefont
  {Ponc{\'e}}}, \bibinfo {author} {\bibfnamefont {E.~R.}\ \bibnamefont
  {Margine}}, \ and\ \bibinfo {author} {\bibfnamefont {F.}~\bibnamefont
  {Giustino}},\ }\href@noop {} {\bibfield  {journal} {\bibinfo  {journal}
  {Physical Review B}\ }\textbf {\bibinfo {volume} {97}},\ \bibinfo {pages}
  {121201} (\bibinfo {year} {2018})}\BibitemShut {NoStop}%
\bibitem [{\citenamefont {Ponc{\'e}}\ \emph {et~al.}(2020)\citenamefont
  {Ponc{\'e}}, \citenamefont {Li}, \citenamefont {Reichardt},\ and\
  \citenamefont {Giustino}}]{ponce2020first}%
  \BibitemOpen
  \bibfield  {author} {\bibinfo {author} {\bibfnamefont {S.}~\bibnamefont
  {Ponc{\'e}}}, \bibinfo {author} {\bibfnamefont {W.}~\bibnamefont {Li}},
  \bibinfo {author} {\bibfnamefont {S.}~\bibnamefont {Reichardt}}, \ and\
  \bibinfo {author} {\bibfnamefont {F.}~\bibnamefont {Giustino}},\ }\href@noop
  {} {\bibfield  {journal} {\bibinfo  {journal} {Reports on Progress in
  Physics}\ }\textbf {\bibinfo {volume} {83}},\ \bibinfo {pages} {036501}
  (\bibinfo {year} {2020})}\BibitemShut {NoStop}%
\bibitem [{\citenamefont {Fr{\"o}hlich}(1937)}]{frohlich1937h}%
  \BibitemOpen
  \bibfield  {author} {\bibinfo {author} {\bibfnamefont {H.}~\bibnamefont
  {Fr{\"o}hlich}},\ }\href@noop {} {\bibfield  {journal} {\bibinfo  {journal}
  {Proc. Roy. Soc.(London)}\ }\textbf {\bibinfo {volume} {160}},\ \bibinfo
  {pages} {230} (\bibinfo {year} {1937})}\BibitemShut {NoStop}%
\bibitem [{\citenamefont {Verdi}\ and\ \citenamefont
  {Giustino}(2015)}]{verdi2015frohlich}%
  \BibitemOpen
  \bibfield  {author} {\bibinfo {author} {\bibfnamefont {C.}~\bibnamefont
  {Verdi}}\ and\ \bibinfo {author} {\bibfnamefont {F.}~\bibnamefont
  {Giustino}},\ }\href@noop {} {\bibfield  {journal} {\bibinfo  {journal}
  {Physical Review Letters}\ }\textbf {\bibinfo {volume} {115}},\ \bibinfo
  {pages} {176401} (\bibinfo {year} {2015})}\BibitemShut {NoStop}%
\bibitem [{\citenamefont {Marzari}\ and\ \citenamefont
  {Vanderbilt}(1997)}]{marzari1997maximally}%
  \BibitemOpen
  \bibfield  {author} {\bibinfo {author} {\bibfnamefont {N.}~\bibnamefont
  {Marzari}}\ and\ \bibinfo {author} {\bibfnamefont {D.}~\bibnamefont
  {Vanderbilt}},\ }\href@noop {} {\bibfield  {journal} {\bibinfo  {journal}
  {Physical review B}\ }\textbf {\bibinfo {volume} {56}},\ \bibinfo {pages}
  {12847} (\bibinfo {year} {1997})}\BibitemShut {NoStop}%
\bibitem [{\citenamefont {Askerov}\ and\ \citenamefont
  {Figarova}(2009)}]{askerov2009thermodynamics}%
  \BibitemOpen
  \bibfield  {author} {\bibinfo {author} {\bibfnamefont {B.~M.}\ \bibnamefont
  {Askerov}}\ and\ \bibinfo {author} {\bibfnamefont {S.}~\bibnamefont
  {Figarova}},\ }\href@noop {} {\emph {\bibinfo {title} {Thermodynamics, Gibbs
  Method and Statistical Physics of Electron Gases}}},\ Vol.~\bibinfo {volume}
  {57}\ (\bibinfo  {publisher} {Springer Science \& Business Media},\ \bibinfo
  {year} {2009})\BibitemShut {NoStop}%
\bibitem [{\citenamefont {Shankland}(1971)}]{shankland1971interpolation}%
  \BibitemOpen
  \bibfield  {author} {\bibinfo {author} {\bibfnamefont {D.~G.}\ \bibnamefont
  {Shankland}},\ }in\ \href@noop {} {\emph {\bibinfo {booktitle} {Computational
  Methods in Band Theory}}}\ (\bibinfo  {publisher} {Springer},\ \bibinfo
  {year} {1971})\ pp.\ \bibinfo {pages} {362--367}\BibitemShut {NoStop}%
\bibitem [{\citenamefont {Koelling}\ and\ \citenamefont
  {Wood}(1986)}]{koelling1986interpolation}%
  \BibitemOpen
  \bibfield  {author} {\bibinfo {author} {\bibfnamefont {D.}~\bibnamefont
  {Koelling}}\ and\ \bibinfo {author} {\bibfnamefont {J.}~\bibnamefont
  {Wood}},\ }\href@noop {} {\bibfield  {journal} {\bibinfo  {journal} {Journal
  of Computational Physics}\ }\textbf {\bibinfo {volume} {67}},\ \bibinfo
  {pages} {253} (\bibinfo {year} {1986})}\BibitemShut {NoStop}%
\end{thebibliography}%


%merlin.mbs apsrev4-1.bst 2010-07-25 4.21a (PWD, AO, DPC) hacked
%Control: key (0)
%Control: author (8) initials jnrlst
%Control: editor formatted (1) identically to author
%Control: production of article title (-1) disabled
%Control: page (0) single
%Control: year (1) truncated
%Control: production of eprint (0) enabled
\begin{thebibliography}{21}%
\makeatletter
\providecommand \@ifxundefined [1]{%
 \@ifx{#1\undefined}
}%
\providecommand \@ifnum [1]{%
 \ifnum #1\expandafter \@firstoftwo
 \else \expandafter \@secondoftwo
 \fi
}%
\providecommand \@ifx [1]{%
 \ifx #1\expandafter \@firstoftwo
 \else \expandafter \@secondoftwo
 \fi
}%
\providecommand \natexlab [1]{#1}%
\providecommand \enquote  [1]{``#1''}%
\providecommand \bibnamefont  [1]{#1}%
\providecommand \bibfnamefont [1]{#1}%
\providecommand \citenamefont [1]{#1}%
\providecommand \href@noop [0]{\@secondoftwo}%
\providecommand \href [0]{\begingroup \@sanitize@url \@href}%
\providecommand \@href[1]{\@@startlink{#1}\@@href}%
\providecommand \@@href[1]{\endgroup#1\@@endlink}%
\providecommand \@sanitize@url [0]{\catcode `\\12\catcode `\$12\catcode
  `\&12\catcode `\#12\catcode `\^12\catcode `\_12\catcode `\%12\relax}%
\providecommand \@@startlink[1]{}%
\providecommand \@@endlink[0]{}%
\providecommand \url  [0]{\begingroup\@sanitize@url \@url }%
\providecommand \@url [1]{\endgroup\@href {#1}{\urlprefix }}%
\providecommand \urlprefix  [0]{URL }%
\providecommand \Eprint [0]{\href }%
\providecommand \doibase [0]{http://dx.doi.org/}%
\providecommand \selectlanguage [0]{\@gobble}%
\providecommand \bibinfo  [0]{\@secondoftwo}%
\providecommand \bibfield  [0]{\@secondoftwo}%
\providecommand \translation [1]{[#1]}%
\providecommand \BibitemOpen [0]{}%
\providecommand \bibitemStop [0]{}%
\providecommand \bibitemNoStop [0]{.\EOS\space}%
\providecommand \EOS [0]{\spacefactor3000\relax}%
\providecommand \BibitemShut  [1]{\csname bibitem#1\endcsname}%
\let\auto@bib@innerbib\@empty
%</preamble>
\bibitem [{\citenamefont {Giustino}(2017)}]{giustino2017electron}%
  \BibitemOpen
  \bibfield  {author} {\bibinfo {author} {\bibfnamefont {F.}~\bibnamefont
  {Giustino}},\ }\href@noop {} {\bibfield  {journal} {\bibinfo  {journal}
  {Reviews of Modern Physics}\ }\textbf {\bibinfo {volume} {89}},\ \bibinfo
  {pages} {015003} (\bibinfo {year} {2017})}\BibitemShut {NoStop}%
\bibitem [{\citenamefont {Ponc{\'e}}\ \emph {et~al.}(2018)\citenamefont
  {Ponc{\'e}}, \citenamefont {Margine},\ and\ \citenamefont
  {Giustino}}]{ponce2018towards}%
  \BibitemOpen
  \bibfield  {author} {\bibinfo {author} {\bibfnamefont {S.}~\bibnamefont
  {Ponc{\'e}}}, \bibinfo {author} {\bibfnamefont {E.~R.}\ \bibnamefont
  {Margine}}, \ and\ \bibinfo {author} {\bibfnamefont {F.}~\bibnamefont
  {Giustino}},\ }\href@noop {} {\bibfield  {journal} {\bibinfo  {journal}
  {Physical Review B}\ }\textbf {\bibinfo {volume} {97}},\ \bibinfo {pages}
  {121201} (\bibinfo {year} {2018})}\BibitemShut {NoStop}%
\bibitem [{\citenamefont {Ponc{\'e}}\ \emph {et~al.}(2020)\citenamefont
  {Ponc{\'e}}, \citenamefont {Li}, \citenamefont {Reichardt},\ and\
  \citenamefont {Giustino}}]{ponce2020first}%
  \BibitemOpen
  \bibfield  {author} {\bibinfo {author} {\bibfnamefont {S.}~\bibnamefont
  {Ponc{\'e}}}, \bibinfo {author} {\bibfnamefont {W.}~\bibnamefont {Li}},
  \bibinfo {author} {\bibfnamefont {S.}~\bibnamefont {Reichardt}}, \ and\
  \bibinfo {author} {\bibfnamefont {F.}~\bibnamefont {Giustino}},\ }\href@noop
  {} {\bibfield  {journal} {\bibinfo  {journal} {Reports on Progress in
  Physics}\ }\textbf {\bibinfo {volume} {83}},\ \bibinfo {pages} {036501}
  (\bibinfo {year} {2020})}\BibitemShut {NoStop}%
\bibitem [{\citenamefont {Fr{\"o}hlich}(1937)}]{frohlich1937h}%
  \BibitemOpen
  \bibfield  {author} {\bibinfo {author} {\bibfnamefont {H.}~\bibnamefont
  {Fr{\"o}hlich}},\ }\href@noop {} {\bibfield  {journal} {\bibinfo  {journal}
  {Proc. Roy. Soc.(London)}\ }\textbf {\bibinfo {volume} {160}},\ \bibinfo
  {pages} {230} (\bibinfo {year} {1937})}\BibitemShut {NoStop}%
\bibitem [{\citenamefont {Verdi}\ and\ \citenamefont
  {Giustino}(2015)}]{verdi2015frohlich}%
  \BibitemOpen
  \bibfield  {author} {\bibinfo {author} {\bibfnamefont {C.}~\bibnamefont
  {Verdi}}\ and\ \bibinfo {author} {\bibfnamefont {F.}~\bibnamefont
  {Giustino}},\ }\href@noop {} {\bibfield  {journal} {\bibinfo  {journal}
  {Physical Review Letters}\ }\textbf {\bibinfo {volume} {115}},\ \bibinfo
  {pages} {176401} (\bibinfo {year} {2015})}\BibitemShut {NoStop}%
\bibitem [{\citenamefont {Pellegrini}\ \emph {et~al.}(2016)\citenamefont
  {Pellegrini}, \citenamefont {Marinelli},\ and\ \citenamefont
  {Reiche}}]{pellegrini2016physics}%
  \BibitemOpen
  \bibfield  {author} {\bibinfo {author} {\bibfnamefont {C.}~\bibnamefont
  {Pellegrini}}, \bibinfo {author} {\bibfnamefont {A.}~\bibnamefont
  {Marinelli}}, \ and\ \bibinfo {author} {\bibfnamefont {S.}~\bibnamefont
  {Reiche}},\ }\href@noop {} {\bibfield  {journal} {\bibinfo  {journal}
  {Reviews of Modern Physics}\ }\textbf {\bibinfo {volume} {88}},\ \bibinfo
  {pages} {015006} (\bibinfo {year} {2016})}\BibitemShut {NoStop}%
\bibitem [{\citenamefont {Bostedt}\ \emph {et~al.}(2016)\citenamefont
  {Bostedt}, \citenamefont {Boutet}, \citenamefont {Fritz}, \citenamefont
  {Huang}, \citenamefont {Lee}, \citenamefont {Lemke}, \citenamefont {Robert},
  \citenamefont {Schlotter}, \citenamefont {Turner},\ and\ \citenamefont
  {Williams}}]{bostedt2016linac}%
  \BibitemOpen
  \bibfield  {author} {\bibinfo {author} {\bibfnamefont {C.}~\bibnamefont
  {Bostedt}}, \bibinfo {author} {\bibfnamefont {S.}~\bibnamefont {Boutet}},
  \bibinfo {author} {\bibfnamefont {D.~M.}\ \bibnamefont {Fritz}}, \bibinfo
  {author} {\bibfnamefont {Z.}~\bibnamefont {Huang}}, \bibinfo {author}
  {\bibfnamefont {H.~J.}\ \bibnamefont {Lee}}, \bibinfo {author} {\bibfnamefont
  {H.~T.}\ \bibnamefont {Lemke}}, \bibinfo {author} {\bibfnamefont
  {A.}~\bibnamefont {Robert}}, \bibinfo {author} {\bibfnamefont {W.~F.}\
  \bibnamefont {Schlotter}}, \bibinfo {author} {\bibfnamefont {J.~J.}\
  \bibnamefont {Turner}}, \ and\ \bibinfo {author} {\bibfnamefont {G.~J.}\
  \bibnamefont {Williams}},\ }\href@noop {} {\bibfield  {journal} {\bibinfo
  {journal} {Reviews of Modern Physics}\ }\textbf {\bibinfo {volume} {88}},\
  \bibinfo {pages} {015007} (\bibinfo {year} {2016})}\BibitemShut {NoStop}%
\bibitem [{\citenamefont {Vogl}(1976)}]{vogl1976microscopic}%
  \BibitemOpen
  \bibfield  {author} {\bibinfo {author} {\bibfnamefont {P.}~\bibnamefont
  {Vogl}},\ }\href@noop {} {\bibfield  {journal} {\bibinfo  {journal} {Physical
  Review B}\ }\textbf {\bibinfo {volume} {13}},\ \bibinfo {pages} {694}
  (\bibinfo {year} {1976})}\BibitemShut {NoStop}%
\bibitem [{\citenamefont {Marzari}\ and\ \citenamefont
  {Vanderbilt}(1997)}]{marzari1997maximally}%
  \BibitemOpen
  \bibfield  {author} {\bibinfo {author} {\bibfnamefont {N.}~\bibnamefont
  {Marzari}}\ and\ \bibinfo {author} {\bibfnamefont {D.}~\bibnamefont
  {Vanderbilt}},\ }\href@noop {} {\bibfield  {journal} {\bibinfo  {journal}
  {Physical review B}\ }\textbf {\bibinfo {volume} {56}},\ \bibinfo {pages}
  {12847} (\bibinfo {year} {1997})}\BibitemShut {NoStop}%
\bibitem [{\citenamefont {Ehrenreich}(1959)}]{ehrenreich1959screening}%
  \BibitemOpen
  \bibfield  {author} {\bibinfo {author} {\bibfnamefont {H.}~\bibnamefont
  {Ehrenreich}},\ }\href@noop {} {\bibfield  {journal} {\bibinfo  {journal}
  {Journal of Physics and Chemistry of Solids}\ }\textbf {\bibinfo {volume}
  {8}},\ \bibinfo {pages} {130} (\bibinfo {year} {1959})}\BibitemShut {NoStop}%
\bibitem [{\citenamefont {Chaves}\ \emph
  {et~al.}(2021{\natexlab{a}})\citenamefont {Chaves}, \citenamefont {Larson},
  \citenamefont {Kaxiras},\ and\ \citenamefont
  {Antonelli}}]{chaves2021microscopic}%
  \BibitemOpen
  \bibfield  {author} {\bibinfo {author} {\bibfnamefont {A.~S.}\ \bibnamefont
  {Chaves}}, \bibinfo {author} {\bibfnamefont {D.~T.}\ \bibnamefont {Larson}},
  \bibinfo {author} {\bibfnamefont {E.}~\bibnamefont {Kaxiras}}, \ and\
  \bibinfo {author} {\bibfnamefont {A.}~\bibnamefont {Antonelli}},\ }\href@noop
  {} {\bibfield  {journal} {\bibinfo  {journal} {Physical Review B}\ }\textbf
  {\bibinfo {volume} {104}},\ \bibinfo {pages} {115204} (\bibinfo {year}
  {2021}{\natexlab{a}})}\BibitemShut {NoStop}%
\bibitem [{\citenamefont {Brooks}(1955)}]{brooks1955theory}%
  \BibitemOpen
  \bibfield  {author} {\bibinfo {author} {\bibfnamefont {H.}~\bibnamefont
  {Brooks}},\ }in\ \href@noop {} {\emph {\bibinfo {booktitle} {Advances in
  electronics and electron physics}}},\ Vol.~\bibinfo {volume} {7}\ (\bibinfo
  {publisher} {Elsevier},\ \bibinfo {year} {1955})\ pp.\ \bibinfo {pages}
  {85--182}\BibitemShut {NoStop}%
\bibitem [{\citenamefont {Chattopadhyay}\ and\ \citenamefont
  {Queisser}(1981)}]{chattopadhyay1981electron}%
  \BibitemOpen
  \bibfield  {author} {\bibinfo {author} {\bibfnamefont {D.}~\bibnamefont
  {Chattopadhyay}}\ and\ \bibinfo {author} {\bibfnamefont {H.}~\bibnamefont
  {Queisser}},\ }\href@noop {} {\bibfield  {journal} {\bibinfo  {journal}
  {Reviews of Modern Physics}\ }\textbf {\bibinfo {volume} {53}},\ \bibinfo
  {pages} {745} (\bibinfo {year} {1981})}\BibitemShut {NoStop}%
\bibitem [{\citenamefont {Chaves}\ \emph
  {et~al.}(2021{\natexlab{b}})\citenamefont {Chaves}, \citenamefont
  {Gonz{\'a}lez-Romero}, \citenamefont {Mel{\'e}ndez},\ and\ \citenamefont
  {Antonelli}}]{chaves2021investigating}%
  \BibitemOpen
  \bibfield  {author} {\bibinfo {author} {\bibfnamefont {A.~S.}\ \bibnamefont
  {Chaves}}, \bibinfo {author} {\bibfnamefont {R.~L.}\ \bibnamefont
  {Gonz{\'a}lez-Romero}}, \bibinfo {author} {\bibfnamefont {J.~J.}\
  \bibnamefont {Mel{\'e}ndez}}, \ and\ \bibinfo {author} {\bibfnamefont
  {A.}~\bibnamefont {Antonelli}},\ }\href@noop {} {\bibfield  {journal}
  {\bibinfo  {journal} {Physical Chemistry Chemical Physics}\ }\textbf
  {\bibinfo {volume} {23}},\ \bibinfo {pages} {900} (\bibinfo {year}
  {2021}{\natexlab{b}})}\BibitemShut {NoStop}%
\bibitem [{\citenamefont {Askerov}\ and\ \citenamefont
  {Figarova}(2009)}]{askerov2009thermodynamics}%
  \BibitemOpen
  \bibfield  {author} {\bibinfo {author} {\bibfnamefont {B.~M.}\ \bibnamefont
  {Askerov}}\ and\ \bibinfo {author} {\bibfnamefont {S.}~\bibnamefont
  {Figarova}},\ }\href@noop {} {\emph {\bibinfo {title} {Thermodynamics, Gibbs
  Method and Statistical Physics of Electron Gases}}},\ Vol.~\bibinfo {volume}
  {57}\ (\bibinfo  {publisher} {Springer Science \& Business Media},\ \bibinfo
  {year} {2009})\BibitemShut {NoStop}%
\bibitem [{\citenamefont {Shankland}(1971)}]{shankland1971interpolation}%
  \BibitemOpen
  \bibfield  {author} {\bibinfo {author} {\bibfnamefont {D.~G.}\ \bibnamefont
  {Shankland}},\ }in\ \href@noop {} {\emph {\bibinfo {booktitle} {Computational
  Methods in Band Theory}}}\ (\bibinfo  {publisher} {Springer},\ \bibinfo
  {year} {1971})\ pp.\ \bibinfo {pages} {362--367}\BibitemShut {NoStop}%
\bibitem [{\citenamefont {Koelling}\ and\ \citenamefont
  {Wood}(1986)}]{koelling1986interpolation}%
  \BibitemOpen
  \bibfield  {author} {\bibinfo {author} {\bibfnamefont {D.}~\bibnamefont
  {Koelling}}\ and\ \bibinfo {author} {\bibfnamefont {J.}~\bibnamefont
  {Wood}},\ }\href@noop {} {\bibfield  {journal} {\bibinfo  {journal} {Journal
  of Computational Physics}\ }\textbf {\bibinfo {volume} {67}},\ \bibinfo
  {pages} {253} (\bibinfo {year} {1986})}\BibitemShut {NoStop}%
\bibitem [{\citenamefont {Chaves}\ \emph {et~al.}(2020)\citenamefont {Chaves},
  \citenamefont {Antonelli}, \citenamefont {Larson},\ and\ \citenamefont
  {Kaxiras}}]{chaves2020boosting}%
  \BibitemOpen
  \bibfield  {author} {\bibinfo {author} {\bibfnamefont {A.~S.}\ \bibnamefont
  {Chaves}}, \bibinfo {author} {\bibfnamefont {A.}~\bibnamefont {Antonelli}},
  \bibinfo {author} {\bibfnamefont {D.~T.}\ \bibnamefont {Larson}}, \ and\
  \bibinfo {author} {\bibfnamefont {E.}~\bibnamefont {Kaxiras}},\ }\href@noop
  {} {\bibfield  {journal} {\bibinfo  {journal} {Physical Review B}\ }\textbf
  {\bibinfo {volume} {102}},\ \bibinfo {pages} {125116} (\bibinfo {year}
  {2020})}\BibitemShut {NoStop}%
\bibitem [{\citenamefont {Zhao}\ \emph {et~al.}(2016)\citenamefont {Zhao},
  \citenamefont {Tan}, \citenamefont {Hao}, \citenamefont {He}, \citenamefont
  {Pei}, \citenamefont {Chi}, \citenamefont {Wang}, \citenamefont {Gong},
  \citenamefont {Xu}, \citenamefont {Dravid}, \citenamefont {Uher},
  \citenamefont {Snyder}, \citenamefont {Wolverton},\ and\ \citenamefont
  {Kanatzidis}}]{zhao2015ultrahigh}%
  \BibitemOpen
  \bibfield  {author} {\bibinfo {author} {\bibfnamefont {L.-D.}\ \bibnamefont
  {Zhao}}, \bibinfo {author} {\bibfnamefont {G.}~\bibnamefont {Tan}}, \bibinfo
  {author} {\bibfnamefont {S.}~\bibnamefont {Hao}}, \bibinfo {author}
  {\bibfnamefont {J.}~\bibnamefont {He}}, \bibinfo {author} {\bibfnamefont
  {Y.}~\bibnamefont {Pei}}, \bibinfo {author} {\bibfnamefont {H.}~\bibnamefont
  {Chi}}, \bibinfo {author} {\bibfnamefont {H.}~\bibnamefont {Wang}}, \bibinfo
  {author} {\bibfnamefont {S.}~\bibnamefont {Gong}}, \bibinfo {author}
  {\bibfnamefont {H.}~\bibnamefont {Xu}}, \bibinfo {author} {\bibfnamefont
  {V.~P.}\ \bibnamefont {Dravid}}, \bibinfo {author} {\bibfnamefont
  {C.}~\bibnamefont {Uher}}, \bibinfo {author} {\bibfnamefont {G.~J.}\
  \bibnamefont {Snyder}}, \bibinfo {author} {\bibfnamefont {C.}~\bibnamefont
  {Wolverton}}, \ and\ \bibinfo {author} {\bibfnamefont {M.~G.}\ \bibnamefont
  {Kanatzidis}},\ }\href@noop {} {\bibfield  {journal} {\bibinfo  {journal}
  {Science}\ }\textbf {\bibinfo {volume} {351}},\ \bibinfo {pages} {141}
  (\bibinfo {year} {2016})}\BibitemShut {NoStop}%
\bibitem [{\citenamefont {Hao}\ \emph {et~al.}(2016)\citenamefont {Hao},
  \citenamefont {Shi}, \citenamefont {Dravid}, \citenamefont {Kanatzidis},\
  and\ \citenamefont {Wolverton}}]{hao2016computational}%
  \BibitemOpen
  \bibfield  {author} {\bibinfo {author} {\bibfnamefont {S.}~\bibnamefont
  {Hao}}, \bibinfo {author} {\bibfnamefont {F.}~\bibnamefont {Shi}}, \bibinfo
  {author} {\bibfnamefont {V.~P.}\ \bibnamefont {Dravid}}, \bibinfo {author}
  {\bibfnamefont {M.~G.}\ \bibnamefont {Kanatzidis}}, \ and\ \bibinfo {author}
  {\bibfnamefont {C.}~\bibnamefont {Wolverton}},\ }\href@noop {} {\bibfield
  {journal} {\bibinfo  {journal} {Chemistry of Materials}\ }\textbf {\bibinfo
  {volume} {28}},\ \bibinfo {pages} {3218} (\bibinfo {year}
  {2016})}\BibitemShut {NoStop}%
\bibitem [{\citenamefont {Yuan}\ \emph {et~al.}(2019)\citenamefont {Yuan},
  \citenamefont {Sun}, \citenamefont {Zhang},\ and\ \citenamefont
  {Tang}}]{yuan2019tailoring}%
  \BibitemOpen
  \bibfield  {author} {\bibinfo {author} {\bibfnamefont {K.}~\bibnamefont
  {Yuan}}, \bibinfo {author} {\bibfnamefont {Z.}~\bibnamefont {Sun}}, \bibinfo
  {author} {\bibfnamefont {X.}~\bibnamefont {Zhang}}, \ and\ \bibinfo {author}
  {\bibfnamefont {D.}~\bibnamefont {Tang}},\ }\href@noop {} {\bibfield
  {journal} {\bibinfo  {journal} {Scientific reports}\ }\textbf {\bibinfo
  {volume} {9}},\ \bibinfo {pages} {1} (\bibinfo {year} {2019})}\BibitemShut
  {NoStop}%
\end{thebibliography}%

\end{document}

%\preprint{Page}

\title{Supplementary Information: Outstanding thermoelectric performance predicted for out-of-plane p-doped GeSe} 

\author{Anderson S. Chaves}
%\affiliation{John A. Paulson School of Engineering and Applied Sciences, Harvard University, Cambridge, Massachusetts, 02138, USA}
\affiliation{Gleb Wataghin Institute of Physics, University of Campinas, PO Box 13083-859, Campinas, SP, Brazil}
\affiliation{John A. Paulson School of Engineering and Applied Sciences, Harvard University, Cambridge, Massachusetts, 02138, USA}
\author{Daniel T. Larson}
\affiliation{Department of Physics, Harvard University, Cambridge, Massachusetts, 02138, USA}
\author{Efthimios Kaxiras}
\affiliation{Department of Physics, Harvard University, Cambridge, Massachusetts, 02138, USA}
\affiliation{John A. Paulson School of Engineering and Applied Sciences, Harvard University, Cambridge, Massachusetts, 02138, USA}
\author{Alex Antonelli}
\affiliation{Gleb Wataghin Institute of Physics and Center for Computing in
Engineering \& Sciences, University of Campinas, PO Box 13083-859, Campinas, SP, Brazil}

\date{\today}

%\pacs{71.15.Nc,36.40.-c,72.80.Ga}
%\abbreviations{DFT, PBE, VASP}
%\keywords{Thermoelectric properties, SnSe, n-type SnSe, p-type SnSe, Scattering mechanisms, First-Principles Simulations}

\maketitle

\section{Details of the theoretical framework}

The hole-phonon (h-p) relaxation time (RT) is directly related to the imaginary part of
the momentum- and band-resolved Fan-Migdal electron
self-energy\cite{giustino2017electron,ponce2018towards,ponce2020first}, $1/\tau_{n,{\bf{k}}} = 2 \Im \Theta_{n,{\bf{k}}}(\epsilon=0,T)$,
where\noindent
\begin{flalign} 
\label{imag}
 \begin{split}
& \Im \Theta_{n,{\bf{k}}}(\epsilon,T) = \pi \sum_{m,\theta} \int_{BZ} \frac{d{\bf{q}}}{\Omega_{BZ}} |g_{mn,\theta}({\bf{k},{\bf{q}}})|^2 \\
&       \times \Bigg[\left[n_{{\bf{q}}\theta}(T) + f_{m{\bf{k}}+{\bf{q}}}\right]\delta(\epsilon - (\epsilon_{m{\bf{k}}+{\bf{q}}} - \epsilon_F) + \omega_{{\bf{q}}\theta}) \\
&        + [n_{{\bf{q}}\theta}(T) + 1 - f_{m{\bf{k}}+{\bf{q}}}]\delta(\epsilon - (\epsilon_{m{\bf{k}}+{\bf{q}}} - \epsilon_F) - \omega_{{\bf{q}}\theta})\Bigg]~,
\end{split}
\end{flalign}
$\epsilon_F$ is the Fermi energy calculated by DFT at 0K,
$g_{mn,\theta}({\bf{k},{\bf{q}}}) = \braket{\Psi_{m{\bf{k}}+{\bf{q}}}|\partial_{{\bf{q}}\theta} V_{KS}(r)|\Psi_{n{\bf{k}}}}$
are the h-p coupling matrix elements that are calculated within DFPT,
$\ket{\Psi_{n{\bf{k}}}}$ are Kohn-Sham (KS) orbitals
and $\partial_{{\bf{q}}\theta} V_{KS}$
corresponds to the change of the KS potential upon a
vibrational perturbation with momentum {\bf{q}} and branch index $\theta$,
$\omega_{{\bf{q}}\theta}$ are the corresponding phonon eigenfrequencies, $n_{{\bf{q}}\theta}(T)$ is the Bose-Einstein
distribution, $\epsilon_{m{\bf{k}}}$ is 
the KS energy of a hole in band $m$ with momentum $\bf{k}$, $f_{m{\bf{k}}+{\bf{q}}}$ is the Fermi-Dirac distribution, and
$\Omega_{BZ}$ is the volume of the Brillouin zone (BZ).

For polar materials, such as GeSe and SnSe, holes can be scattered by
a multipolar electric polarization caused by longitudinal optical (LO) phonons and described by
the Fr{\"o}hlich coupling~\cite{frohlich1937h}.
We employ a first-principles approach to the Fr{\"o}hlich coupling\cite{verdi2015frohlich}
which allows for an analytical treatment of the long-range polar singularity. 
%% and solves the problem of attaining convergence that is not amenable to Wannier-Fourier (W-F) interpolation.
To accomplish this, the h-p matrix element is separated into
short-range ($g_{mn,\theta}^{\mathcal{S}}$) and long-range ($g_{mn,\theta}^{\mathcal{L}}$) parts.
The $\mathcal{S}$ part is well behaved and is amenable to W-F interpolation. On the other hand, the $\mathcal{L}$ part is determined
analytically using the Vogl formula~\cite{pellegrini2016physics,bostedt2016linac,vogl1976microscopic}:
\begin{multline}
\label{long}
g_{mn,\theta}^{\mathcal{L}}({\bf{k}},{\bf{q}}) = i \frac{e^2}{\Omega\epsilon_0} \sum_{\kappa}
 \left(\frac{\hbar}{2 N{M_{\kappa} \omega_{{\bf {q}}\theta}}}\right)^{\!\!\frac{1}{2}} \\
 \times \sum_{{\bf G}\ne -{\bf q}} 
 \frac{ ({\bf {q}}+{\bf {G}})\cdot{\bf {Z}}^*_\kappa \cdot {\bf {e}}_{\kappa\theta}({\bf {q}}) } 
 {({\bf {q}}+{\bf {G}})\cdot\bm\zeta_\infty\!\cdot({\bf {q}}+{\bf {G}})} \\
 \times \braket{\Psi_{m{\bf{k}}+{\bf{q}}}|e^{i({\bf{k}}+{\bf{q}})\cdot{\bf{r}}}|\Psi_{n{\bf{k}}}}~.
\end{multline}
Here $\braket{\Psi_{m{\bf{k}}+{\bf{q}}}|e^{i({\bf{k}}+{\bf{q}})\cdot{\bf{r}}}|\Psi_{n{\bf{k}}}} = \left[ U_{{\bf {k}}+{\bf {q}}}\:U_{{\bf {k}}}^{\dagger} \right]_{mn}$ are phase factors given in terms of rotation matrices, $U_{{\bf {k}}+{\bf {q}}}$,
that appear in the definition of maximally localized Wannier functions (MLWFs).\cite{marzari1997maximally}
$\Omega$ is the volume of the unit cell, $M_{\kappa}$ corresponds to the mass of atom $\kappa$,
$N$ is the number of unit cells in the Born-von K{\'a}rm{\'a}n supercell,
${\bf{G}}$ is a reciprocal lattice vector, ${\bf{Z}}^*=Z^*_{\alpha,\beta}$
is the Born effective charge tensor, ${\bf {e}}_{\kappa\theta}({\bf {q}})$
is a phonon eigenmode normalized within the unit cell, $\bm\zeta_\infty = \zeta^{\infty}_{\alpha,\beta}$
corresponds to the high-frequency dielectric constant tensor, $\epsilon_0$ is the vacuum
permittivity, $\hbar$ is the reduced Planck constant, and $e$ is the absolute electron charge.

By using the quasi-static Ehrenreich theory~\cite{ehrenreich1959screening} and
following Ref. \citenum{chaves2021microscopic}, we generalize the first-principles
approach to the Fr{\"o}hlich coupling
by including screening effects due to free carriers in the material
that efficiently screen out the electric field produced by optical vibrations.
This is accomplished by dividing the RT by the following band-dependent screening factor:
\begin{multline}
\label{screen_pol}
% \begin{split}
F_\mathrm{pol}(n,{\bf{k}}) = \Bigg[1 -\frac{1}{2(r_{\infty}(n,{\bf{k}})\cdot{\bf{k}})^2}\ln[1+4(r_{\infty}(n,{\bf{k}})\cdot{\bf{k}})^2] 
+ \frac{1}{1+4(r_{\infty}(n,{\bf{k}})\cdot{\bf{k}})^2}\Bigg]^{-1}~,
%\end{split} 
\end{multline}
where $r_{\infty}$ is the screening radius given by
\begin{equation}
\label{r0inf}
r{_{\infty}}^{-2}(n,{\bf{k}}) = \frac{4\pi{e^2}}{\zeta{_{\infty}}}\int{\left(-\frac{\partial f_{\mu}(T,\epsilon)}{\partial \epsilon_{n,{\bf{k}}}}\right) g(\epsilon) d\epsilon}~.
\end{equation}
Here $g(\epsilon)$ is the density of states (DOS), which can be evaluated
over an energy grid with spacing $d\epsilon$ as follows:
\begin{equation}
\label{dos2}
g(\epsilon) = \int \sum_n \delta(\epsilon - \epsilon_{n,{\bf{k}}}) \frac{d{\bf{k}}}{8\pi^3} = \frac{1}{\Omega N_{{\bf{k}}}}\sum_{n,{\bf{k}}} \frac{\delta(\epsilon - \epsilon_{n,{\bf{k}}})}{d\epsilon}~, 
\end{equation}
where $N_{{\bf{k}}}$ the number of {\bf{k}} points sampled.
Thus from Eqs.~\ref{imag}, \ref{long},~and~\ref{screen_pol} we arrive at the final expressions for the RTs due to
nonpolar ($\tau_\mathrm{npol}$) and screened polar ($\tau_\mathrm{pol}$) scattering:
\begin{equation}
\label{npol}
\frac{1}{\tau_\mathrm{npol}} \equiv \frac{1}{\tau^{npol}_{n,{\bf{k}}}} = 2 \Im \Theta_{n,{\bf{k}}}[\epsilon = 0,T,g^{{\mathcal{S}}}_{mn,\theta}({\bf{k},{\bf{q}}})]~,
\end{equation}

\begin{equation}
\label{pol}
\frac{1}{\tau_\mathrm{pol}} \equiv \frac{1}{\tau^{pol}_{n,{\bf{k}}}} = 2 \Im \Theta_{n,{\bf{k}}}[\epsilon = 0,T,g^{{\mathcal{L}}}_{mn,\theta}({\bf{k},{\bf{q}}})]\times F_\mathrm{pol}(n,{\bf{k}})~.
\end{equation}

Extrinsic scattering by ionized impurities
is included using the theory developed by
Brooks and Herring (B-H).\cite{brooks1955theory,chattopadhyay1981electron}
Following Refs.~\citenum{chaves2021investigating},~\citenum{askerov2009thermodynamics},~and~\citenum{chaves2021microscopic},
the RT due to the scattering of holes by ionized impurities can be written as
\begin{equation}
\label{tau_imp}
\tau_\mathrm{imp} \equiv \tau^{imp}(n,{\bf{k}}) = \frac{\hbar\zeta{_0}{^2}}{2{\pi}{e^4}{n_\mathrm{ii}}F_\mathrm{imp}(n,{\bf{k}})}{\bf{k}}^2 \left\vert\frac{\partial \epsilon_{n,{\bf{k}}}}{\partial {\bf{k}}}\right\vert~,
\end{equation}
where $n_\mathrm{ii}$ is the concentration of ionized impurities,
\begin{equation}
F_\mathrm{imp}(n,{\bf{k}}) = \ln(1+\eta) - \frac{\eta}{1+\eta}
\end{equation}
is the screening function, with $\eta = (2{\bf{k}}\cdot r_0(n,{\bf{k}}))^2$, and
the static screening radius $r_0$ is given by Eq.~\ref{r0inf},
now screened by the static dielectric constant, $\zeta_0$.
It is important to note that the expression above for $\tau_\mathrm{imp}$
extends the parabolic band approximation
in such a way that the average electronic group velocities in the expression
are given analytically in terms of Fourier interpolation
of the DFT band structure through star functions\cite{chaves2021investigating,shankland1971interpolation,koelling1986interpolation,chaves2020boosting}. Namely,
\begin{equation}
\label{velocities2}
\frac{\partial \epsilon_{n,{\bf{k}}}}{\partial {\bf{k}}} \equiv v({n,{\bf{k}}}) 
\approx \frac{i}{n_s}\sum_{m=1}^M a_m \sum_{\{s\}}(s {\bf{R}}_{m}) \exp [i(s {\bf{R}}_{m})\cdot{\bf{k}}]~, 
\end{equation}
with the sum running over all $n_s$ point group symmetry operations $\{s\}$
acting on the direct lattice ${\bf{R}}_m$. Here $M$ is the number of star functions per ${\bf{k}}$ point and
$a_m$ are the Fourier coefficients
of the expansion of the band structure in terms of star functions.

%\bibliography{Ref2.bib}

\cleardoublepage

\newpage
\section{Additional figures}
%Additional figures has been added to this Supplementary Information 

\renewcommand{\thefigure}{S1}
\begin{figure*}[h!]
\centering
\includegraphics[width=1\textwidth]{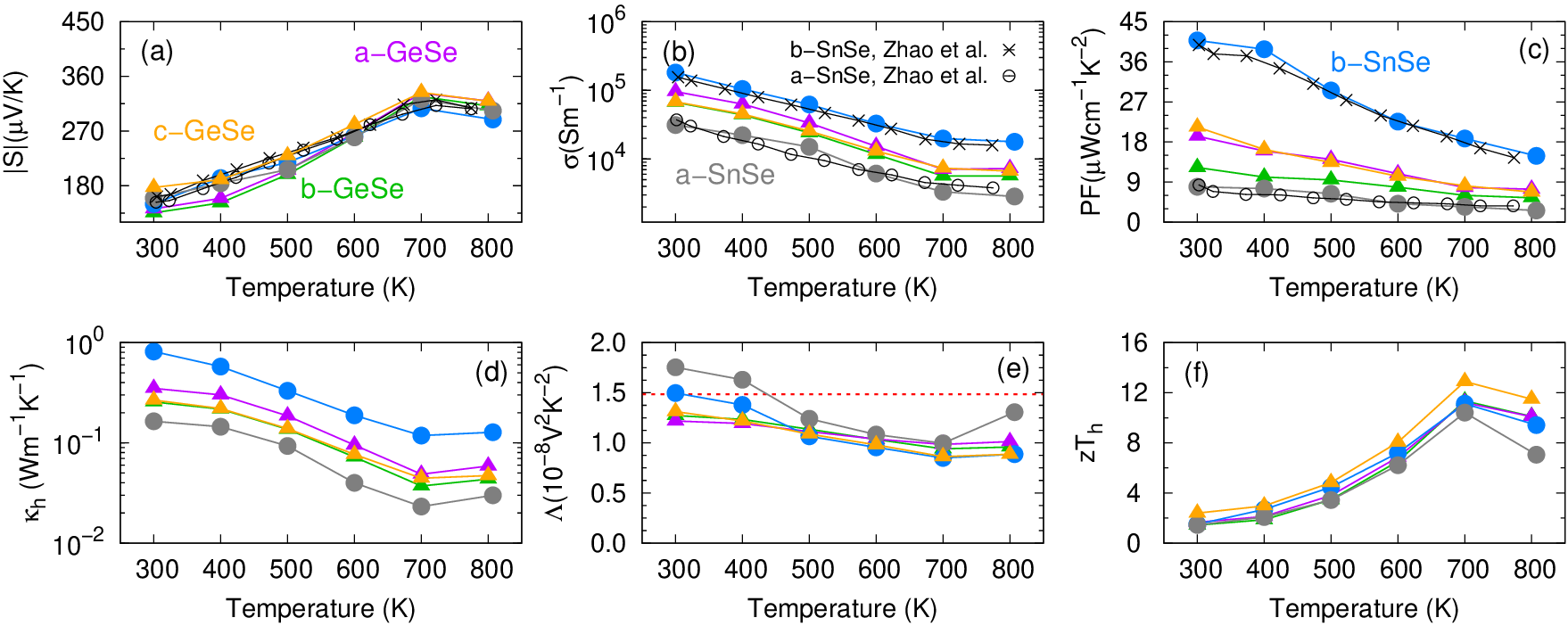}
\vspace*{6mm}\caption{\label{figureS1}
                 The same as Fig.~3 but including c-GeSe. Calculated thermoelectric transport properties of p-doped GeSe (all axes) and SnSe (a- and b-axis) 
                 as functions of temperature,
                 along with available experimental data on p-doped SnSe reported
                 by Zhao \textit{et al.}\cite{zhao2015ultrahigh}
                 The GeSe transport properties were calculated by using the temperature dependent 
                 carrier density and the ionized impurities concentrations derived from a-axis p-doped SnSe, also reported
                 by Zhao \textit{et al.}\cite{zhao2015ultrahigh}
                 $(a)$ Seebeck coefficient $S$, $(b)$ electrical conductivity, $\sigma$, $(c)$ power factor, PF,
                 $(d)$ thermal conductivity due to the hole transport, ${\kappa}_\mathrm{h}$, $(e)$ Lorenz function, $\Lambda$,
                 where the horizontal dashed red line indicates the non-degenerate limit for semiconductors,
                 $\Lambda_\mathrm{nd} = 1.485\times 10^{-8}$V$^2$K$^{-2}$,
                 and $(f)$ the hole figure of merit, zT$_\mathrm{h}$. }
\end{figure*}

\renewcommand{\thefigure}{S2}
\begin{figure*}[h!]
\centering
\includegraphics[width=1\textwidth]{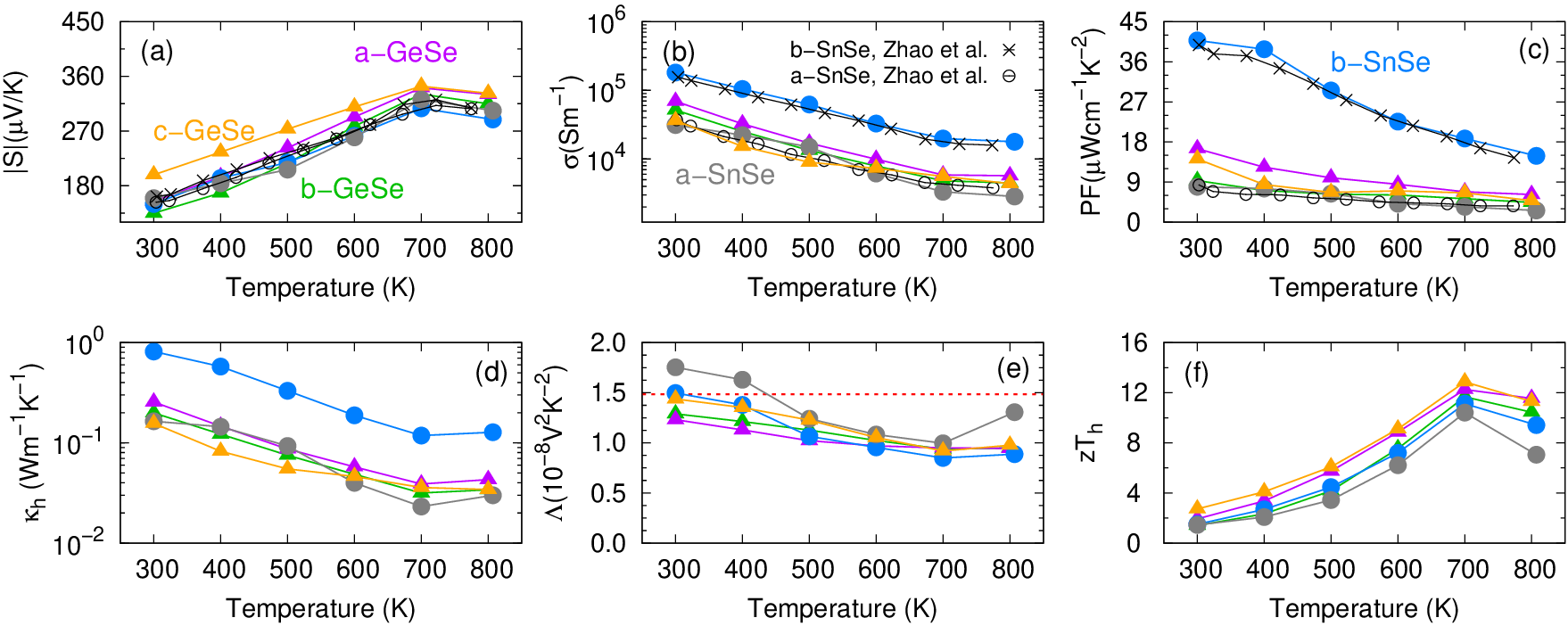}
\vspace*{6mm}\caption{\label{figureS2}
                 Calculated thermoelectric transport properties of p-doped GeSe (all axes) and SnSe (a- and b-axis) 
                 as functions of temperature,
                 along with available experimental data on p-doped SnSe reported
                 by Zhao \textit{et al.}\cite{zhao2015ultrahigh}
                 The GeSe transport properties were calculated by using the temperature dependent
                 carrier density and the ionized impurities concentrations derived from b-axis p-doped SnSe, also reported
                 by Zhao \textit{et al.}\cite{zhao2015ultrahigh}
                 $(a)$ Seebeck coefficient $S$, $(b)$ electrical conductivity, $\sigma$, $(c)$ power factor, PF,
                 $(d)$ thermal conductivity due to the hole transport, ${\kappa}_\mathrm{h}$, $(e)$ Lorenz function, $\Lambda$,
                 where the horizontal dashed red line indicates the non-degenerate limit for semiconductors,
                 $\Lambda_\mathrm{nd} = 1.485\times 10^{-8}$V$^2$K$^{-2}$,
                 and $(f)$ the hole figure of merit, zT$_\mathrm{h}$. }
\end{figure*}

\renewcommand{\thefigure}{S3}
\begin{figure*}[h!]
\centering
\includegraphics[width=1\textwidth]{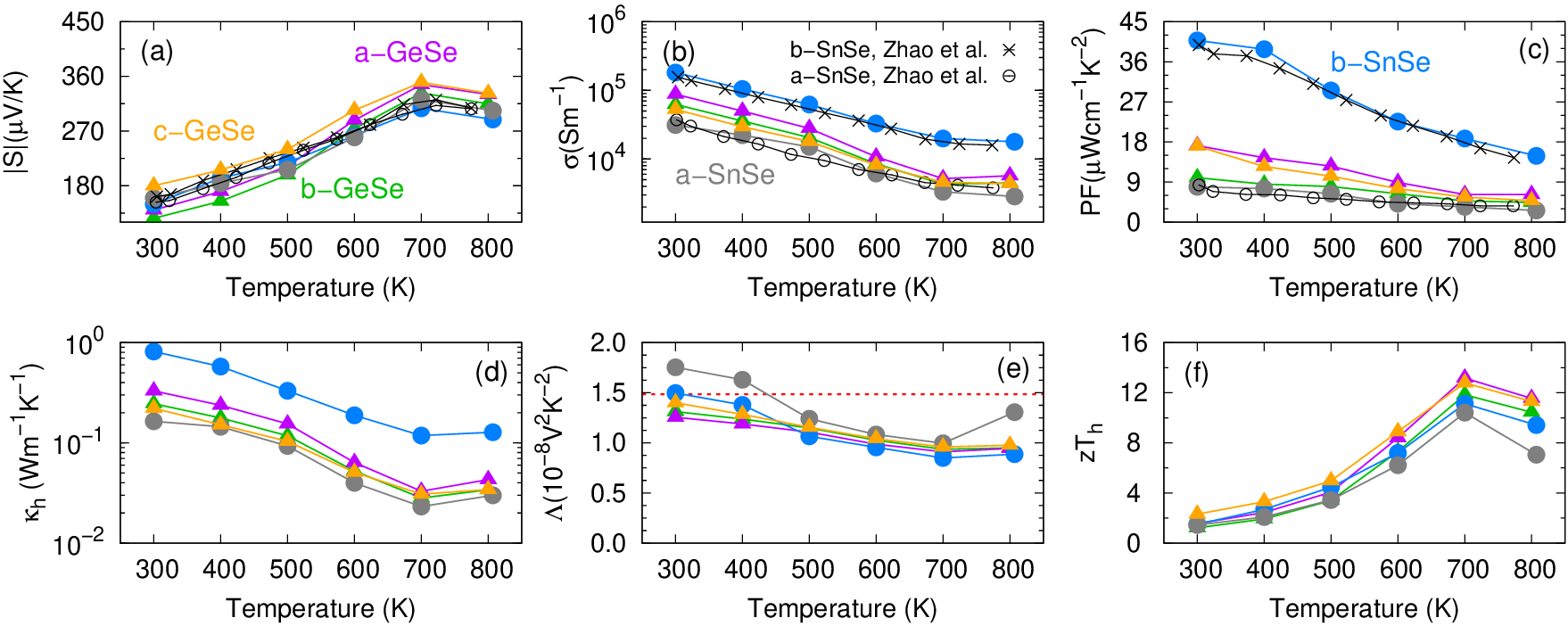}
\vspace*{6mm}\caption{\label{figureS3}
                 Calculated thermoelectric transport properties of p-doped GeSe (all axes) and SnSe (a- and b-axis)
                 as functions of temperature,
                 along with available experimental data on p-doped SnSe reported
                 by Zhao \textit{et al.}\cite{zhao2015ultrahigh}
                 The GeSe transport properties were calculated by using the temperature dependent
                 carrier density and the ionized impurities concentrations derived from c-axis p-doped SnSe, also reported
                 by Zhao \textit{et al.}\cite{zhao2015ultrahigh}
                 $(a)$ Seebeck coefficient $S$, $(b)$ electrical conductivity, $\sigma$, $(c)$ power factor, PF,
                 $(d)$ thermal conductivity due to the hole transport, ${\kappa}_\mathrm{h}$, $(e)$ Lorenz function, $\Lambda$,
                 where the horizontal dashed red line indicates the non-degenerate limit for semiconductors,
                 $\Lambda_\mathrm{nd} = 1.485\times 10^{-8}$V$^2$K$^{-2}$,
                 and $(f)$ the hole figure of merit, zT$_\mathrm{h}$. }
\end{figure*}

\renewcommand{\thefigure}{S4}
\begin{figure*}[h!]
\centering
\includegraphics[width=0.5\textwidth]{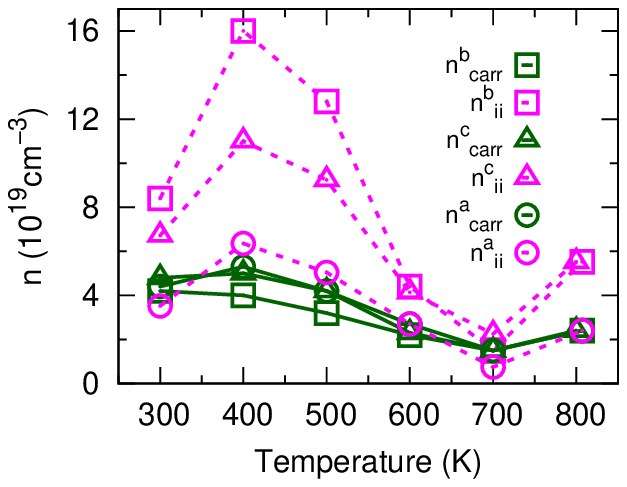}
\vspace*{6mm}\caption{\label{figureS4}
Carrier density, $n_\mathrm{carr}$ (green solid line), and ionized impurity concentration, $n_\mathrm{ii}$ (dashed magenta line), determined by
self-consistently matching the experimental results for a-, b- and c-axis p-doped SnSe reported in Zhao et al.\cite{zhao2015ultrahigh}}
\end{figure*}

\renewcommand{\thefigure}{S5}
\begin{figure*}[h!]
\centering
\includegraphics[width=1\textwidth]{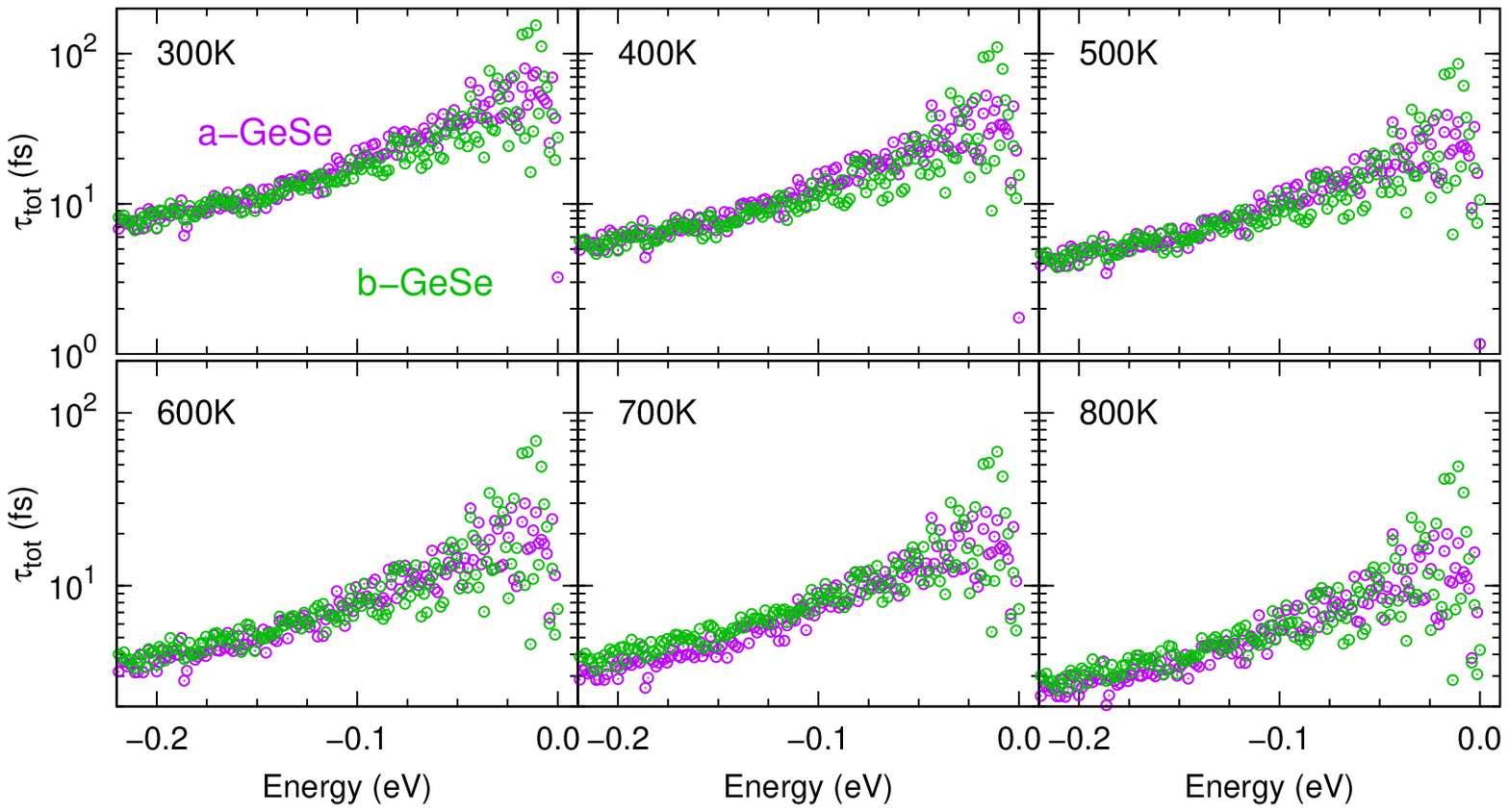}
\vspace*{6mm}\caption{\label{figureS5}
Temperature dependence ($T =$ 300-800~K) of the total relaxation time ($\tau_\mathrm{tot}$) as a function of the hole  
energies for out-of-plane (a-axis) and in-plane (b-axis) p-doped GeSe.}
\end{figure*}

\renewcommand{\thefigure}{S6}
\begin{figure*}[h!]
\centering
\includegraphics[width=1\textwidth]{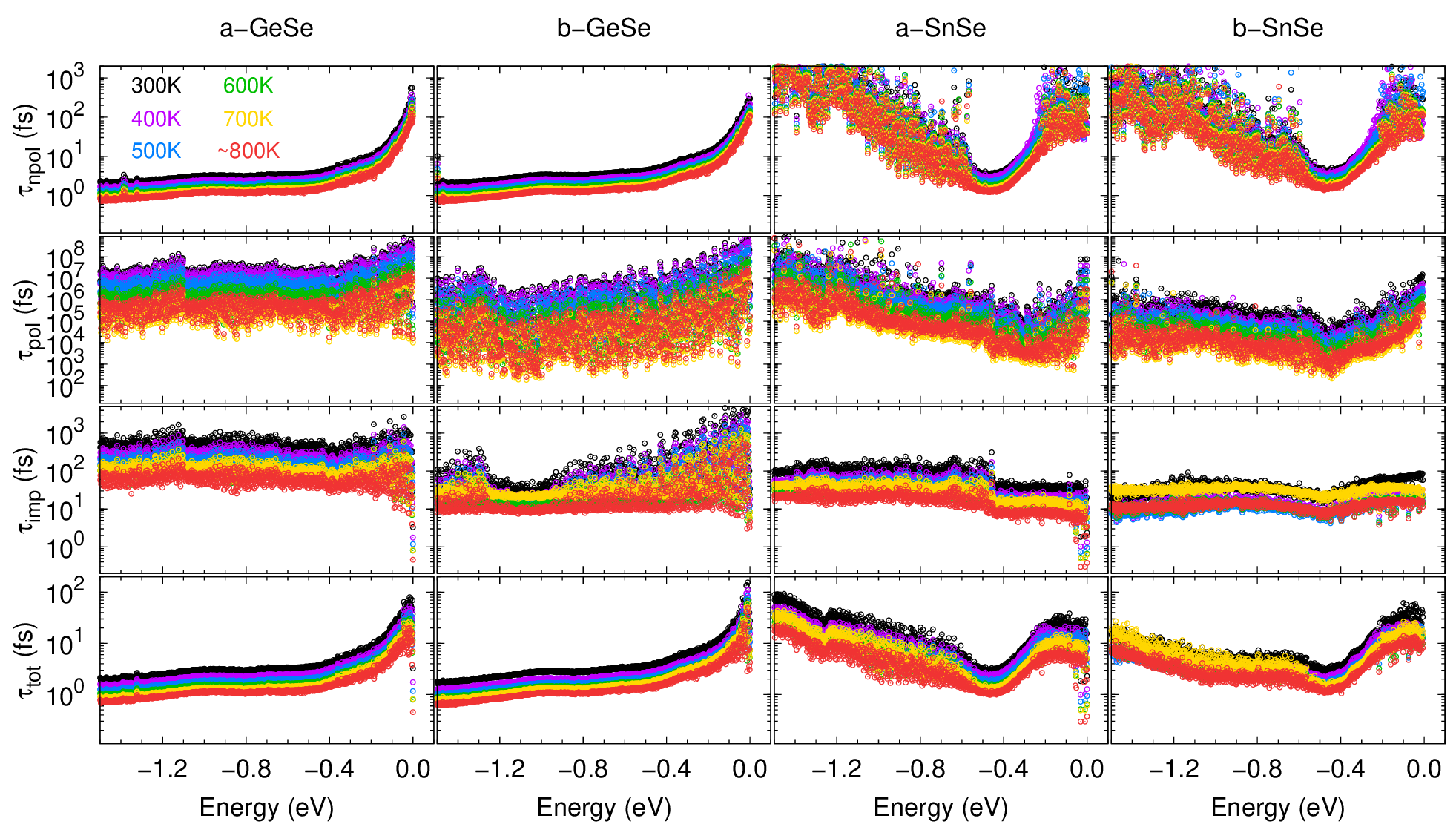}
\vspace*{6mm}\caption{\label{figureS6}
Temperature dependence of the relaxation time 
as a function of hole energy for each of the scattering mechanisms: nonpolar scattering by phonons ($\tau_\mathrm{npol}$),
screened polar scattering by optical phonons ($\tau_\mathrm{pol}$), screened scattering by ionized impurities ($\tau_\mathrm{imp}$) 
and total scattering ($\tau_\mathrm{tot}$). Results are shown for p-doped GeSe (a- and b-axis) and p-doped SnSe (a- and b-axis). The hole density was derived 
from experimental transport data for the a-axis of p-doped SnSe as reported by Zhao \textit{et al.}\cite{zhao2015ultrahigh}, as explained in the main text.}
\end{figure*}

\renewcommand{\thefigure}{S7}
\begin{figure*}	
	\centering
	\includegraphics[width=1\textwidth]{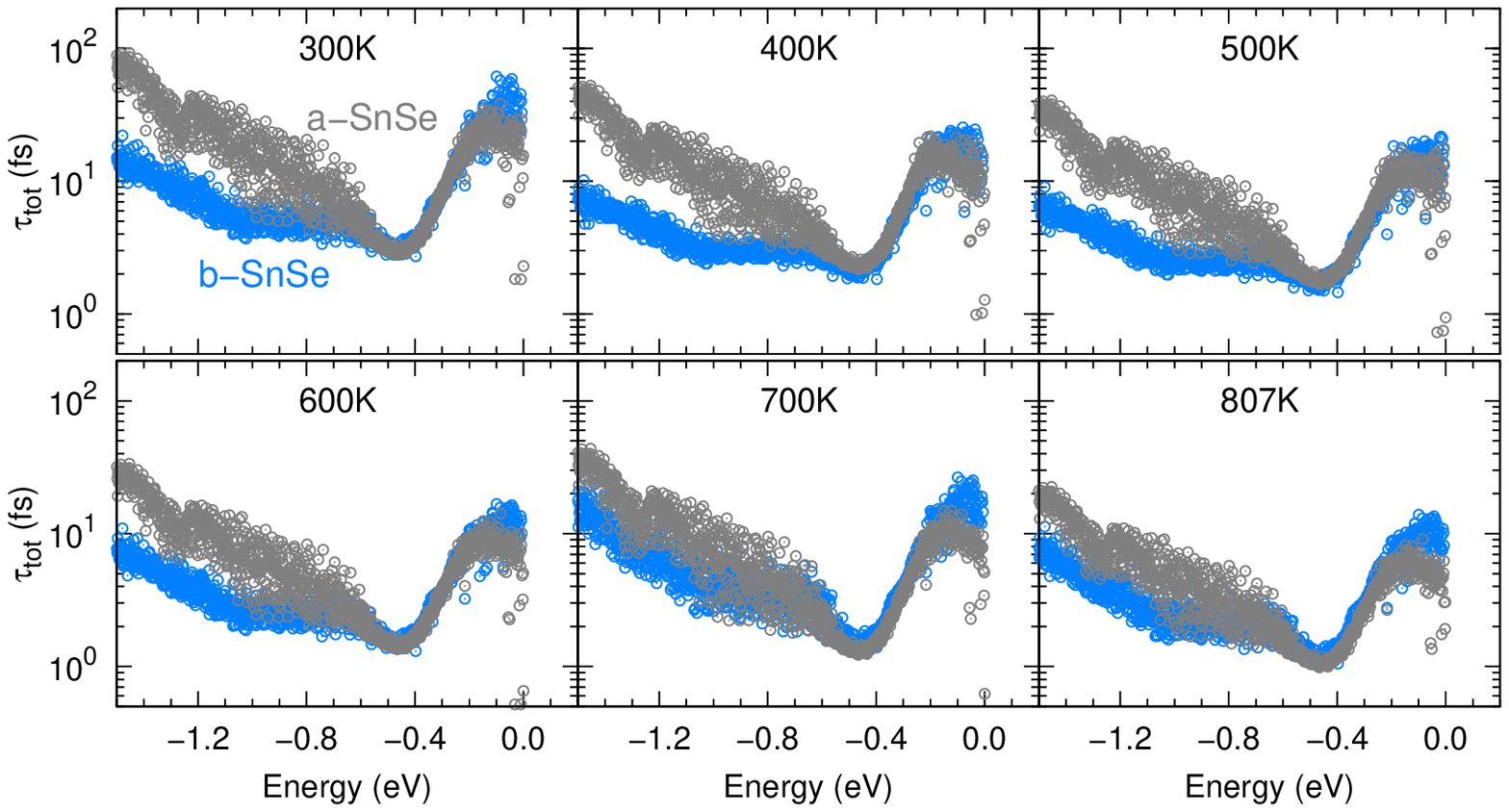}
	\vspace*{6mm}\caption{\label{figureS7}
         Total relaxation time ($\tau_\mathrm{tot}$) as a function of hole
         energy for out-of-plane (a-axis) and in-plane (b-axis) p-doped SnSe at six temperatures between 300 and 807~K.}
\end{figure*}

\renewcommand{\thefigure}{S8}
\begin{figure*}
\centering
\includegraphics[width=1\textwidth]{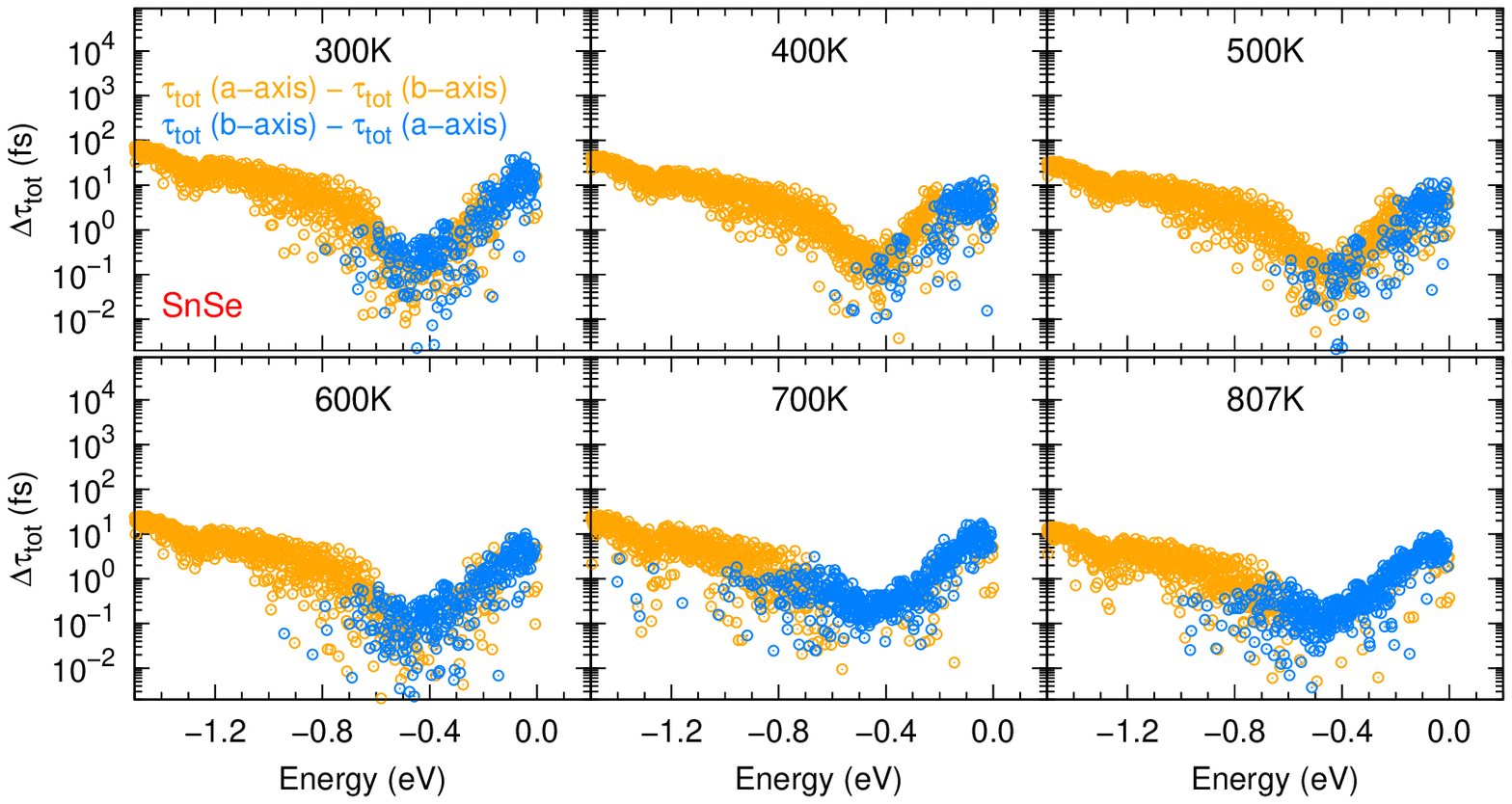}
\vspace*{6mm}\caption{\label{figureS8}
                Absolute value of the difference in total relaxation time, $\Delta \tau_\mathrm{tot}$, between out-of-plane (a-axis)
                and in-plane (b-axis) p-doped SnSe as a function of hole energy for six different temparatures. Yellow (blue) points indicate that the a-axis RT is larger (smaller) than the b-axis RT.}
\end{figure*}

\renewcommand{\thefigure}{S9}
\begin{figure*}
        \centering
        \includegraphics[width=1\textwidth]{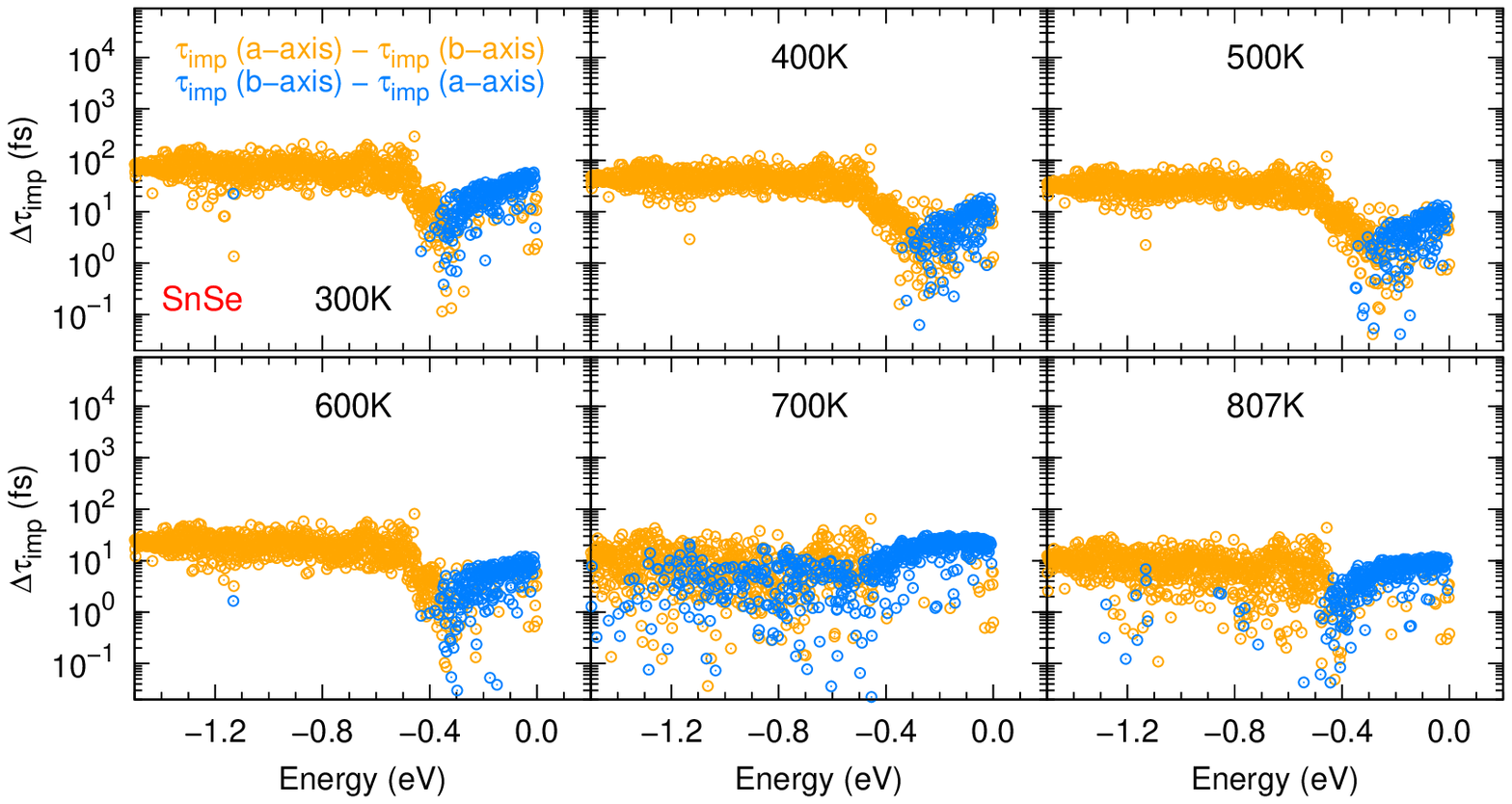}
        \vspace*{6mm}\caption{\label{figureS9}
        Absolute value of the difference in relaxation time due to impurity scattering, $\Delta \tau_\mathrm{imp}$, between out-of-plane (a-axis)
        and in-plane (b-axis) p-doped SnSe as a function of hole energy for six different temparatures. Yellow (blue) points indicate that the a-axis RT is larger (smaller) than the b-axis RT.
}
\end{figure*}

\renewcommand{\thefigure}{S10}
\begin{figure*}
        \centering
        \includegraphics[width=1\textwidth]{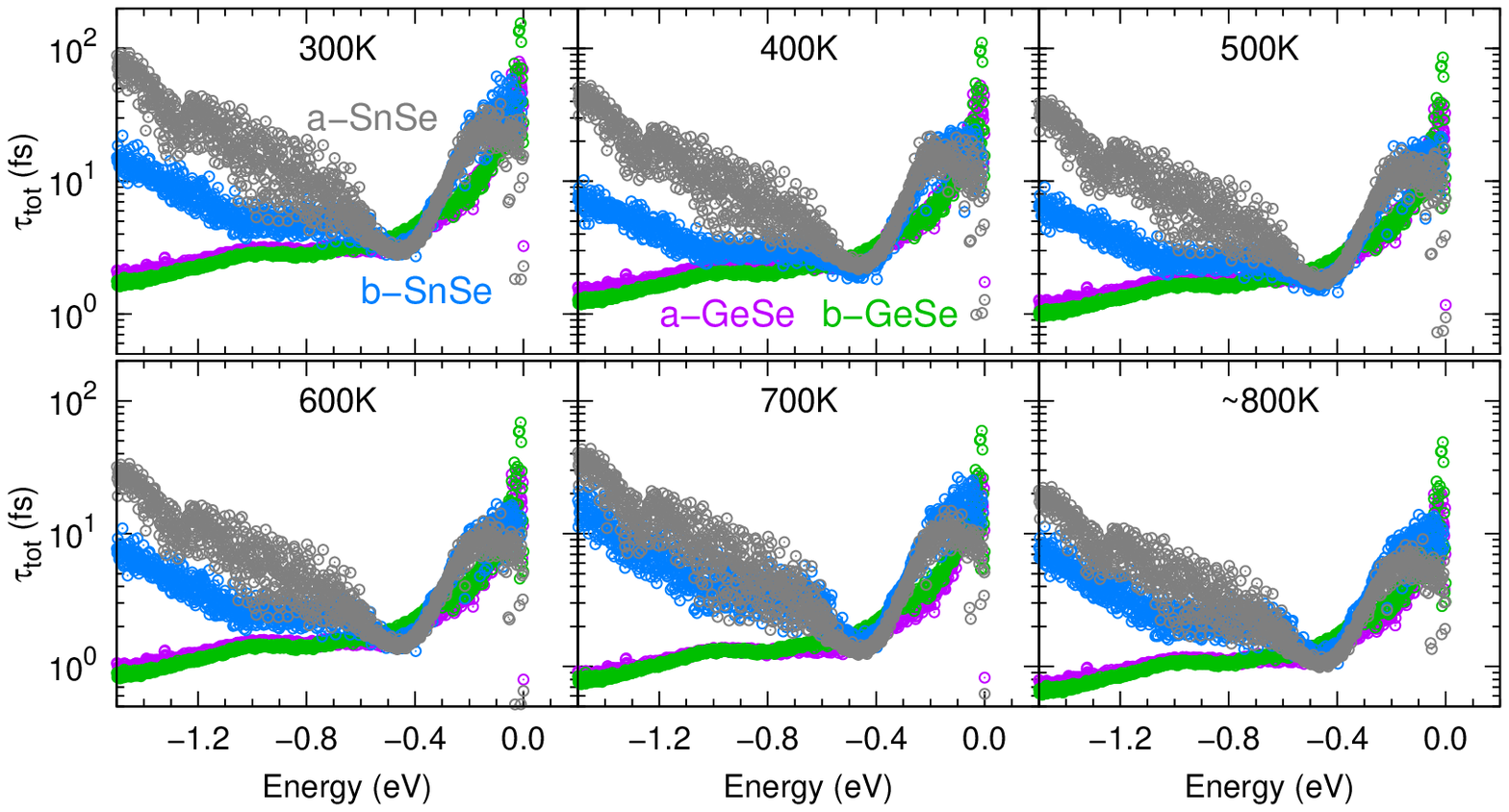}
        \vspace*{6mm}\caption{\label{figureS10}
        Total relaxation time ($\tau_\mathrm{tot}$) as a function of hole
        energy for p-doped GeSe (a- and b-axis) and p-doped SnSe (a- and b-axis), for six different temperatures between 300 and 800~K. The hole density was derived
        from experimental transport data for the a-axis of p-doped SnSe as reported by Zhao \textit{et al.}\cite{zhao2015ultrahigh}}
\end{figure*}

\renewcommand{\thefigure}{S11}
\begin{figure*}
        \centering
        \includegraphics[width=1\textwidth]{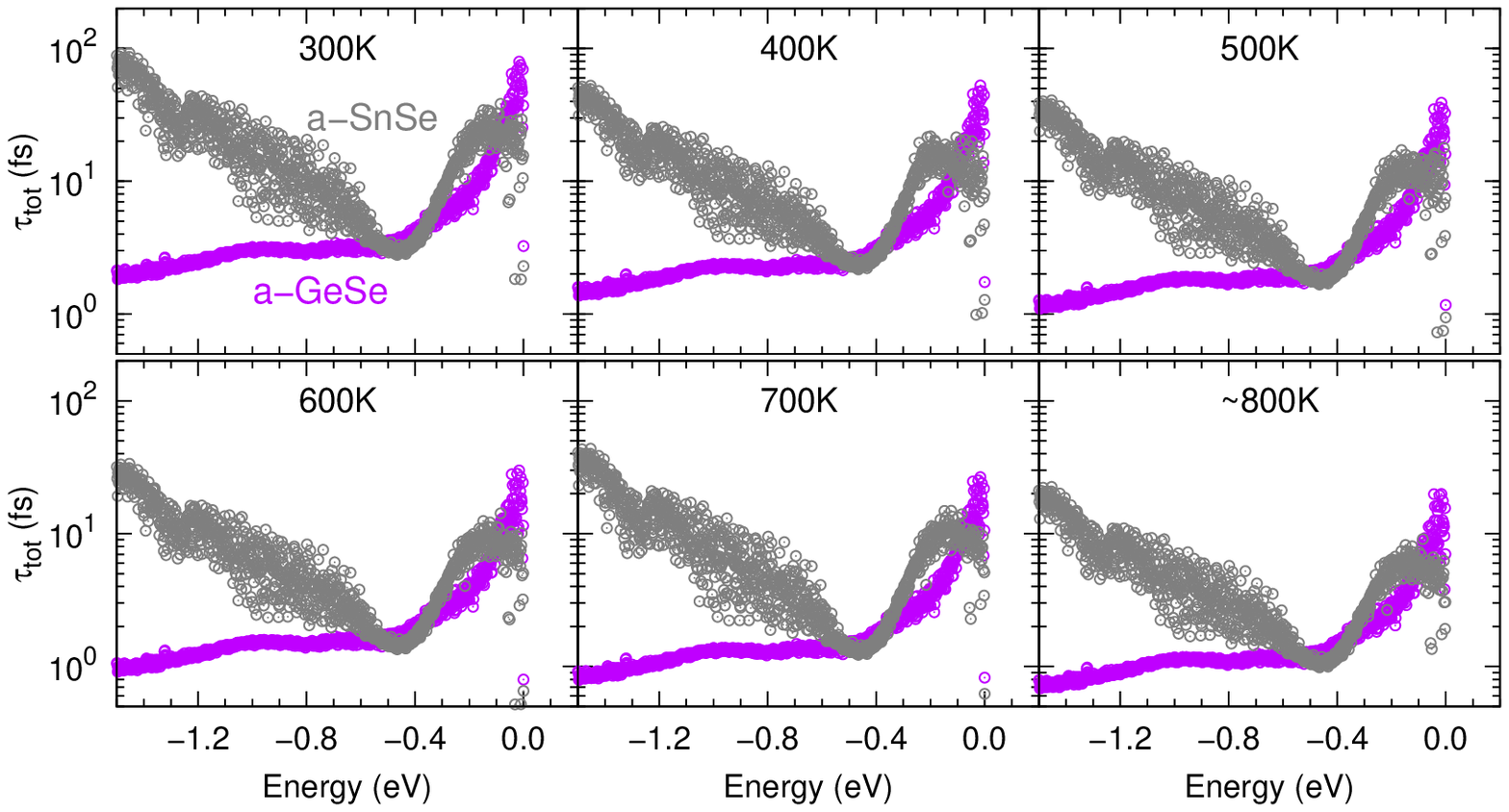}
        \vspace*{6mm}\caption{\label{figureS11}
        Total relaxation time ($\tau_\mathrm{tot}$) as a function of hole
        energy for out-of-plane (a-axis) p-doped GeSe and SnSe for six different temperatures between 300 and 800~K.}
\end{figure*}

\renewcommand{\thefigure}{S12}
\begin{figure*}
        \centering
        \includegraphics[width=1\textwidth]{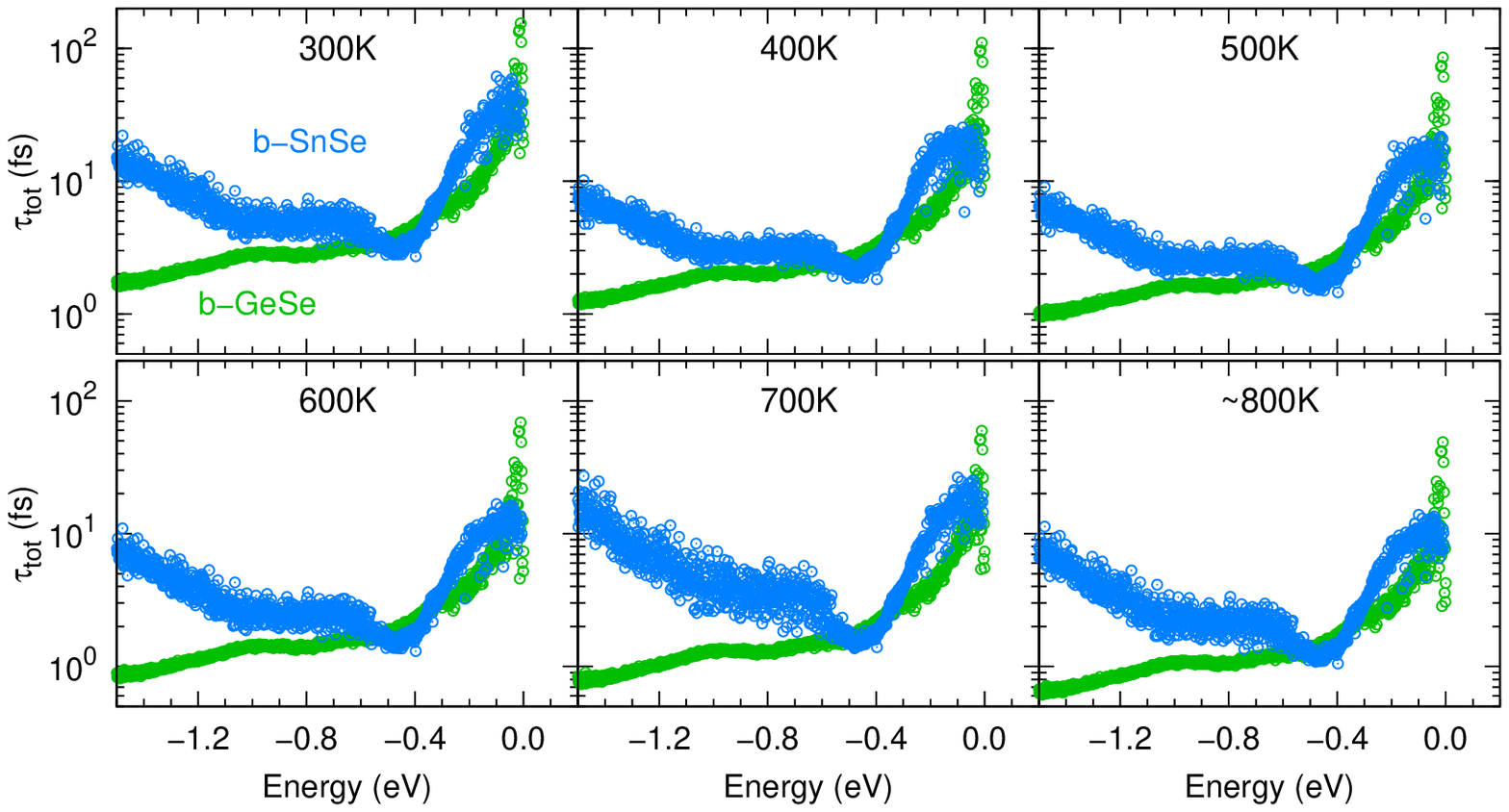}
        \vspace*{6mm}\caption{\label{figureS12}
        Total relaxation time ($\tau_\mathrm{tot}$) as a function of hole
        energy for in-plane (b-axis) p-doped GeSe and SnSe for six different temperatures between 300 and 800~K.}
\end{figure*}

\renewcommand{\thefigure}{S13}
\begin{figure*}
        \centering
        \includegraphics[width=1\textwidth]{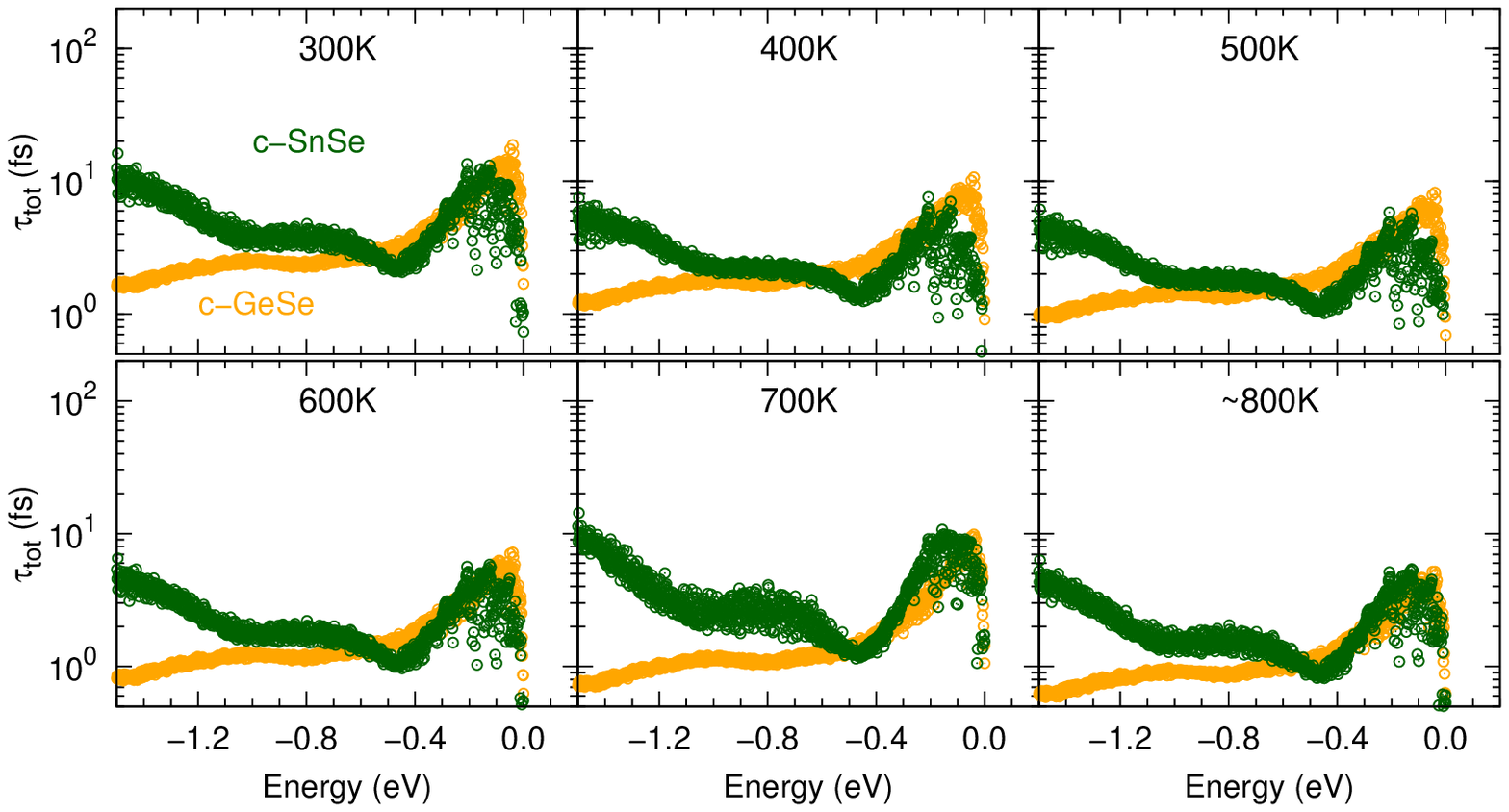}
        \vspace*{6mm}\caption{\label{figureS13}
        Total relaxation time ($\tau_\mathrm{tot}$) as a function of hole
        energy for in-plane (c-axis) p-doped GeSe and SnSe for six different temperatures between 300 and 800~K.}
\end{figure*}

\renewcommand{\thefigure}{S14}
\begin{figure*}
        \centering
        \includegraphics[width=0.8\textwidth]{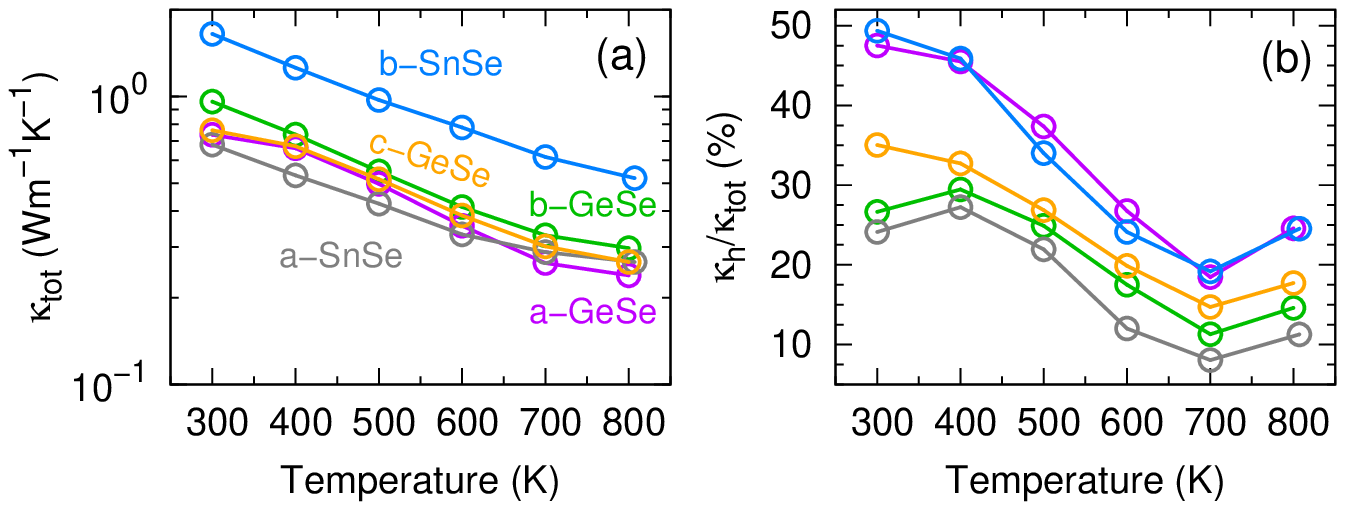}
        \vspace*{6mm}\caption{\label{figureS14}
        (a) Total thermal conductivity, $\kappa_\mathrm{tot}$, as a function of temperature for 
        p-doped GeSe (all axes) and SnSe (a- and b-axis). The lattice thermal conductivities were 
        derived from Hao et al.\cite{hao2016computational} (b) Fraction of the total thermal conductivity $\kappa_\mathrm{tot}$ contributed by holes, as a function of temperature.}
\end{figure*}

\renewcommand{\thefigure}{S15}
\begin{figure*}[h!]
\centering
\includegraphics[width=0.6\textwidth]{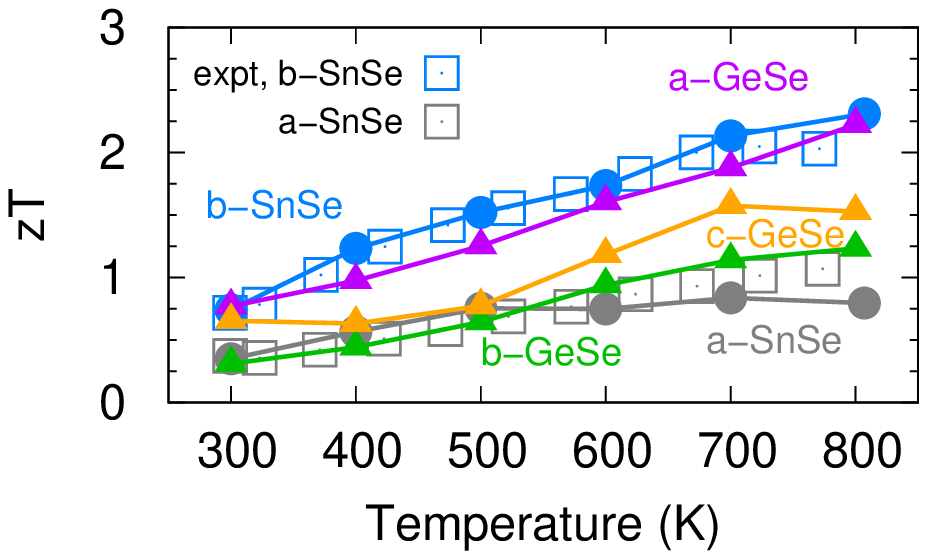}
\vspace*{6mm}\caption{\label{figureS16}
        Thermoelectric figure of merit ($zT$) as a function of temperature for p-doped GeSe (all axes) and SnSe (a- and b-axis).
        The experimental data reported by Zhao \textit{et al.}\cite{zhao2015ultrahigh} for both axes of p-doped
        SnSe are shown for comparison. The GeSe transport properties were calculated using the temperature dependent
        carrier density and ionized impurity concentration derived from b-axis p-doped SnSe, also reported
        by Zhao \textit{et al.}\cite{zhao2015ultrahigh}
}
\end{figure*}

\renewcommand{\thefigure}{S16}
\begin{figure*}[h!]
\centering
\includegraphics[width=0.6\textwidth]{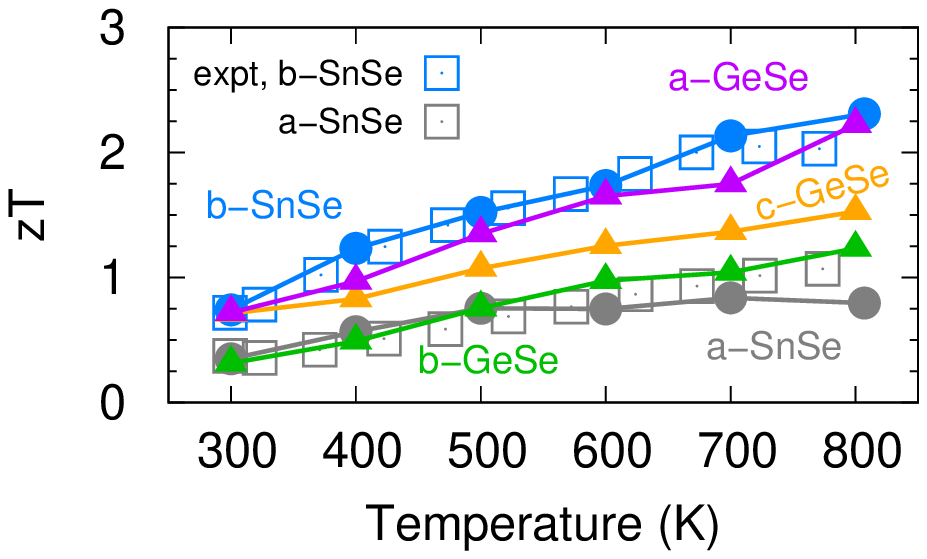}
\vspace*{6mm}\caption{\label{figureS17}
        Thermoelectric figure of merit ($zT$) as a function of temperature for p-doped GeSe (all axes) and SnSe (a- and b-axis).
        The experimental data reported by Zhao \textit{et al.}\cite{zhao2015ultrahigh} for both axes of p-doped
        SnSe are shown for comparison. The GeSe transport properties were calculated using the temperature dependent
        carrier density and ionized impurity concentration derived from c-axis p-doped SnSe, also reported
        by Zhao \textit{et al.}\cite{zhao2015ultrahigh}
}
\end{figure*}

\renewcommand{\thefigure}{S17}
\begin{figure*}
        \centering
        \includegraphics[width=1\textwidth]{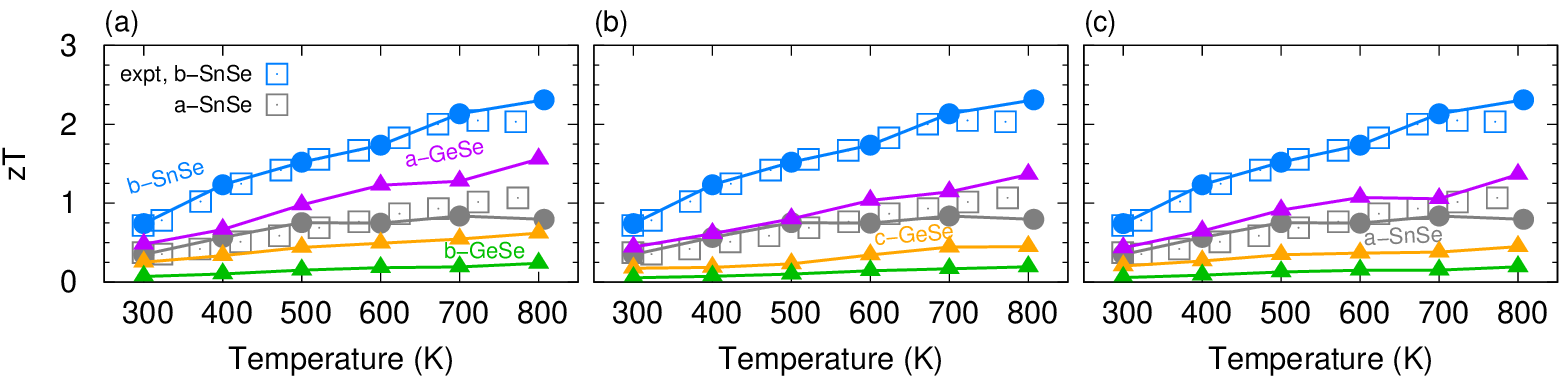}
        \vspace*{6mm}\caption{\label{figureS15}
        Thermoelectric figure of merit ($zT$) as a function of temperature for p-doped GeSe (all axes) and SnSe (a- and b-axis). The results
        for GeSe use values of $\kappa_\mathrm{latt}$ calculated by Yuan et al.\cite{yuan2019tailoring}
        The experimental data reported by Zhao \textit{et al.}\cite{zhao2015ultrahigh} for both axes of p-doped
        SnSe are shown for comparison. The GeSe transport properties were calculated using the temperature dependent
        carrier density and ionized impurity concentration derived from (a) a-axis (b) b-axis and (c)
        c-axis p-doped SnSe, also reported by Zhao \textit{et al.}\cite{zhao2015ultrahigh}}
\end{figure*}

\renewcommand{\thefigure}{S18}
\begin{figure*}
        \centering
        \includegraphics[width=0.45\textwidth]{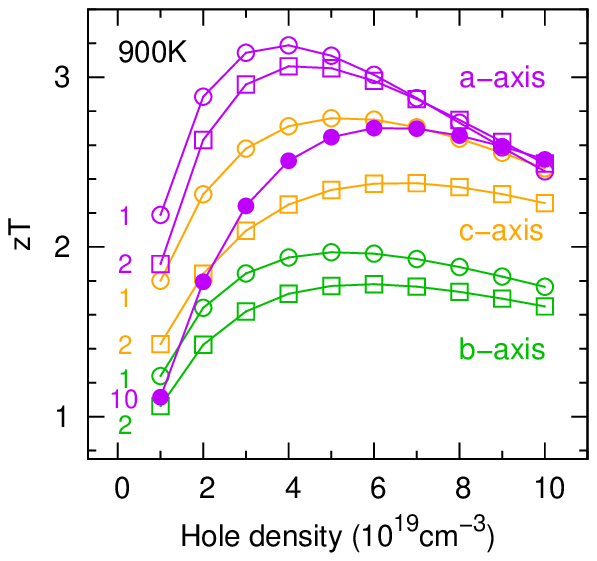}
        \vspace*{6mm}\caption{\label{figureS18}
         Thermoelectric figure of merit, $zT$, as a function of hole density for fixed ratios $n_\mathrm{ii}/n_\mathrm{carr}= 1$, 2, 10, which label the left side of each curve corresponding to the a-, b- and c-axis of
         p-doped GeSe at 900~K.}

\end{figure*}

\cleardoublepage

\bibliography{Ref2.bib}

%\bibliographystyle{apsrev4-1} 
%\bibliography{Ref2.bib}
%{\footnotesize
%\bibliography{Ref.bib}%}